\documentclass[a4paper,12pt]{article}

\usepackage{lineno}
\usepackage[auth-sc]{authblk}

\usepackage{amsmath,amssymb,amsthm} 	% mathmode
\usepackage{accents}
\usepackage[nolist]{acronym} 			% abbreviations
\usepackage{adjustbox}
\usepackage{booktabs}					% tables
\usepackage{caption}
\usepackage{comment}					% for commenting entire blocks
\usepackage{color,soul}
\usepackage{crimson}                    % Crimson font
\usepackage[at]{easylist}               % multi level listing with 1, 1a, etc.
\usepackage[official]{eurosym}			% for € sign
\usepackage{tabularx}					% tables align with page width
\usepackage{float}
\usepackage{hyperref}
\usepackage{lscape}
\usepackage{tablefootnote} 				% for table footnotes
\usepackage{scrextend}  				% for table footnotes
\usepackage{setspace}
\usepackage[superscript,biblabel]{cite} % make subscript citations
\usepackage{subfig}						% for subfloats
\usepackage{tikz}						% to compile tree figures
\usetikzlibrary{trees}					% predefined functionality for organizational charts
\usepackage{url}
\usepackage[lof,lot]{etoc}

\usepackage[long]{optidef}

\makeatletter
\newcounter{manualsubequation}
\renewcommand{\themanualsubequation}{\alph{manualsubequation}}
\newcommand{\startsubequation}{%
  \setcounter{manualsubequation}{0}%
  \refstepcounter{equation}\ltx@label{manualsubeq\theequation}%
  \xdef\labelfor@subeq{manualsubeq\theequation}%
}
\newcommand{\tagsubequation}{%
  \stepcounter{manualsubequation}%
  \tag{\ref{\labelfor@subeq}\themanualsubequation}%
}
\let\subequationlabel\ltx@label
\makeatother

\usepackage{anysize}
\marginsize{2.5cm}{2.4cm}{1.5cm}{2cm}

\usepackage{parskip}

\begin{acronym}
 \acro{DIETER}{Dispatch and Investment Evaluation Tool with Endogenous Renewables}
 \acro{DACCS}{direct air carbon capture and storage }
 \acro{ERAA}{European Resource Adequacy Assessment}
 \acro{PV}{photovoltaics}
 \acro{TYNDP}{Ten Year Network Development Plan}
 \acro{VRE}{variable renewable energy}
 \acro{VREDA}{Variable Renewable Energy Drought Analyzer}
 \acro{VMBT}{varible-duration Mean-Below-Threshold}
\end{acronym}

\title{Long-duration electricity storage needs for coping with Dunkelflaute events in Europe}
\author[1,2,*]{Martin Kittel}
\author[1,3]{Alexander Roth}
\author[1]{Wolf-Peter Schill}
\affil[1]{DIW Berlin, Department of Energy, Transportation, Environment, Mohrenstra{\ss}e 58, 10117 Berlin, Germany}
\affil[2]{Technical University Berlin, Digital Transformation in Energy Systems, Einsteinufer 25 (TA 8), 10587 Berlin, Germany}
\affil[3]{Bruegel, Rue de la Charité 33, 1210 Saint-Josse-ten-Noode, Belgium}
\affil[*]{Corresponding author: \url{mkittel@diw.de}}

\date{}

\begin{document}

%\begin{bibunit}
\sloppy

\maketitle
\thispagestyle{empty}% Reset page style to 'empty'

\begin{abstract}

% NATURE GUIDELINES: 150 words or fewer. Begin with the background and rationale for the work. The final sentence must begin with a phrase like “In this work” or “Here, we show”, and contain a brief summary of the  major results and conclusions of the current work, written in  the present tense. Do not include references, acronyms or abbreviations. We do not allow a graphical abstract.

\noindent Coping with prolonged periods of low availability of wind and solar power, also referred to as variable renewable energy droughts or ``Dunkelflaute'', emerges as a key challenge for realizing decarbonized energy systems based on renewable energy. Here we investigate the role of long-duration electricity storage and geographical balancing through transmission in dealing with such events in Europe, combining a time series analysis of renewable availability with power sector modeling of 35~historical weather years. We find that extreme droughts define long-duration storage operation and investment. Assuming policy-relevant interconnection, the least-cost system in our model capable of coping with the most extreme event requires 351~terawatt hours long-duration storage capacity, corresponding to 7\% of yearly European electricity demand. While nuclear power can partially reduce storage needs, the storage-mitigating effect of fossil backup plants in combination with carbon removal is limited. Policymakers and system planners should prepare for a rapid expansion of long-duration storage to safeguard the renewable energy transition in Europe.
\end{abstract}

%\begin{keyword}
%variable renewable energy \sep variable renewable energy droughts \sep long-duration storage \sep power sector modeling

%\MSC[2010] 62H30
%\end{keyword}

%\end{frontmatter}
%%%%%%%%%%%%%%%%%%%%%%%%%%%%%%%%%%%%%%%%%%%%%%%%%%%%%%%

%\onehalfspacing

\newpage

% all links to SI are corrupted

%%%%%%%%%%%%%%%%%%%%%%%%%%%%%%%%%%%%%%%%%%%%%%%%%%%%%%%
\section*{Introduction}
%%%%%%%%%%%%%%%%%%%%%%%%%%%%%%%%%%%%%%%%%%%%%%%%%%%%%%%

% Motivation: Climate neutrality \& increasing importance of power sector, %Importance of RES, % Droughts
To mitigate climate change and to meet international commitments, the European Union aims to achieve net zero emissions of greenhouse gases by 2050. The power sector will play a central role in a decarbonized economy. Massively expanding renewable energy sources would allow for a rapid substitution of fossil fuels in the power sector as well as in the industry, transport, and heating sector via electrification, a widely recognized decarbonization strategy also referred to as ``sector coupling'' \cite{jacobson_100_2017,de_coninck_global_2018,davis_net-zero_2018,bogdanov_full_2021,IPCC_2022_WGIII_Ch_2}. The potentials of firm renewable energy sources, such as hydro- or bioenergy, are limited in most European countries. In contrast, wind and solar power have vast expansion potentials and promise declining costs \cite{way_empirically_2022,lee_ipcc_2023}. Therefore, these \ac{VRE} technologies are likely to form the backbone of transitioning to net zero in most European countries \cite{jacobson_100_2017,brown_response_2018}. With a rising share of \ac{VRE}, the European power sector becomes increasingly exposed to weather variability \cite{staffell_increasing_2018,collins_impacts_2018}. This has spurred a debate in the energy policy domain about the security of supply \cite{dawkins_weather_2019,dawkins_characterising_2020,huneke_f_kalte_2017,dhu_hintergrundpapier_2021,deutscher_bundestag_sicherstellung_2019,wood_go_2021,novacheck_evolving_2021,entso-e_tyndp_2022,the_economist_defying_2023,bloomberg_europes_2024}. Of particular concern are long-lasting extreme weather events, also referred to as ``\ac{VRE} droughts'' or ``Dunkelflaute'', which are characterized by an overall very low availability of \ac{VRE} sources \cite{raynaud_energy_2018,kittel_measuring_2024,van_der_most_temporally_2024}.

% Long-duration storage, % Interconnection
Previous research has shown that for growing shares of \ac{VRE} increasing flexibility would not only enable a least-cost energy system, but also required to deal with imbalances in the power sector \cite{paulus_potential_2011,zerrahn_long-run_2017,shaner_geophysical_2018,brown_synergies_2018,child_flexible_2019,victoria_early_2020}. This can be provided by different types of electricity storage, demand response, or cross-border transmission. Spatial flexibility allows balancing of regional variations in demand and variable renewable supply by interconnecting countries via electricity and hydrogen grids \cite{fursch_role_2013,schaber_parametric_2012}. In Europe, such cross-border interconnection can mitigate energy storage needs, particularly by balancing differences in wind power availability across countries \cite{roth_geographical_2023}. Yet, it is unclear to what extent this storage-mitigating effect of geographical balancing persists during extreme pan-European \ac{VRE} droughts of synoptic scale affecting many countries simultaneously~\cite{kittel_multi-threshold_2026}. Besides interconnection, a wide range of flexibility technologies can help to cope with \ac{VRE} droughts. Particularly long-duration storage, sometimes also referred to as ``ultra-long-duration storage'' in the literature \cite{brown_ultra_2023}, is a central option as it can shift renewable surplus energy to periods of low wind and solar availability over long time scales, i.e.,~from several days up to seasons \cite{schill_2020,dowling_role_2020,brown_ultra_2023}. Low energy capital costs are key to make long-duration storage viable in energy systems that heavily rely on wind and solar power \cite{ziegler_2019,albertus_long-duration_2020,sepulveda_design_2021,chu_2025}. From a techno-economic perspective, the production of hydrogen or its derivatives using renewable electricity, followed by storage and reconversion to electricity, currently appears most promising for enabling long-duration energy storage \cite{iea_future_2019,victoria_role_2019,dowling_role_2020,hunter_techno-economic_2021}.

% Teaser der research questions
In this paper, we analyze how \ac{VRE} droughts impact the operation and sizing of long-duration storage in a fully renewable European power sector. We further investigate how varying levels of geographical balancing via interconnection affect long-duration storage investments in the least-cost system. We also examine how different technologies interact during extreme \ac{VRE} droughts and assess the role of nuclear power and fossil fuel-based backup capacity with emission abatement.

% Literature on VRE droughts, PRL events, impact of \ac{VRE} droughts on energy system
While there is growing research interest in \ac{VRE} droughts and their power sector implications, respective analyses vary in subject as well as temporal and spatial scope~\cite{kittel_measuring_2024}. One strand of the literature focuses on \ac{VRE} drought characterization based on time series of wind speeds, solar irradiation, or normalized renewable availability factors, also referred to as capacity factors. This includes analyses of historic or future wind droughts for Germany \cite{ohlendorf_frequency_2020}, the North Sea \cite{patlakas_low_2017}, Ireland \cite{leahy_persistence_2013}, or the UK \cite{cannon_using_2015,potisomporn_spatial_2022,potisomporn_evaluating_2023,abdelaziz_assessing_2024}. These studies typically focus on frequency-duration distributions, return periods, or spatio-temporal correlations. Specifically, wind droughts have been studied as \ac{VRE} anomalies, i.e.,~cumulative deviations from climatological means or other reference profiles, on a global scale \cite{antonini_identification_2024} or for the Netherlands \cite{stoop_climatological_2024}. Combining wind and solar power in renewable technology portfolios can mitigate drought characteristics within regions. This portfolio effect has been studied for Europe \cite{raynaud_energy_2018,kies_critical_2021,kapica_potential_2024,hu_implications_2023,breyer_reflecting_2022,kittel_multi-threshold_2026}, the US \cite{rinaldi_wind_2021}, or individual countries, such as Germany \cite{kaspar_climatological_2019,mockert_meteorological_2023}, Hungary \cite{mayer_probabilistic_2023}, India \cite{gangopadhyay_role_2022}, and Japan \cite{ohba_climatology_2022}. In addition, the complementary of wind and solar power across regions further reduces extreme drought severity. This balancing effect has been illustrated for Europe \cite{kaspar_climatological_2019,breyer_reflecting_2022,hu_implications_2023,kittel_multi-threshold_2026} and the US \cite{rinaldi_wind_2021}.
% Relevant insights from Kittel and Schill \cite{kittel_quantifying_2024}: Most extreme droughts with significant inter-annual variability predominantly occurring in winter.
Another literature strand explores positive residual load events, which relate to periods where \ac{VRE} generation falls short of electric demand 
%and may occur during peak load and/or \ac{VRE} drought periods 
\cite{kittel_measuring_2024}. These periods indicate the need for system flexibility in general or, specifically, for a technology that supplies energy during such events. Positive residual load events have been studied for Europe \cite{raynaud_energy_2018,van_der_wiel_influence_2019,otero_characterizing_2022,otero_copula-based_2022,allen_standardised_2023}, Norway, France, Italy, Spain and Sweden \cite{van_der_most_temporally_2024}, Germany \cite{ruhnau_storage_2022,kondziella_techno-economic_2023}, Northern Italy \cite{francois_statistical_2022}, Africa \cite{plain_accounting_2019}, and the US \cite{bracken_standardized_2024}. Further, \ac{VRE} droughts that cause power sector stress events can be detected in energy system models through shadow prices reflecting the value of stored energy or transmission grid capacity, extreme electricity prices, resource adequacy metrics, or emission patterns, as shown for Europe \cite{grochowicz_using_2024} or the US \cite{su_compound_2020,wessel_technology_2022,akdemir_assessing_2022,sundar_2023}.

% Literature on inter-annual variability in energy modeling
Energy system models increasingly include multiple sectors and energy carriers to explore future scenarios of sector-coupled energy systems \cite{fodstad_next_2022,levin_energy_2023}. Due to increasing complexity and computational burden, these models are often solved for only a single or a limited number of weather years. However, optimal energy model outcomes vary substantially across years in terms of optimal dispatch and investment in generation and storage capacities, transmission grid capacities, electricity demand, prices, levelized costs of electricity, or emissions. This is well-documented for single countries such as the UK \cite{bloomfield_quantifying_2016,pfenninger_dealing_2017,zeyringer_designing_2018,staffell_increasing_2018,diesing_exploring_2024} or Ireland \cite{diesing_exploring_2024}, for overall Europe \cite{staffell_using_2016,pfenninger_long-term_2016,collins_impacts_2018,schlott_impact_2018,grochowicz_intersecting_2023,gotske_designing_2024}, or for the US \cite{dowling_role_2020,hill_effects_2021,ruggles_planning_2024}. For Europe, the total system costs of a future net-zero energy system vary between $\pm10$\% depending on the year used for system design \cite{gotske_designing_2024}. A general finding of this literature strand is that energy systems modeling based on only one or a few weather years identifies system configurations that may lead to suboptimal capacity choices or operational inadequacies. Extreme renewable drought events, which vary substantially across years and regions, are likely to contribute significantly to these inter-annual variations in energy system modeling~\cite{kittel_multi-threshold_2026}.

While there is growing interest in the meteorology and energy systems analysis domains on the impact of weather variability on renewable energy systems \cite{craig_overcoming_2022,sundar_2023}, the literature lacks a distinct analysis of the impact of extreme renewable droughts on energy storage needs, and how spatial system flexibility may alleviate these.

% Research questions & results
Here we combine two open-source methods to investigate how \ac{VRE} droughts impact optimal long-duration storage investments in a least-cost configuration of a fully decarbonized European power sector. We use renewable time series analysis for \ac{VRE} drought identification and a power sector model for determining least-cost operation and deployment decisions across a summer-to-summer planning horizon. We further quantify how much long-duration storage could be avoided with different degrees of electricity and hydrogen exchange between countries. We determine the long-duration storage need for dealing with extreme renewable droughts considering policy-oriented European interconnection levels, what ``no-regret'' long-duration storage capacity remains for a scenario with unconstrained geographical balancing of such events, and how much these results vary across years in our model setup. In doing so, we also shed light on appropriate weather year selection for modeling weather-resilient energy system scenarios in Europe. We additionally illustrate how different types of flexibility options interact for coping with extreme renewable droughts and explore what role nuclear or fossil power plants with emission abatement could play in mitigating storage needs.

%%%%%%%%%%%%%%%%%%%%%%%%%%%%%%%%%%%%%%%%%%%%%%%%%%%%%%%
\section*{Results}
%%%%%%%%%%%%%%%%%%%%%%%%%%%%%%%%%%%%%%%%%%%%%%%%%%%%%%%

% IM werden nicht gezogen in Text einarbeiten

% Here, we discuss our model results. In Section~\ref{sec:discussion}, we discuss the relevance of our model results to the real world.

\subsection*{Extreme renewable energy droughts coincide with major discharging periods of long-duration storage}

% Intuition droughts and storage
One of the principal insights from our time series and model-based analysis is that the operation of long-duration storage coincides with \ac{VRE} droughts. Figure~\ref{fig:figure_1} illustrates drought patterns of simulated renewable technology portfolios with policy-relevant capacity mixes from the \ac{TYNDP} 2022, which are retrieved from the renewable time series analysis. The figure further shows the least-cost long-duration storage operation for the weather year 1996/97, in which we find the most extreme pan-European drought in the data (Supplementary Fig.~\ref{fig:figure_si1}). Notably, several European countries are affected simultaneously \cite{kittel_multi-threshold_2026}. The highlighted extreme drought events consist of sequences of shorter but more severe droughts within contiguous periods of well-below-average renewable availability. Such events may last up to several months and span across the turn of years.

%%%%% Figure 1 %%%%%%

% Storage operation within droughts
During these long-lasting events, average renewable availability is well below its long-run mean, but still remains well above zero. This means that long-duration storage is not necessarily required to continuously discharge during the entire drought event. However, these longer events comprise shorter periods with significantly lower availability, lasting multiple days or even two to three weeks. During these periods, storage is often discharged at full capacity and the storage state-of-charge declines strongly, particularly when co-occurring with peak-demand periods.

% Winter vs. summer droughts, demand, storage
Extreme events typically occur in winter, leading to significant long-duration discharge periods. Countries that heavily rely on wind power, such as the UK, may face the most extreme renewable technology portfolio events also in summer, driven by wind droughts that are generally more pronounced in summer in Europe \cite{cannon_using_2015,kittel_multi-threshold_2026}. Yet, electricity demand in the UK is much higher in winter than in summer, similar to other central and northern European countries. This seasonal demand effect by far outweighs the differences between summer and winter droughts. Consequently, UK's major storage discharging period coincides with the winter drought, while the slightly larger summer drought hardly affects the storage state-of-charge.

\subsection*{Droughts drive long-duration storage energy capacity}

% Positive correlation between droughts and storage
For each modeled weather year, there is also a clear positive correlation between the most extreme renewable droughts in a given country and that country's long-duration storage energy need (Figure~\ref{fig:figure_2}a). We find that this applies to nearly all European countries when modeled as energy islands and also to the pan-European copperplate scenario.
% ggf. Regressionsergebnisse für Szenario (3) in SI einfügen und hier drauf verweisen

%%%%%% Figure 2

\begin{comment}
\begin{figure}[t]
\centering
\subfloat[Long-duration storage only.\label{fig:regression_lds}]
    {{\includegraphics[width=.47\textwidth]{figures/Figure_2a.pdf}}}
\qquad
\subfloat[Mid- and long-duration storage.\label{fig:regression_mds_lds}]
    {{\includegraphics[width=.47\textwidth]{figures/Figure_2b.pdf}}}
    \caption{Correlation of the drought mass of most extreme winter drought events and normalized storage energy capacity.\\ \scriptsize{For comparison, we normalize the least-cost storage energy with the annual demand for electricity (including electrified heating) and hydrogen. For illustration, we exclude countries with least-cost storage energy below 5~TWh and countries with binding storage expansion potential constraints. We further include the pan-European copperplate scenario (CP). Supplementary Fig.~\ref{fig:figure_si3} shows the unfiltered regression results.}}%
\label{fig:regression_yearly_dem}%
\end{figure}
\end{comment}

% Differences between countries i)
The correlation between the most extreme winter drought events and least-cost storage capacities differs between countries. In Central European countries such as France or the UK, storage capacity increases substantially in years with more severe droughts, illustrated by the steep slope of the fitted regression lines in Figure~\ref{fig:figure_2}. This effect is because electricity demand has a pronounced winter peak in these countries (Figure~\ref{fig:figure_3}). In contrast, more severe droughts hardly increase storage capacities in other countries, particularly in Romania or Spain, driven by a less seasonal or even summer-peaking electricity demand in such countries. Supplementary Fig.~\ref{fig:figure_si4} shows how the storage-defining effect of extreme droughts slightly diminishes when removing the effect of demand seasonality in a stylized setting of the Germany power sector. Normalized long-duration storage energy needs are generally much lower in Spain than those in Central or Northern European countries. One driving factor for this are significant hydro reservoir and pumped-hydro storage capacities in Spain, which substitute long-duration storage capacities for dealing with extreme droughts to some extent. A similar finding applies to the pan-European copperplate scenario. These effects lead to a lower sensitivity of long-duration storage needs to increasingly severe droughts, visible as lower regression line intercepts and, in the case of Spain, a relatively moderate slope of the regression line. Considering both hydro and long-duration storage capacities for the regression reduces these effects (Figure~\ref{fig:figure_2}b).

%%%% Figure 3

% Copperplate
In the pan-European copperplate scenario, the lower drought mass show that droughts are notably less severe than in individual countries, resulting in significantly lower normalized storage needs. This is due to a geographical balancing effect, which spatially smooths renewable generation and demand patterns across all of Europe, thereby mitigating extreme droughts \cite{roth_geographical_2023,kittel_multi-threshold_2026}.

% Exceptions can be explained by various factors
In Section~\nameref{ssec:additional_regression} in the Supplementary Information, we discuss why the correlation between the most extreme winter drought event and the least-cost long-duration storage size for the same year is not perfect. We also provide results based on alternative capacity mixes with lower shares of solar \ac{PV} for drought mass computation.
% In the latter, the drought mass of the most extreme events tends to decrease in some countries and the copperplate scenario. This is due to the lower solar \ac{PV} share, which alleviates the impact of solar seasonality on compound portfolio droughts in winter. However, regression slopes hardly change, indicating a limited impact of the alternative capacity mix on drought mass results. The most extreme drought for the copperplate scenario occurs in both scenarios in the winter of 1996/97.

\subsection*{Least-cost long-duration storage capacities decrease with geographical balancing}

% possible analyses: (i) Storage cycles as energy discharge divided by energy capacity, (ii) variability E2P, (iii) discharging capacity plus its relative sizing compared to peak demand (iv) how much LDS energy remains without flachen H2 Bedarf anderer Sektoren
% "LDS needed to compensate for resource gaps"

% Storage energy and value of interconnection
Interconnection between countries via electricity and hydrogen grids provides spatial flexibility to the power sector for coping with \ac{VRE} droughts. To systematically assess this relationship, we analyze four scenarios with varying degrees of interconnection. Figure~\ref{fig:figure_4} shows that increasing degrees of interconnection reduce aggregate storage energy needs across all scenarios. 

%%%%% Figure 4

% storage needs in our scenarios
In scenario (1), we model a counterfactual in which every country operates as an energy island, excluding the possibility of exchanging electricity or hydrogen with its neighbors. The resulting median (maximum) long-duration storage energy capacity aggregated across all modeled countries is 232~(378)~terawatt hours (TWh). This corresponds to around 4.7 (7.6)\% of the yearly European electricity demand. Scenario (2) allows for policy-oriented \ac{TYNDP} 2022 cross-border exchange of electricity, which reduces (increases) the median (maximum) long-duration storage energy capacity to 206~(385)~TWh, or 4.2 (7.8)\% of yearly demand. The slight increase in maximum storage investments is mainly driven by an increase in offshore wind generation in Belgium, which is transmitted to and stored in additional long-duration storage capacities in the Netherlands, while in Belgium the long-duration storage energy capacity remains at its expansion limit. Additionally including hydrogen exchange within policy-oriented limits in scenario (3) further decreases the median (maximum) long-duration storage energy to 170~(351)~TWh, equivalent to 3.4~(7.1)\% of yearly demand. Scenario (4) models the counterfactual of unconstrained exchange of electricity and hydrogen between countries. This results in the lowest storage energy investments with a median (maximum) value of 91~(159)~TWh, corresponding to 1.8 (3.2)\% of yearly demand. Increasing interconnection capacity enables more pronounced geographical balancing, which smooths the impact of storage-defining renewable droughts, resulting in decreasing least-cost storage capacity \cite{kittel_multi-threshold_2026}. Higher levels of interconnection capacities reduce the need for long-duration storage energy capacity both on an aggregate level and in most countries (Supplementary Fig.~\ref{fig:figure_si12}).
% storage needs if grid expansion cannot be realized
If the expansion of the electricity or hydrogen networks in scenarios (2) or (3), as envisaged by European transmission system operators, cannot be realized, our results suggest that least-cost long-duration storage energy capacities will be between those of scenarios (1) and (2).

% Inter-annual variation
In our model, least-cost storage deployment shows a substantial degree of inter-annual variation. This is driven by differences in renewable energy droughts and demand patterns across years (compare Figure~\ref{fig:figure_3}). Compared to the island setting in scenario (1), the inter-annual variation is higher in scenarios (2) and (3) with policy-oriented geographical balancing. This is because the correlation of storage-defining droughts in individual countries varies across space and weather years \cite{kittel_multi-threshold_2026}. When drought events do not occur simultaneously across countries, more pronounced geographical balancing can be leveraged. This reduces long-duration storage energy investments, which decreases the lower bound of the range. This effect is even more pronounced when including cross-border exchange hydrogen in scenario (3). As all drought events are balanced to the fullest extent possible in the copperplate scenario, we find the lowest inter-annual variation in this case.

% Ranking of years
Among the weather years in our data, 1996/97 marks the period with the highest aggregated long-duration storage energy investments across all interconnection scenarios in the least-cost system configurations. This is in line with our renewable availability time series analysis, which finds that the most extreme pan-European drought affects many countries simultaneously (Supplementary Fig.~\ref{fig:figure_si1}). In our model, the reducing effect of geographical balancing on long-duration storage investments remains limited for policy-oriented interconnection levels in scenarios (2) and (3) in 1996/97. This is because the storage-defining drought event in this year is temporally highly correlated across many countries and also coincides with demand peaks in several countries (see teal boxes and gray lines in Figure~\ref{fig:figure_5}). In contrast, the unconstrained balancing in the copperplate scenario (4) substantially mitigates overall storage investments, as energy can be imported from countries that are not or less affected by renewable droughts. In other years, the effects of interconnection may play out differently. For instance, geographical balancing has a particularly strong storage-mitigating effect in 1988/89, as the largest droughts in major countries hardly overlap and co-occur with periods of high demand only to a limited extent (Supplementary Fig.~\ref{fig:figure_si6}). In contrast, the storage needs in 1987/88 seem less affected by interconnection. Its relative position in the ranking of years even slightly increases with increasing interconnection. This can be explained by a large temporal overlap of droughts in individual countries (Supplementary Fig.~\ref{fig:figure_si7}). Supplementary Fig.~\ref{fig:figure_si8} additionally illustrates how higher interconnection levels increasingly affect the ranking of weather years.

%%%% Figure 5 

% Einordnung scenarios (1), (3), (4).
In the least-cost system configuration, the maximum long-duration storage energy capacity of 159~TWh in scenario (4) occurs in 1996/97, which comprises the most extreme \ac{VRE} drought in the data. This storage capacity is required to balance the remaining energy deficit of this event under the assumption of unlimited cross-country exchange of electricity and hydrogen. That is, storage capacity needed to deal with this event cannot be compensated any further by additional interconnection, given that geographical balancing has already been leveraged across all of Europe to the greatest extent possible. Hence, in our model, 159~TWh or 3.2\% of the yearly European electricity demand, is the minimum need for long-duration storage capacity for a reliable, fully renewable European energy system. However, since unconstrained geographical balancing cannot be realized in reality, scenario (3) shows the most policy-relevant long-duration storage need that is required for dealing with the most extreme simulated Dunkelflaute event in the data. In this scenario, long-duration storage capacities are much higher, amounting to 351~TWh or 7.1\% of the yearly demand.

\subsection*{Interactions of different flexibility options for coping with extreme droughts}

\begin{figure}[h]
\centering
    \includegraphics[width=\textwidth]{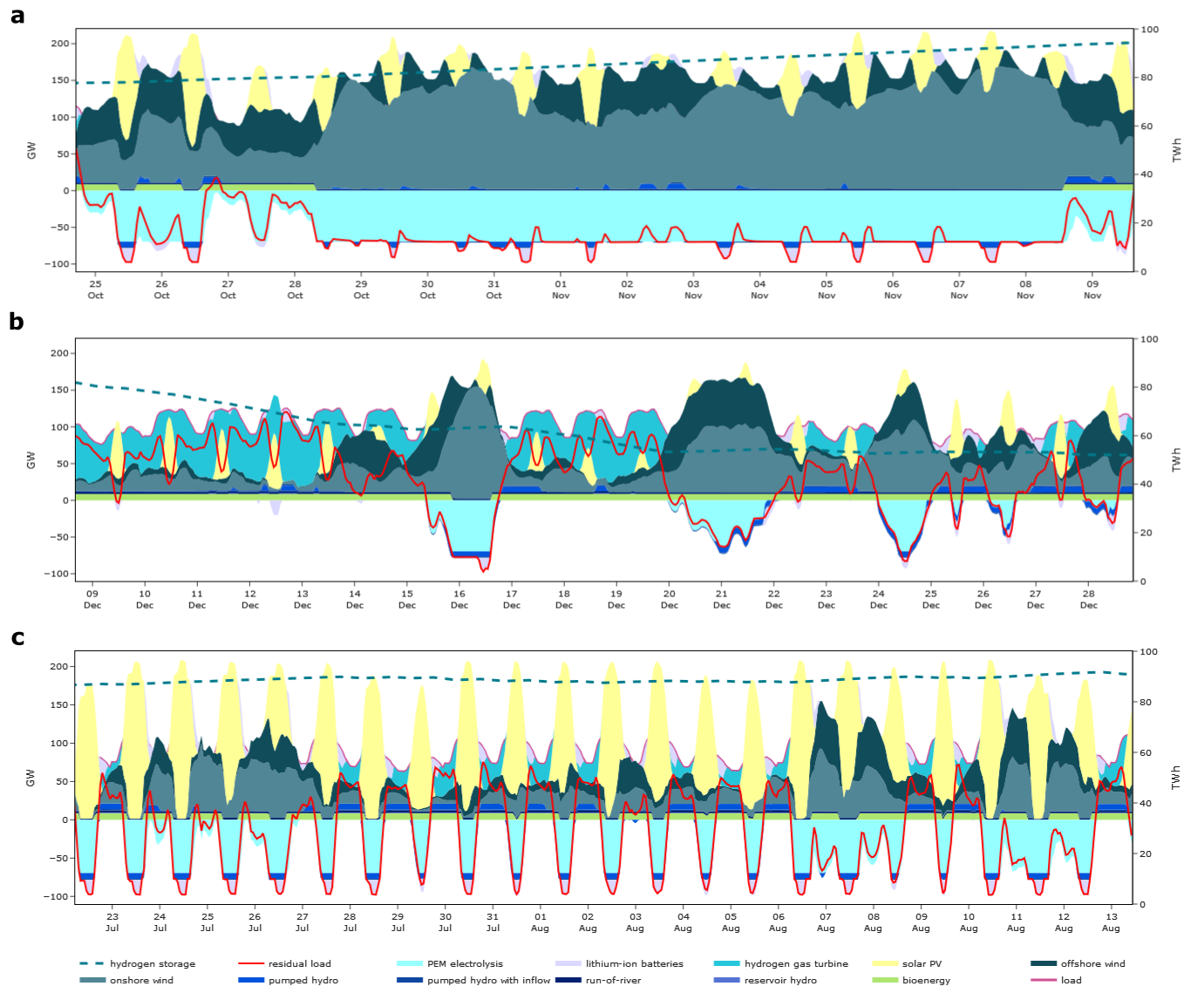}
    \caption{\textbf{Least-cost power sector operation in Germany for the weather year 1996/97.} The positive part of the left y-axis relates to generation and storage discharge, and its negative part to electricity demand and storage charge. The right y-axis refers to the long-duration storage state-of-charge. For illustration, we focus on scenario (1) excluding cross-border exchange of electricity or hydrogen. \textbf{a} Long-duration charging period before an extreme drought. \textbf{b} Long-duration discharge period within an extreme drought. \textbf{c} Long-duration storage follows diurnal solar \ac{PV} pattern during summer.}
    \label{fig:figure_6}
\end{figure}

In our model, we find complex interactions between long-duration storage and other flexibility options in the power sector. The least-cost dispatch patterns in exemplary weeks in Germany and Spain for the weather year 1996/97 in Figure~\ref{fig:figure_6} illustrate such interactions. 

Short-duration flexibility options, such as batteries and pumped hydro storage, generally have low costs for power capacity but high costs for energy capacity. This means that supplying power is relatively inexpensive, but storing energy is costly. In contrast, long-duration flexibility options, such as hydrogen-based electricity storage, exhibit opposite cost structures: electrolysis and hydrogen turbines, used for long-duration storage charging and discharging, incur high power capacity costs, while long-duration energy capacity costs are low. Hydro reservoirs and bioenergy have similar cost structures and serve comparable roles as long-duration storage. Given these differences in cost structures, the dispatch patterns of these flexibility options show different effects in our model-based scenarios that enable coping with extreme \ac{VRE} droughts at the lowest possible cost.

%%% Figure 6

\begin{comment}
\begin{figure}[htbp]
\centering
\subfloat[Long-duration charging period before an extreme drought.\label{fig:timeseries_1}]
{{\includegraphics[width=.99\textwidth]{figures/Figure_6a.png}}}
\qquad
\subfloat[Long-duration discharge period within an extreme drought.\label{fig:timeseries_2}]
{{\includegraphics[width=.99\textwidth]{figures/Figure_6b.png}}}
\qquad
%\subfloat[Long-duration discharge period in Spain within an extreme drought.\label{fig:timeseries_3}]
%{{\includegraphics[width=.99\textwidth]{figures/timeseries_lds_discharge_sds+mds_charge_discharge_pos_rl_ES_no_leg.png}}}
%\qquad
\subfloat[Long-duration storage follows diurnal solar \ac{PV} pattern during summer. \label{fig:timeseries_4}]
{{\includegraphics[width=.99\textwidth]{figures/Figure_6c.png}}}\\
\caption{Least-cost power sector operation in Germany for the weather year 1996/97. \\ \scriptsize{The positive part of the left y-axis relates to generation and storage discharge, and its negative part to electricity demand and storage charge. The right y-axis refers to the long-duration storage state-of-charge. For illustration, we focus on scenario (1) excluding cross-border exchange of electricity or hydrogen.}}
\label{fig:figure_6}%
\end{figure}
\end{comment}

% short-duration flexibility limits electrolysis capacity needs
Our model finds that during charging periods of the long-duration through electrolysis, batteries may charge and discharge, or bioenergy and hydro reservoirs generate electricity to maximize electrolyzer utilization, even if residual load is negative (Figure~\ref{fig:figure_6}a). In doing so, these short-duration flexibility options enable a continuous operation of electrolysis, which limits the need for expensive electrolysis capacity.

% short-duration flexibility limits reconversion capacity needs
While long-duration storage typically provides most of the energy needed to cope with a renewable drought, it may not necessarily provide all the power capacity. To reduce the build-out of costly hydrogen gas turbines for covering peak residual load events, battery storage is sometimes used in our least-cost system. Despite the additional energy losses from conversion, this may lead to charging and discharging cycles of batteries or pumped hydro while hydrogen gas turbines generate electricity (e.g.,~on December 12, 16, and 17 in Figure~\ref{fig:figure_6}b).

% short-duration flexibility limits reconversion capacity needs
Short-duration flexibility options are also used in our model to balance brief events of renewable surplus generation within longer drought events. In such cases, the discharging of long-duration storage is interrupted, and batteries and pumped hydro storage are used instead for balancing (e.g., visible on December 16, 22, 25, or 26 in Figure~\ref{fig:figure_6}b). Here, the model favors shorter-duration storage over hydrogen-based storage because of lower roundtrip energy losses, indicating a ``merit order'' of storage technologies.

% LDS helps batteries --> lass es uns rausnehmen, gemeinsam mit Panel c
%In countries with less pronounced solar seasonality in winter and significant reservoir capacity, such as Spain, the least-cost discharge capacity of batteries is relatively large compared to hydrogen gas turbines. During extreme droughts, long-duration storage discharge enables battery charging before and after \ac{PV} peaks, even in positive residual load periods (e.g., visible on January 24 in Figure~\ref{fig:timeseries_3}). Although this operation incurs high conversion losses, it leverages lower-cost power capacity of short-duration flexibility options that is anyway required to balance diurnal \ac{PV} variations at other times of the year, which minimizes higher-cost long-duration storage discharge capacity. 

% LDS also used for other services
Finally, long-duration storage may also be used for diurnal cycling in summer in the least-cost system (Figure~\ref{fig:figure_6}c). Here, hydrogen-based storage integrates solar \ac{PV} surplus energy, which reduces the need for short-duration battery capacity that otherwise would be required for this purpose. Importantly, the least-cost long-duration storage capacity is not defined by these diurnal variability patterns in summer but rather by major winter droughts. In other words, the long-duration storage capacity that is built to cope with winter droughts can have repercussions on the least-cost battery capacity and dispatch decisions for dealing with summertime solar \ac{PV} variability.

These model-based findings complement those of a previous study focused on the US context, which demonstrated that long-duration storage can also provide short-term flexibility \cite{li_influence_2024}.

\subsection*{Impact of firm zero-emission generation}\label{sec:nuc}

\begin{figure}[h]
\centering
    \includegraphics[width=\textwidth]{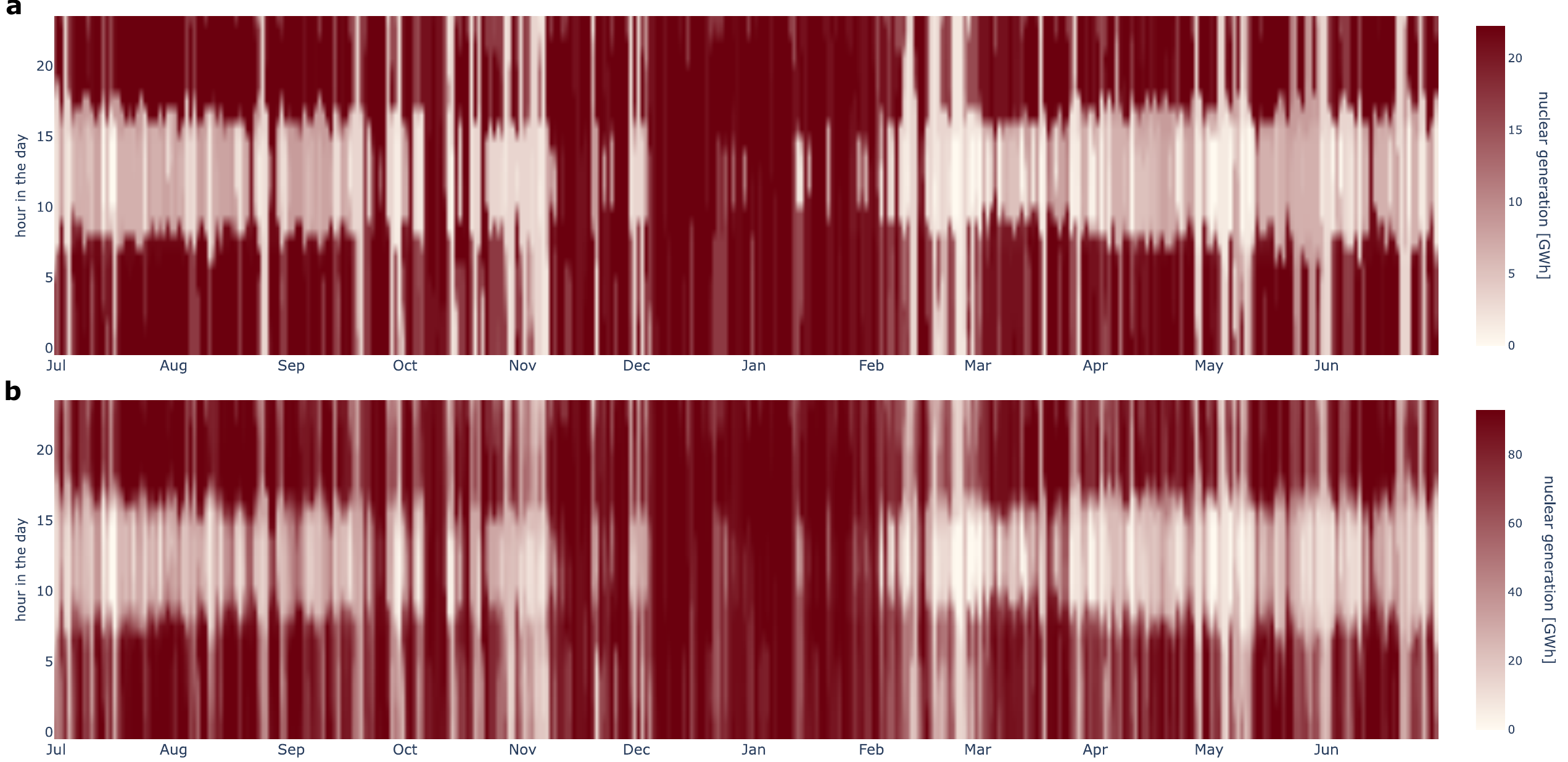}
    \caption{\textbf{Hourly and daily generation patterns of nuclear power aggregated across all countries in scenario (3) with policy-oriented exchange of electricity and hydrogen for 1996/97.} Supplementary Fig.~\ref{fig:figure_si13} and Supplementary Fig.~\ref{fig:figure_si14} show the operational patterns for all interconnection scenarios for low and high levels of nuclear power, respectively. \textbf{a} Low levels of nuclear power. \textbf{b} High levels of nuclear power.}
    \label{fig:figure_7}
\end{figure}

% Impact of nuclear power in Europe
In our default scenario, we allow for firm and variable renewable technologies only. However, nuclear power is considered a valid decarbonization option in some European countries. To reflect such country-specific energy policy strategies, we introduce two complementary scenarios with exogenous low or high levels of nuclear power, amounting to 24~ gigawatts (GW) or 102~GW, respectively. We assume that nuclear power serves as a firm zero-emission generation capacity that does not face inter-hourly operational constraints but minor costs for ramping up or down that relate to wear and tear costs and potentially also to energy losses. In our model, nuclear electricity generation complements particularly diurnal solar \ac{PV} variations and is operated at full capacity during extended periods in winter during extreme renewable portfolio droughts, which reduces the need for long-duration storage discharging. This is visible in the least-cost operational patterns of both low and high levels of nuclear power in the extreme winter of 1996/97 (Figure~\ref{fig:figure_7}), in which many countries are simultaneously affected by severe renewable droughts (Figure~\ref{fig:figure_5}). These operational patterns generally apply across all investigated weather years (Supplementary Fig.~\ref{fig:figure_si15} and Supplementary Fig.~\ref{fig:figure_si16}). Additionally, the exogenous firm generation capacity generally decreases least-cost investments in \ac{VRE} technologies (Supplementary Fig.~\ref{fig:figure_si9}, Supplementary Fig.~\ref{fig:figure_si10}, Supplementary Fig.~\ref{fig:figure_si11}), which mitigates the overall system's need for flexibility and long-duration storage.

%%%% Figure 7

\begin{comment}
\begin{figure}[htbp]
\centering
\subfloat[Low levels of nuclear power.\label{fig:low_nuc_TYNDP-eh_main}]
{{\includegraphics[width=\textwidth]{figures/Figure_7a.pdf}}}
\quad
\subfloat[High levels of nuclear power.\label{fig:high_nuc_TYNDP-eh_main}]
{{\includegraphics[width=\textwidth]{figures/Figure_7b.pdf}}}\\
\caption{Hourly and daily generation patterns of nuclear power aggregated across all countries in scenario (3) with policy-oriented exchange of electricity and hydrogen for 1996/97. Supplementary Fig.~\ref{fig:figure_si13} and~\ref{fig:figure_si14} show the operational patterns for all interconnection scenarios for low and high levels of nuclear power, respectively.}
\label{fig:figure_7}%
\end{figure}
\end{comment}

\begin{figure}[h]
    \centering\includegraphics[width=.99\textwidth,keepaspectratio]{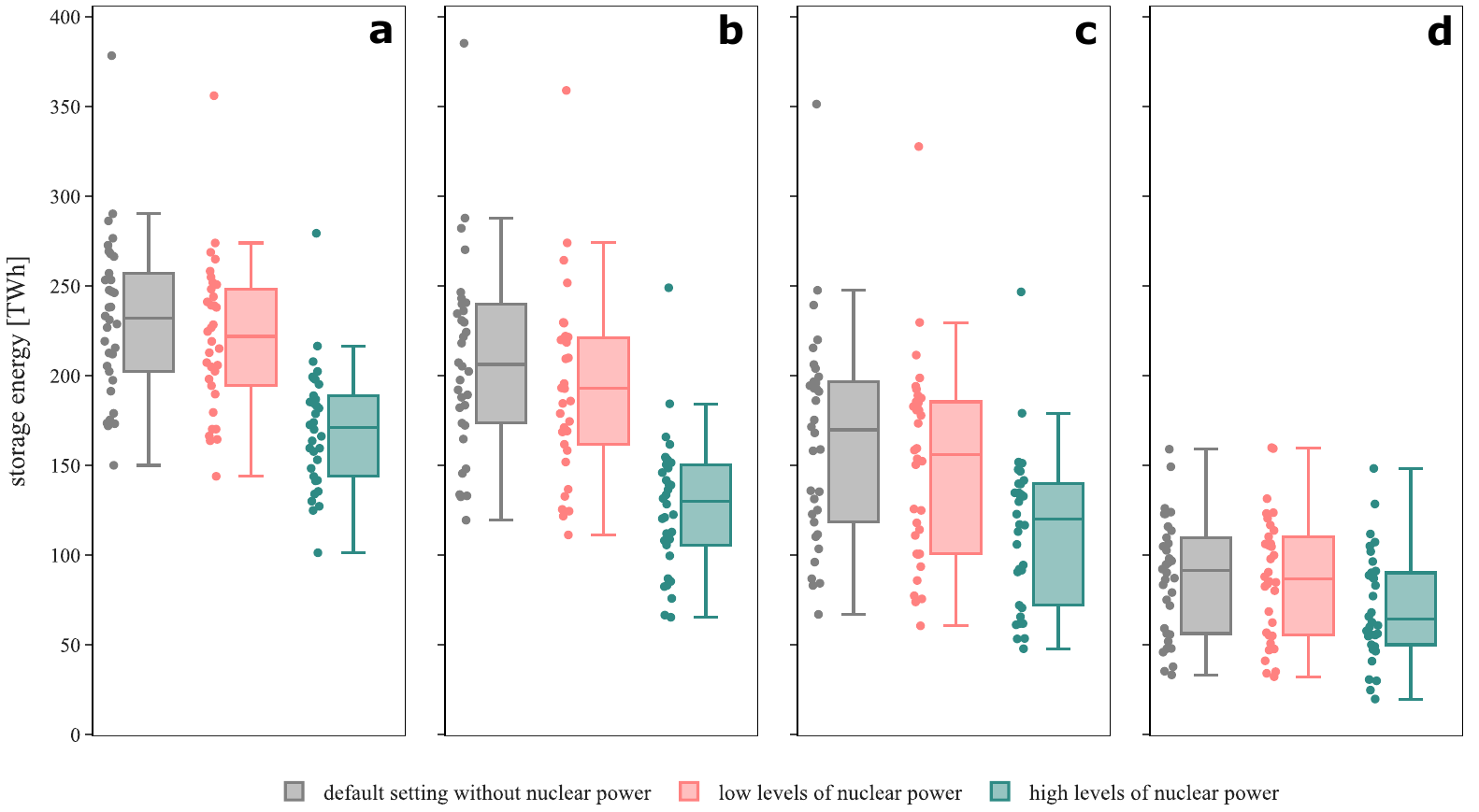}
    \caption{\textbf{Least-cost long-duration storage energy capacity aggregated across all countries for all modeled weather years and interconnection scenarios for different levels of nuclear capacities.} Each dot refers to one independently modeled weather year. The center line denotes the median, box limits indicate the interquartile range (Q1–Q3), whiskers extend to 1.5× the interquartile range, and points beyond the whiskers represent outliers. \textbf{a} Scenario (1): no exchange of electricity nor hydrogen (island systems). \textbf{b} Scenario (2): policy-oriented exchange of electricity, no exchange of hydrogen. \textbf{c} Scenario (3): policy-oriented exchange of electricity and hydrogen. \textbf{d} Scenario (4): pan-European copperplate assuming unconstrained exchange of electricity and hydrogen.}
    \label{fig:figure_8}
\end{figure}

Low levels of nuclear power reduce the least-cost long-duration storage energy capacity across all investigated interconnection scenarios only to a minor extent (Figure~\ref{fig:figure_8}). Without geographical balancing in scenario (1), low nuclear capacities mitigate least-cost long-duration storage investments only in those countries where nuclear power is deployed, decreasing the aggregated median (maximum) long-duration storage energy capacity by 4~(6)\%. With policy-oriented cross-border exchange in scenarios (2) and (3), the mitigating effects of low levels of nuclear power on the aggregated median (maximum) storage energy capacity are slightly larger with 6~(7)\% or 8~(7)\%, respectively. This is because the firm generation technology mitigates renewable droughts both domestically and in other countries. Under the assumption of unconstrained geographical balancing in scenario (4), median long-duration storage energy reduces by 5\%, while the maximum storage energy even slightly increases by less than 1\% because of interactions with bioenergy in the least-cost system. 
%In this case, nuclear power disproportionally displaces bioenergy, i.e.,~one gigawatt of nuclear power substitutes more than one gigawatt of bioenergy. Electricity generation from bioenergy, which serves as a renewable baseload technology in the previous scenarios, is therefore only partially replaced by nuclear power during renewable droughts, which in turn can necessitate a slightly higher long-duration storage energy capacity for coping with these events. 
Long-duration storage further remains necessary to continuously meet the flat hydrogen demand from coupled sectors, as assumed in our model (see Section~\nameref{ssec:hydrogen_model} in the Supplementary Information).

%%%% Figure 8 

In the scenario with high levels of nuclear power, higher shares of demand during extreme droughts can be met by nuclear, which reduces least-cost long-duration storage investments in our model-based analysis. This effect is complemented by nuclear disproportionately displacing least-cost investments into wind and solar \ac{PV}, which in turn further lowers long-duration storage needs. Compared to the default scenario without nuclear power, these effects result in a substantial decrease of the median (maximum) storage energy capacity by 26~(26)\% in scenario~(1). Allowing cross-border electricity trade in scenario~(2) enables nuclear power exports to balance renewable droughts domestically and in other countries, which further reduces long-duration storage energy capacity by 37~(35)\% in our model. In scenario (3), the reduction of 29~(30)\% is less pronounced because more renewable surplus is integrated, converted to electrolytic hydrogen, and shared across countries, leading to a more beneficial role of long-duration hydrogen storage. In scenario (4), nuclear power can balance renewable droughts across Europe without grid constraints, lowering least-cost mean storage energy by 30\%. The maximum least-cost storage energy capacity declines by only 7\%. One reason for this is that nuclear displaces bioenergy and some battery discharging capacity, resulting in a net firm capacity gain far less than the 78~GW additional nuclear capacity. Further, even during extended extreme droughts, renewable generation potential substantially above zero remains \cite{kittel_multi-threshold_2026}. The substitution of wind and solar capacity by nuclear power thus yields a higher residual demand, which is met by firm technologies including long-duration storage. %Third, renewable capacity shifts to regions with superior generation potentials, e.g., \ac{PV} to Iberia, resulting in higher \ac{VRE} full-load hours, which increases the system value of long-duration storage. These effects partially offset the storage-mitigating effect of high nuclear levels, particular in weather years with severe droughts. 
To conclude, our model-based analysis suggests that also in cases with high nuclear capacities, substantial long-duration storage investments remain optimal in the least-cost solution. In an additional sensitivity for Germany, we show that significant long-duration storage capacities remain part of the least-cost solution even in cases with much higher capacities of nuclear (or other firm zero-carbon technologies) exceeding by far the peak nuclear generation capacity ever realized in Germany (see Section~\nameref{ssec:substitution_si} in the Supplementary Information).

\subsection*{Sensitivity analyses}\label{sec:sensitivity}

\begin{figure}[h]
    \centering
    \includegraphics[width=\linewidth]{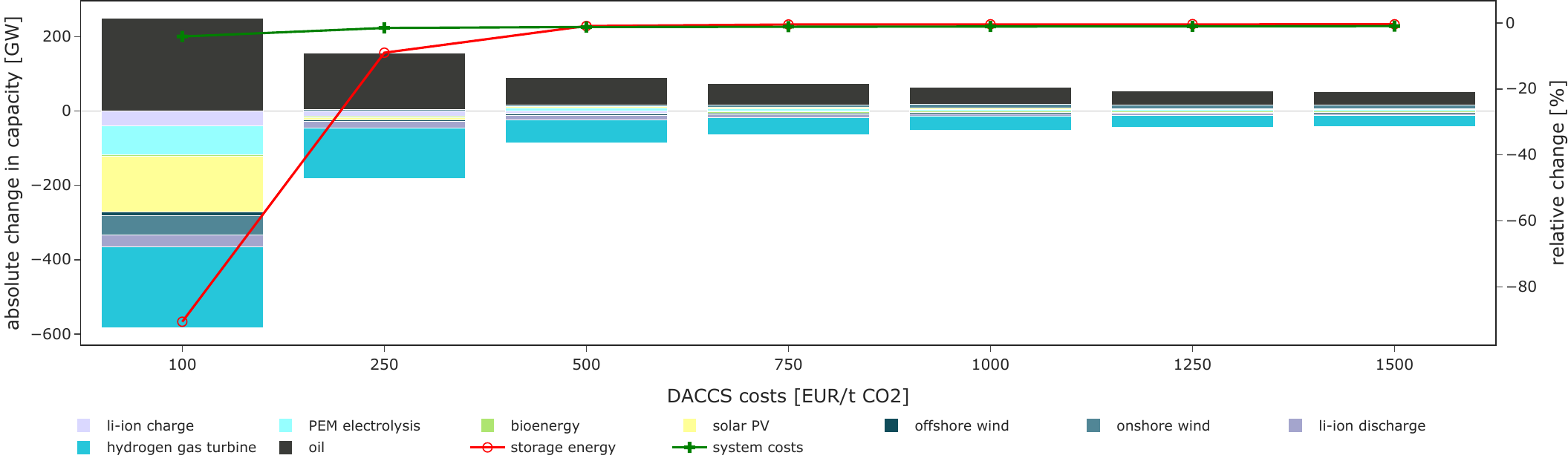}
    \caption{\textbf{Changes in least-cost capacities and system costs for varying costs of direct air capture and storage of emissions from the operation of oil-fired backup capacity.} Absolute change in generation and flexibility capacity (bars, left y-axis) and relative change of long-duration energy storage capacity and system costs (lines, right y-axis) aggregated across all countries in Europe in the interconnection scenario (3) compared to our default setting in 1996/97 for varying costs of direct air capture and storage of emissions from the operation of oil-fired backup capacity. For readability, we show the zero line of the left y-axis in gray.}
    \label{fig:figure_9}
\end{figure}

% impact of back-up technology: setting
Our analysis highlights the crucial role of long-duration storage for coping with pronounced renewable droughts in a climate-neutral European energy system based on renewable or zero-emission technologies. Alternatively, conventional fossil power plants combined with carbon dioxide (CO$_2$) removal could serve load during extreme drought events. A potential option for this is \ac{DACCS}, which removes CO$_2$ emissions directly from the atmosphere and stores them underground without requiring proximity to emission sources \cite{young_cost_2023}. In a sensitivity analysis of interconnection scenario (3) and weather year 1996/97, we allow for the endogenous deployment of oil-fired backup capacity and a corresponding use of \ac{DACCS}. Unlike gas, oil can be easily stored above ground at low costs and transported via trucks, avoiding the need for extensive capital-intensive infrastructure. Since \ac{DACCS} has not yet achieved maturity, its cost remains highly uncertain and depends on policy support, economies of scale, technological learning curves, and the profile of its electricity demand \cite{young_cost_2023}. We account for these uncertainties by varying the CO$_2$ removal cost from 100 to 1,500 EUR per tonne (t) CO${_2}$ (Figure~\ref{fig:figure_9}). The lower cost estimates in this range require a combination of optimistic assumptions, including continuous \ac{DACCS} operation at low electricity price levels
%, which can only be realized with electricity supply from firm zero- or low-emission technologies. Coupling \ac{DACS} to \ac{VRE} sources would result in substantially higher costs 
\cite{young_cost_2023}.

% impact of back-up technology: results
Only at very low \ac{DACCS} costs of 100 EUR per t CO$_2$, 250~GW of oil-fired backup capacities in Europe are deployed in the least-cost system. These substitute 90\% of the long-duration storage energy capacity and partially also wind, solar \ac{PV}, batteries, electrolysis, and hydrogen turbines compared to our default scenario. However, significant energy storage capacity needs of over 30~TWh remain required in the least-cost solution. For \ac{DACCS} costs above 100 EUR per t CO$_2$, the substitution effect for the long-duration storage discharging component diminishes substantially, while all other capacities are not affected. Likewise, only for very low levels of \ac{DACCS} costs of 100 EUR per t CO$_2$, system costs reduce considerably by 4.0\%. For all the other \ac{DACCS} costs, system cost savings range only around 1.0\%. That is, our results are very robust against introducing conventional electricity generation and \ac{DACCS}, except if this option could be realized at extremely low costs. We find similar results for a related sensitivity that introduces load shedding, drawing on a wide range of different values of lost load (see Section~\nameref{ssec:voll_sensitivity_si} in the Supplementary Information).

%%%%% Figure 9

\begin{figure}[h]
\centering
    \includegraphics[width=\textwidth]{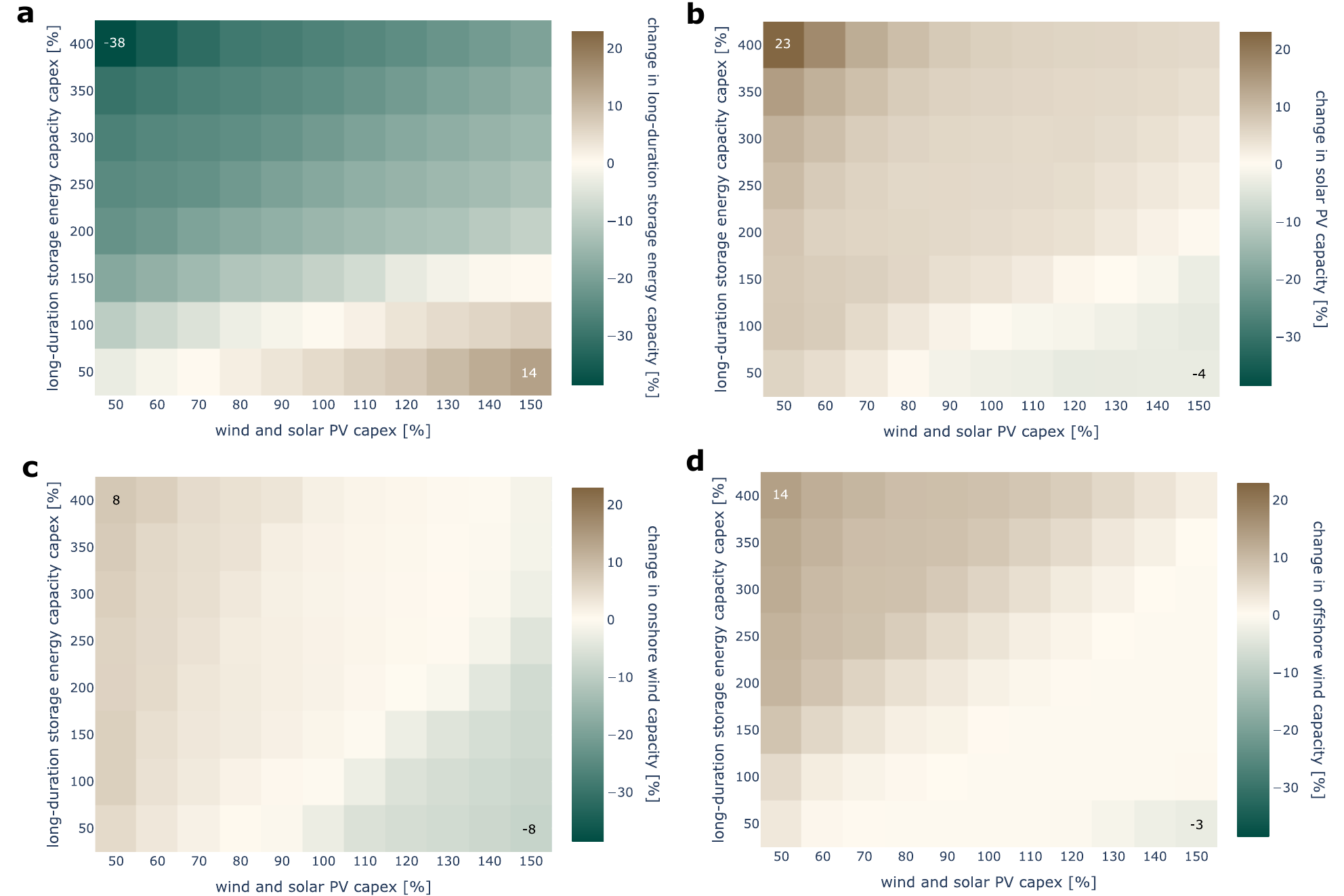}
    \caption{\textbf{Relative changes in least-cost investment decisions for varying levels of investment costs of onshore wind, offshore wind, solar \ac{PV}, and long-duration storage energy capacity in 1996/97 aggregated across all countries.} Cost and deployment variations are denoted in percentage. A cost variation of 100\% refers to the original parameterization. The annotated scenarios refer to the most pronounced changes across all scenarios. \textbf{a} Change in long-duration storage energy. \textbf{b} Change in solar \ac{PV} capacity. \textbf{c} Change in onshore wind capacity. \textbf{d} Change in offshore wind capacity.}
    \label{fig:figure_10}
\end{figure}

% Cost sensitivity: setup
The costs of onshore and offshore wind, as well as solar \ac{PV}, have declined continuously over the last decades. They are projected to decrease further, although uncertainties remain concerning possible economies of scale and technological progress \cite{creutzig_underestimated_2017,jaxa-rozen_sources_2021,way_empirically_2022}. Similarly, hydrogen-based underground storage has so far been hardly deployed, also resulting in cost uncertainty. To address this, we analyze the sensitivity of least-cost deployment to variations in overnight investment costs of these technologies (Supplementary Table~\ref{tab:cost_assumptions}) for the policy-relevant interconnection scenario (3) and the year 1996/97, which contains the most extreme renewable drought in our data. The energy capacity costs of underground hydrogen storage are orders of magnitude lower than those of wind and solar \ac{PV}. To reflect these cost asymmetries, we vary storage energy costs across a broader range, with a focus on high-cost scenarios (Figure~\ref{fig:figure_10}).

% Cost sensitivity: results
In our model, lower \ac{VRE} or higher storage energy costs lead to overbuilding wind and \ac{PV} to reduce reliance on storage. Although the latter reduces the potential to shift renewable surplus energy to periods with low \ac{VRE} availability or high demand, the additional renewable capacity decreases residual demand that otherwise would require long-duration storage discharge. %Offshore wind and \ac{PV} are overbuilt more than onshore wind, which may be due to lower solar \ac{PV} costs and higher offshore wind availability. 
Conversely, lower storage energy or higher \ac{VRE} costs favor long-duration storage deployment over \ac{VRE} capacity expansion. This substitution particularly affects onshore wind. These patterns generally apply across all investigated cost scenarios. Moderate deviations from the default setting of future wind and solar \ac{PV} costs appear most relevant \cite{way_empirically_2022}. For these scenarios, least-cost storage deployment is hardly affected in our parameterization.

\section*{Discussion}\label{sec:discussion}
%%%%%%%%%%%%%%%%%%%%%%%%%%%%%%%%%%%%%%%%%%%%%%%%%%%%%%%

% Summary, research questions
This paper analyzes how extreme renewable energy droughts impact least-cost investments and operations of long-duration electricity storage in a future renewable European power sector by combining renewable availability time series analysis for drought identification and energy system modeling.
%To this end, we combine a renewable time series analysis for \acl{VRE} drought identification with power sector modeling to analyze scenarios with different interconnection levels across European countries for 35~weather years.

% Storage-defining droughts
Our analysis yields several key insights. Extreme renewable energy droughts identified for the policy-oriented renewable capacity portfolios simulated here, which may last several weeks or even months, define the least-cost operational and investment needs for long-duration storage determined in our model. 
%Major discharge periods of the least-cost long-duration storage coincide with the most pronounced drought periods identified in the data. Further, our results reveal a positive correlation between simulated extreme drought events and least-cost storage energy capacity investments for many European countries and historical weather years. 
Firm renewable energy sources such as hydro reservoirs or bioenergy reduce least-cost long-duration storage capacities for dealing with extreme droughts, while co-occurring high-demand periods exacerbate them. %The most pronounced pan-European renewable drought identified in the \ac{VRE} drought analysis leads to the highest long-duration storage need.
Our model-based findings highlight that long-duration storage is an indispensable technology for coping with very pronounced renewable droughts in fully renewable energy systems with significant power supply and demand seasonality.
% is this the case of lower renewable shares

% Storage-mitigating interconnection
% auch irgendwie nur eine wiederholung vorheriger ergebnisse, oder? vorschlag, darstellen, dass diese effekte auch in realität gelten dürften
% könnte man die beiden paragraphen nicht vereinen?
Interconnection among European countries can reduce least-cost long-duration energy storage capacity. While previous literature has found a similar effect in case studies covering parts of the US or Europe and a limited selection of weather years \cite{staadecker_2024,brown_value_2021,roth_geographical_2023,ihlemann_2022}, we show that this finding also holds for the pan-European renewable energy system including 33~countries optimized here, particularly considering the most pronounced simulated drought events that occur in a large data set of 35~weather years. Yet, the storage-mitigating effect of policy-relevant interconnection levels of the \ac{TYNDP} 2022 remains limited and could only be substantially reduced in scenarios with interconnection levels far beyond envisaged grid expansion plans. 
%As inferred from our copperplate scenario, the least-cost storage capacity required for coping with the most extreme events in the data could only be substantially reduced in scenarios with interconnection levels far beyond envisaged grid expansion plans. In contrast, policy-oriented interconnection levels, as foreseen in the \ac{TYNDP} 2022 scenarios, mitigate least-cost storage investments only to a limited extent. 
Our analysis shows that additional cross-border exchange of renewable hydrogen, as planned in the European Union, contributes to reducing least-cost long-duration storage needs.

% Substantial storage needs, %Vergleich zu bestehender Literature 
Our model results suggest that there is a sizable need for long-duration storage in a fully renewable European power sector, irrespective of the extent of the assumed interconnection. Even in a perfectly interconnected Europe, which allows for unconstrained spatial smoothing of the most extreme drought in the data, 159~TWh of long-duration storage energy capacity result in the least-cost solution of our baseline parameterization, corresponding to around 3\% of yearly electric load. This number indicates that there is a lower bound for storage energy capacity required in a fully renewable European energy system. Its magnitude appears sizable, particularly considering that barely any hydrogen-based long-duration storage capacity is currently installed in Europe~\cite{talukdar_techno-economic_2024}. In more policy-relevant \ac{TYNDP} interconnection scenarios between European countries, the least-cost long-duration storage investments substantially increase in our model to 351~TWh, or more than 7\% of yearly European electricity demand. This finding leads to the conclusion that transitioning to a renewable European energy system may require significant long-duration storage capacities in the order of several hundred TWh, particularly for coping with extreme, yet rarely occurring, Dunkelflaute events. Our results extend previous analyses that focus on droughts in single countries, such as Germany, where long-duration storage capacities of 9\% \cite{kondziella_techno-economic_2023} or 10\% \cite{ruhnau_storage_2022} of annual electricity demand are required in a least-cost system configuration for dealing with extreme drought events. The slight differences in storage needs are likely driven by the storage-mitigating effect of geographical balancing across countries, which we include in our modeling.
%\cite{gotske_designing_2024}: also 1996/97, but additionally XXX
% most extrem ePRL in ruhnau & qvist occurred in winter 1996/97
% Sollen wir noch weitere politikrelevante nennen, z.B. Langfristszenarien, dort ähnliche Größenordnung

% Feasibility of LDS capacity
Overall, realizing the energy capacity levels determined by our model appears technically feasible, considering that around 1700~TWh of natural gas storage is currently installed in Europe \cite{gie_2021}, even when taking into account the lower volumetric energy density of hydrogen compared to methane. Yet, experience with converting existing gas storage to hydrogen or building new underground hydrogen storage is so far limited \cite{talukdar_techno-economic_2024}. The same applies to hydrogen grids as well as hydrogen generation and re-conversion infrastructure.

% Policy action I, II
Importantly, large-scale adoption of hydrogen-based long-duration storage, comprising electrolyzers, caverns, and hydrogen turbines for reconversion, will likely have long lead times because of supply chain and permitting bottlenecks \cite{odenweller_probabilistic_2022,odenweller_green_2025}. To safeguard the renewable energy transition in Europe, system planners and policymakers should thus consider early action to enable rapid scaling for realizing hydrogen storage investments. Furthermore, the maximum long-duration storage energy capacity in Europe identified in the least-cost system configuration, driven by the most extreme renewable drought in the winter of 1996/97, exceeds the next highest storage need determined for the weather year 1984/85 by 42\%. Market actors are unlikely to invest in such rarely utilized long-duration storage capacity without additional deployment incentives. Therefore, targeted support instruments or capacity mechanisms may be necessary to ensure the realization of sufficient storage capacity.
% Alternatively, unconventional industrial load shedding over extended time scales, such as multiple days or even weeks, could help mitigate long-duration storage needs for dealing with these very rare yet extreme Dunkelflaute events. Enhancing the flexibility of industrial consumers would require appropriate incentivization, such as compensation schemes, and merits future research.

% Interaction of flex options
We also observe complex interactions of long-duration storage with short-duration batteries and other flexibility options. A complementary operation of these technologies can mitigate least-cost long-duration storage and battery charging and discharging capacities to some extent, hence decreasing total system costs.
%We also show that long-duration storage does not continuously discharge during prolonged drought periods, and instead, shorter-duration flexibility options are used. Further, hydrogen storage needed for coping with winter droughts can complement batteries in balancing diurnal \ac{PV} variability in summer.
This finding complements those of a previous study focused on the US context, which demonstrated that long-duration storage can also provide short-term flexibility \cite{li_influence_2024}.
%Detailed numerical power sector modeling with an adequate representation of different flexibility technologies and a high temporal resolution appears indispensable for analyzing such interactions. System planners should consider these potential interactions of a wide range of complementary flexibility options for realizing least-cost renewable energy systems.

% Impact of nuclear power, % contentious role of nuclear power
Our findings prove robust against the addition of moderate levels of flexible nuclear generation, which mitigate the least-cost long-duration storage investments in our model only to a minimal extent. In a setting with substantially higher nuclear generation capacity, the storage-mitigating effect is more pronounced. Yet, substantial long-duration storage investments remain optimal in the least-cost system configuration. As we largely abstract from inter-hourly flexibility constraints, nuclear's potential to substitute long-duration storage needs as found here can be interpreted as an upper bound of this storage-mitigating effect. Including more specific operational constraints would likely reduce it. Notably, the role of nuclear in a climate-neutral European energy system is contentious due to scalability challenges, exceptionally long construction times, and high uncertainty in final investment costs as evident in ongoing expansion projects in Europe and globally \cite{goke_flexible_2025}. Likewise, nuclear waste management faces substantial economic, environmental, and societal challenges. Additionally, the future deployment of other advanced firm zero-emission generation technologies remains highly uncertain, as these require very optimistic cost assumptions to be competitive against variable renewables in combination with long-duration storage~\cite{stoecker_2025}.

% Sensitivity analysis: oil plus DACS -> eigentlich müssten wir schauen, ob das in allen jahren gezogen wird
% Wenn Analyse in allen Jahren, dann auf 42% gap zu 1984/85 abstellen
A sensitivity analysis for the weather year 1996/97, which includes the most extreme simulated renewable drought we find in the data, shows that the least-cost deployment of fossil backup capacity in combination with carbon abatement via \ac{DACCS} could result in minor system cost decreases, but only if \ac{DACCS} costs are very low. For higher \ac{DACCS} costs, which appear more likely given current knowledge \cite{young_cost_2023}, the system cost effect is only marginal. Under these scenarios, such backup capacity would only proportionally displace long-duration storage discharge capacity while storage energy capacity remains hardly affected in the least-cost system. Considering the significant uncertainty concerning realizable \ac{DACCS} costs and implementation levels, our analysis suggests that the role of fossil fuel-based backup capacity with emission abatement for mitigating long-duration storage remains limited in a climate-neutral European energy system. 
% Previous literature has shown the benefit of integrating green methanol into backup generation since methanol storage has a low capacity cost and can be stored and dispatched with interannual fluctuation

% Weather year selection, % TYNDP critique
Next, our analysis shows that the selection of weather years has a notable impact on least-cost long-duration storage use, which confirms previous research \cite{dowling_role_2020,brown_ultra_2023,diesing_exploring_2024,gotske_designing_2024}. Different weather years result in widely varying least-cost long-duration storage investments, particularly in scenarios with constrained interconnection.
%These inter-annual variations emphasize the significance of accounting for extreme renewable droughts when planning renewable energy systems, particularly long-duration storage sizing. 
However, due to computational limitations, many policy-relevant studies rely on only one or a limited set of weather years. For instance, the \ac{TYNDP} 2022 \cite{entso-e_tyndp_2022}, which is a crucial European planning tool to determine cross-border infrastructure needs, draws on three weather years (1995, 2008, and 2009) for its long-term scenarios for 2050. Our analysis shows that the least-cost long-duration storage capacity considering \ac{TYNDP} interconnection levels in the corresponding summer-to-summer weather years (1994/95, 1995/96, 2007/08, 2008/09, and 2009/10) ranges from 103 to 239~TWh. Notably, we find the least-cost storage capacity required to balance the most extreme drought in winter 1996/97 is 47\% higher than this range's upper bound. This underscores the importance of considering multiple weather years for identifying weather-resilient system configurations and long-duration storage sizing, particularly those that include the most pronounced drought events.

% VRE drought analysis
%To select such years, we propose using \ac{VRE} drought analysis based on renewable availability time series and multi-threshold indicators such as the drought mass metric. Importantly, respective time series analyses should rely on multiple drought thresholds and account for sequences of contiguous droughts with varying severity to adequately capture the most extreme events \cite{kittel_measuring_2024,kittel_quantifying_2024}. 

% Planning horizon
For Europe, the power sector impacts of \ac{VRE} droughts are most pronounced in winter, particularly in the winter of 1996/97 \cite{kittel_multi-threshold_2026}. We thus argue that our summer-to-summer modeling approach better captures the effects of compound drought events spanning across the turn of years, as compared to models using single calendar years.

% sensitivity: LDS energy and VRE costs -> doppelung mit results
%Another sensitivity analysis shows that least-cost long-duration storage and \ac{VRE} capacity depend on investment costs. Given tremendous recent cost decreases for renewable technologies, moderate deviations from the wind and solar costs in our default setting appear most plausible. For these cases, least-cost long-duration storage capacity is hardly affected. %this underpins our finding that lds is an indespensible technology for realizing a decarbonized european energy system heavily reling on wind and solar

% Sensitivity analysis: VOLL -> vllt SI, dann hier nicht relevant, sondern diskussion in SI 
% discussion: VOLLs für DE schwanken zwischen 7000 und 18000, wobei diese Findings tw. schon über 15 Jahre alt sind und inflationsbereinigt deutlich darüber liegen dürften, dh richtig relevant sind vor allem die Balken auf der rechten Seite.
% range from 1000 to 100000 EUR/MWh
% https://www.sciencedirect.com/science/article/pii/S0306261924022797: 7000-18000 EUR/MWh, but also  higher estimates, e.g., the social cost of unserved load (87,000$ MWh−1) estimated for the 2021 blackout in Texas, or 100000EUR/MWh in https://www.nature.com/articles/s41560-022-00994-y
% cost savings for unrealistically low VOLL arounds 1%, for more realistic VOLLs near zero. We conclude that load shedding is not a viable option for dealing with extreme renewable droughts

% High-level Einordung DF
Finally, there is no consensus on the definition of variable renewable energy droughts \cite{kittel_measuring_2024}. In the energy policy debate, events lasting from just a few hours to one or two weeks have been labeled as Dunkelflaute \cite{huneke_f_kalte_2017,the_economist_defying_2023,ronsch_problem_2024,bloomberg_europes_2024}. Based on our analysis, we propose refining the Dunkelflaute notion to focus on events with the most significant implications for long-duration power sector flexibility: extended (winter) periods where renewable energy falls short of electricity demand, which ultimately define the energy capacity and the operation of long-duration storage. We suggest not using the term Dunkelflaute for very short periods of low wind and solar availability, especially not for a few hours within a day.
% While such shorter events will become more frequent as \ac{VRE} penetration increases \cite{kittel_quantifying_2024}, dealing with them particularly requires short-duration flexibility options that rather focus on power and not on energy, such as battery storage. Yet, such short-duration technologies are likely to be inherently necessary for systems with high shares of wind and solar to balance regular diurnal demand and solar variations.

% Limitation: electrified space heating
Like in any model-based analysis, we made several assumptions and simplifications, whose qualitative effects we discuss briefly in the following. First, space heating will likely be electrified to a substantial extent in the future in Europe \cite{european_commission_repowereu_2022}. This will not only increase annual electricity demand but also lead to a more pronounced demand seasonality with higher load in winter. While we account for the projected additional electricity demand via linear scaling, our demand time series, which we chose to ensure consistency with the renewable energy time series, only have moderately seasonal profiles as projected for 2025 (see Section~\nameref{ssec:methods_tech_cap}). Alternative demand profiles with more pronounced heating-related seasonality are likely to cause higher least-cost long-duration electricity storage investments than modeled here. However, these could in turn be mitigated by long-duration thermal energy storage \cite{schmidt_2025,schmidt_mix_2025}, which we abstract from here. Similarly, winter heat demand peaks could be addressed by building insolations \cite{zeyen_mitigating_2021}. We expect that these factors will balance each other to some degree with respect to the needed long-duration electricity storage capacities. Yet, analyzing the interactions of different types of thermal storage, heating technologies, and building renovation with the power sector in detail is a complex task \cite{zeyen_mitigating_2021,roth_power_2024}, requires a dedicated research design beyond the scope of this paper, and merits future research.

% Limitation: electrified space cooling
Similarly, electricity demand is likely to increase in summer in many European countries because of a growing demand for cooling \cite{colelli_air_2023}. While this may increase flexibility requirements of the future energy system, we expect the additional demand from cooling would affect short-duration flexibility options rather than long-duration storage, considering typical diurnal cooling patterns and their partial coincidence with solar \ac{PV} generation peaks. Our analysis also indicates that the most pronounced renewable energy droughts, which also drive long-duration storage needs, primarily occur in winter and not in summer in an interconnected European energy system.

% Limitation: Other flex options, conventional backup, and alternative hydrogen storage technologies
Next, our analysis excludes industrial and commercial load shifting or shedding as well as optimized grid interactions of battery-electric vehicles. While these flexibility options would generally enhance system flexibility, their impact on long-duration storage capacity remains likely limited because of their short duration. Similarly, we do not consider long-duration storage technologies other than hydrogen cavern and porous storage in combination with electrolyzers and hydrogen turbines. For example, methanol-based long-duration storage in combination with Allam cycle turbines could be a potential alternative \cite{brown_ultra_2023}. This technology does not depend on highly localized underground sites but allows for aboveground storage that can generally be placed anywhere \cite{caglayan_technical_2020,talukdar_techno-economic_2024}, which could lead to spatial redistribution of least-cost storage capacity. Further, if this alternative storage technology came with higher roundtrip efficiencies or with lower costs than assumed here for hydrogen-based long-duration storage, least-cost storage capacities aggregated across all countries are likely to increase compared to our results. Importantly, we do not argue that hydrogen-based underground storage in combination with electrolyzers and hydrogen turbines is the only or best technology to serve as long-duration storage. We have chosen this technology as it seems the very promising given current knowledge. The least-cost long-duration storage capacities found in this analysis can also be considered as the energy amounts that need to be shifted or covered by any alternative technology.

% MGA implementation ladder level 1
Importantly, our analysis focuses exclusively on least-cost solutions and does not account for near-optimal system configurations, which could, for instance, exhibit lower implementation barriers or greater societal acceptability. Future work could employ a modeling-to-generate-alternatives approach to explore the extent to which such near-optimal capacity layouts exist and at what additional system costs \cite{lombardi_near-optimal_2025}. Likewise, it would be of interest to investigate how uncertainties over future technology cost or demand developments impact storage outcomes, e.g.,~by using Monte Carlo simulations or other techniques for optimization under uncertainty.

Further, the model used in this analysis is deterministic and assumes perfect foresight. Under limited foresight, long-duration storage might be operated more precautiously compared to our analysis, i.e.,~stockpiling more energy over longer time spans and avoiding very low states of charges to hedge against extreme drought events that may occur later in planning horizon \cite{schmidt_2025}. 
%In consequence, we argue that the least-cost storage energy capacities we find in this analysis under perfect foresight probably define a lower bound when considering a more realistic operation of long-duration storage under limited foresight.
Accordingly, the least-cost operational and capacity decisions of the long-duration storage found in this analysis should be interpreted as benchmark solutions. Under imperfect foresight, which applies to operators in the real world, long-duration storage needs may exceed those determined here. Likewise, energy system models that incorporate stochasticity may reach different conclusions regarding the role of technologies dominated by operational costs, such as \ac{DACCS}, warranting future research.

% hydropower
Moreover, we do not account for hydropower droughts, neither in the time series analysis nor in the capacity expansion model. Analyzing hydropower droughts and their interactions with wind and solar droughts appears to be a promising avenue for future research. At the same time, it appears unlikely that including hydro droughts would qualitatively alter our main findings, as the number of affected countries and the overall contribution of hydro power to European electricity supply remains limited.

% Outlook: heating
% Future work could also explore the effects of additional electricity demand in winter, driven by more pronounced electrification of space heating. In such a setting, low temperatures compounding extreme \ac{VRE} droughts would be of particular interest.
% A promising extension of our research design would be to investigate how various forms of thermal energy storage and heating technologies in combination with different levels of building renovation \cite{bloess_power_2018,zeyen_mitigating_2021,roth_power_2024} impact long-duration electricity storage needs, and how parameter assumptions like investment costs or standing losses influence results.

% Outlook: oil
%Future work could also explore the economics and applicability of dual use turbines in renewable energy systems, which could run on both oil and hydrogen.

% Outlook: time horizons, climate change
Future work could investigate the effects of different time horizons for energy modeling, ranging from single years to several years \cite{dowling_role_2020}, yet with a particular focus on renewable drought events. This also includes quantifying the distortions of modeling either summer-to-summer periods, as done in our study, or winter-to-winter periods. And while perfect foresight within a year results in relevant benchmark outcomes, considering imperfect foresight for long-duration storage operations within a year may offer complementary insights into real-world storage use. Further, exploring the impact of future changes in renewable energy droughts and storage needs driven by climate change appears desirable. This could build on an emerging body of research on climate-driven changes in wind and solar generation and datasets such as \cite{formayer_secures_2023,antonini_weather_2024}. %This strand of research would benefit from interdisciplinary collaboration at the intersection of climate and energy modeling \cite{craig_overcoming_2022}.

%%%%%%%%%%%%%%%%%%%%%%%%%%%%%%%%%%%%%%%%%%%%%%%%%%%%%%%
\section*{Methods}
%%%%%%%%%%%%%%%%%%%%%%%%%%%%%%%%%%%%%%%%%%%%%%%%%%%%%%%

In this study, we combine two methods: a \ac{VRE} drought analysis based on availability time series of wind and solar power, and a cost-minimizing capacity expansion model of the European power sector.
% The latter also considers additional technologies such as bioenergy and energy demand patterns.

\subsection*{Method and data for identification of variable renewable energy droughts}\label{sec:vre_analysis_method}

% VREDA intro
We use the open-source tool \ac{VREDA} to identify and evaluate \ac{VRE} drought patterns based on availability time series. \ac{VREDA} has been designed to implement good practices of multi-threshold drought identification as outlined by Kittel and Schill \cite{kittel_measuring_2024} and has been applied before to characterize drought patterns in Europe \cite{kittel_multi-threshold_2026}. It employs the \ac{VMBT} method for drought identification, which varies the permissible drought duration between two full calendar years and one hour. This method searches for periods where renewable availability has a moving average below a specific drought qualification threshold by iteratively decreasing the drought duration. In each iteration, the algorithm sets the averaging interval to the respective event duration. Initially, it searches for drought events that last two full years and iteratively decreases the averaging interval to identify shorter events. A time series section with a moving average below the drought threshold identifies a drought event. It is then excluded from subsequent iterations, in which the averaging interval decreases further and additional (shorter) events are identified.
%This routine ensures that no drought events are fully and uniquely accounted for in the search, without undue overlap \cite{kittel_measuring_2024}. 

% VMBT motivation
The iterative procedure of \ac{VMBT} overcomes shortcomings of previous research \cite{kittel_measuring_2024}. The method allows for pooling of adjacent periods that independently may not qualify as \ac{VRE} drought to capture longer-lasting events, i.e., intermediate periods with a renewable availability above the drought threshold. It further identifies unique events, avoids double counting as well as overlaps with adjacent events, and captures the full temporal extent of drought periods (``event'' definition). 

% Data
We use country-level \ac{VRE} availability time series provided by the Pan-European Climate Database, including 35~weather years from 1982 to 2016 for on- and offshore wind as well as solar \ac{PV} \cite{de_felice_entso-e_2022}. Using a summer-to-summer planning horizon, this means that we cover 34 complete winter seasons.

% Thresholds
For comparability across regions, drought thresholds are scaled relative to specific fractions of country-specific long-run mean availability factors for the period 1982 to 2016 \cite{kittel_measuring_2024}. These fractions range from 10 to 100\% of mean availability, increasing in 5\% increments and reflecting different levels of drought severity. We refer to these fractions to as ``relative thresholds''. Periods that qualify as drought events based on lower thresholds are likely brief and severe with very low availability. In contrast, events identified by higher thresholds may last substantially longer, i.e.,~up to several weeks or months \cite{kittel_multi-threshold_2026}.
% Note that our relative thresholds are time-invariant, i.e.,~they do not vary across seasons. While seasonally differentiated thresholds could help to distinguish between regular seasonal variations and extreme weather events, as demonstrated by \cite{stoop_climatological_2024, antonini_identification_2024}, they come with methodological challenges, as seasons differ across countries and weather variables. Further, the usefulness of seasonal thresholds for identifying extreme events that have the largest power sector impacts remains unclear. For example, similar deviations from seasonal averages might lead to substantially higher needs for long-duration storage in winter where demand is high than in summer where it is low.

% Composite time series
We analyze drought patterns for renewable technology portfolios, comprising on- and offshore wind power as well as solar \ac{PV}, for two cases that differ regarding the assumed electricity transmission between countries \cite{kittel_measuring_2024}: completely isolated countries (``energy islands'') or perfect interconnection across all countries (pan-European ``copperplate''). For the energy islands scenario, we combine all technology-specific time series into a portfolio time series using capacity-weighted averages. The respective weights are based on policy-relevant assumptions on renewable capacity mixes from the \ac{TYNDP} 2022 ``Distributed Energy'' scenario (default) and ``Global ambition'' scenario in a sensitivity analysis~\cite{entso-e_tyndp2022_2022}. We update these assumptions for Germany according to the latest government targets \cite{bundesamt_fur_justiz_gesetz_2024}. For the copperplate scenario, we combine all country-level portfolio time series into a single pan-European composite time series, using weights according to the \ac{TYNDP} 2022 (scenario ``Distributed Energy'').
%This composite time series represents the renewable portfolio availability of a pan-European copperplate. 
% While time series-based \ac{VRE} drought analysis does not allow for geographical balancing based on policy-oriented interconnection levels \cite{kittel_measuring_2024}, identified drought patterns of isolated countries appear to be a reasonable approximation for drought patterns including policy-oriented interconnection \cite{kittel_quantifying_2024}.

% Drought mass metric
We use the ``drought mass'' metric devised by Kittel and Schill \cite{kittel_multi-threshold_2026} to identify extreme drought events by aggregating the drought patterns of numerous single-threshold analyses, ranging in 5\%-increments from the 10\% to 75\% threshold. To compute the drought mass score, we first modify these patterns by assigning the value $1$ to drought hours and the value $0$ to those hours that do not qualify as drought for each threshold. Next, we equally weigh the resulting drought patterns across all thresholds. We then accumulate the hourly scores up to the cut-off threshold 75\%, excluding the drought patterns based on higher thresholds. This approach determines the multi-threshold event duration according to the 75\%-analysis, while the event drought mass aggregates the drought patterns identified by included thresholds. The highest cumulative score identifies the most extreme event per summer-to-summer planning horizon.

\begin{figure}[h]
\centering
    \includegraphics[width=\textwidth]{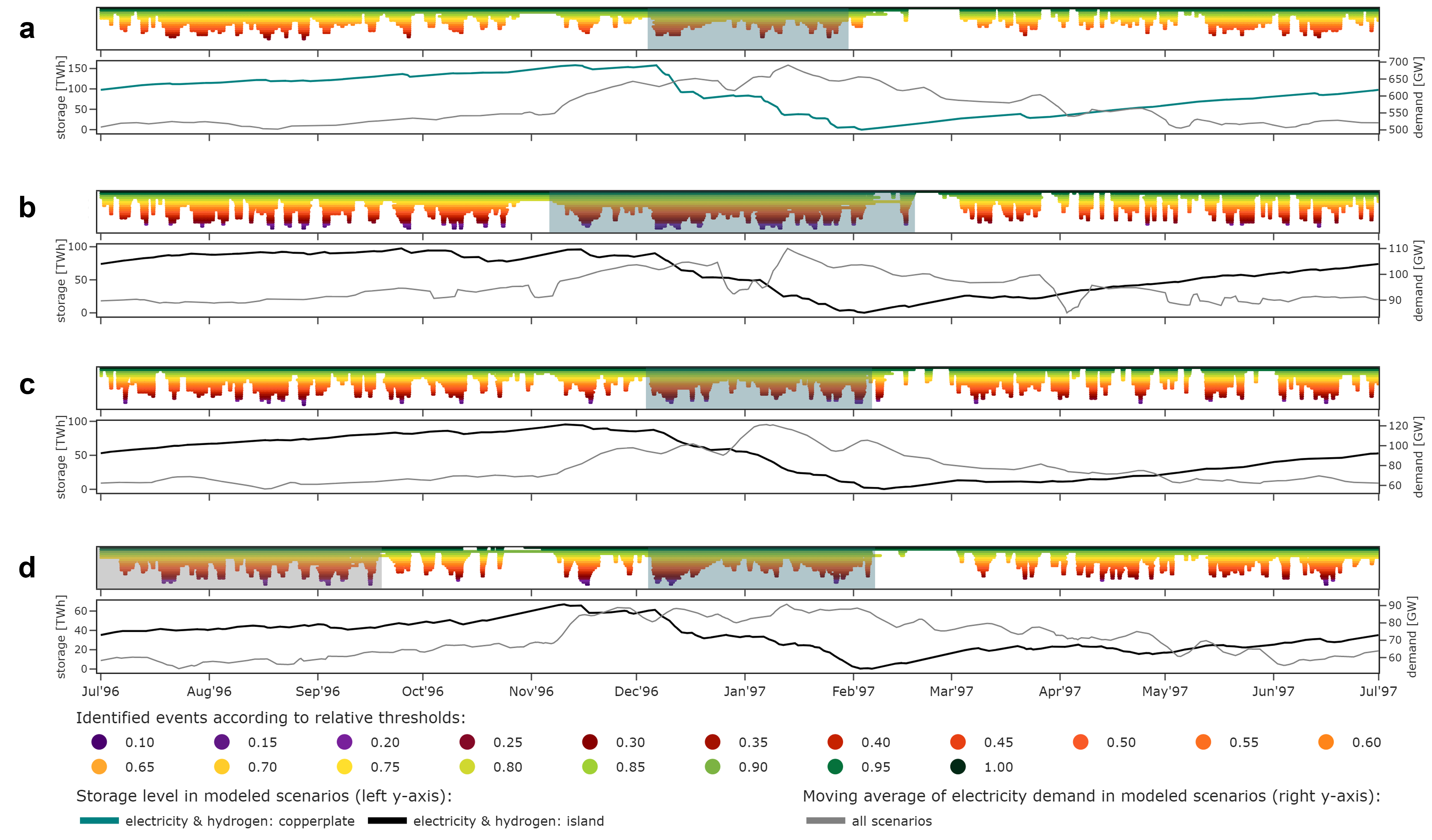}
    \caption{\textbf{Simulated drought events, electricity demand, and least-cost state-of-charge of long-duration storage in winter 1996/97.} The top row of each panel shows the identified drought patterns lasting longer than 12 hours across all color-coded thresholds, with the most extreme drought events occurring in winter (teal boxes) or throughout the year (gray boxes). The bottom row of each panel displays the associated exogenous smoothed demand profiles used in the optimization and the resulting least-cost storage state-of-charge levels. Panel \textbf{a} corresponds to the pan-European copperplate scenario with unconstrained energy exchange across Europe, \textbf{b} to Germany as an island system, \textbf{c} to France as an island system, and \textbf{d} to the United Kingdom as an island system.}
    \label{fig:figure_1}
\end{figure}

% Winter droughts
Wind droughts are more frequent and severe in summer than in winter \cite{cannon_using_2015,ohlendorf_frequency_2020,kittel_multi-threshold_2026}. In countries with high wind shares in their capacity portfolio, the most extreme renewable portfolio drought events may thus occur in summer. Since peak electricity demand periods usually occur in winter in Central and Northern European countries, summer droughts generally matter less than winter droughts. To account for this, we compute a yearly drought mass score for droughts occurring throughout the summer-to-summer planning horizon and a winter drought mass score for drought events between October and March. When illustrating the relation of drought patterns and long-duration storage operation, we display both the most extreme summer and winter droughts if the highest drought mass score relates to summer drought (compare gray and teal boxes in Figure~\ref{fig:figure_1}). Conversely, we mark only one event if the highest yearly drought mass score refers to a winter drought (compare teal boxes only in Figure~\ref{fig:figure_1}).

\subsection*{Power sector model}

% DIETER intro \& temporal horizon
We use the open-source dispatch and capacity expansion model of the European power sector \ac{DIETER} \cite{gaete-morales_dieterpy_2021,zerrahn_long-run_2017} to analyze the interaction between \ac{VRE} droughts and long-duration storage capacity investments in a fully renewable European power sector. The model features a simple transport model for exchanging electricity across countries, abstracting from grid constraints within countries. Different model versions have been applied to study various aspects of \ac{VRE} integration and their interaction with flexibility options or sector coupling technologies \cite{zerrahn2018,schill_long-run_2018,schill_electricity_2020,schill_flexible_2020,kittel_renewable_2022,gils_model-related_2022,roth_geographical_2023,kirchem_power_2023,gaete-morales_power_2024,gueret_impacts_2024,roth_power_2024,schmidt_2025}. \ac{DIETER} is a linear program that determines least-cost capacity and dispatch decisions, optimizing over all contiguous hours of a full year under perfect foresight.

% General model setup
Exogenous model inputs entail techno-economic parameters such as investment and variable costs, availability time series of wind and solar \ac{PV}, as well as price-inelastic demand time series for electricity and hydrogen. Model results can be interpreted as the outcomes of a perfect, frictionless European electricity market, where all power generators maximize their profits. Costs are minimized for the overall system, depending on interconnection assumptions. In the energy islands case, costs are minimized for every country in isolation.

%%% Figure 11

\begin{figure}[h]
    \centering
    \includegraphics[width=0.8\linewidth, keepaspectratio]{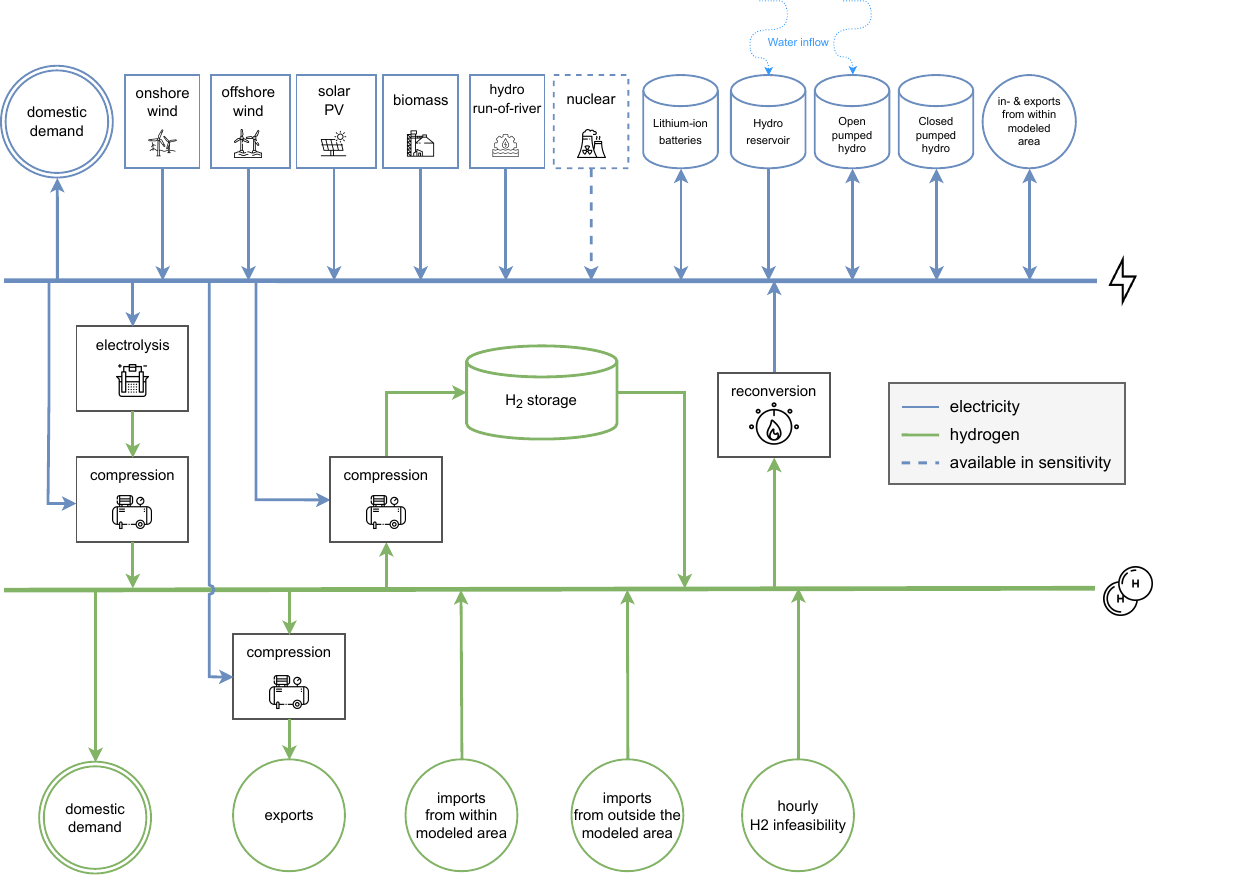}
    \caption{\textbf{Schematic overview of the model \ac{DIETER}.} The figure depicts a schematic representation of the energy system model DIETER, which has been used in this study. Blue elements refer to electricity-related elements, while green to hydrogen-related. They are connected by electrolysis and reconversion. This cover has been designed using resources from Flaticon.com.}
    \label{fig:figure_11}
\end{figure}

Figure~\ref{fig:figure_11} provides an overview of the model's structure, how and with which technologies electricity and hydrogen can be generated, stored, and exchanged.

% Formal definition of the hydrogen module.
While the general model formulation of \ac{DIETER} has been described extensively in the papers mentioned above, we use a version that features an improved representation of hydrogen technologies in this study. The model includes the generation, storage, and transport of renewable hydrogen technologies as well as its re-conversion to electricity. The hydrogen-based long-duration storage energy capacity denotes the lower heating value of the storage working gas. A formal definition of the additional equations is available in Section~\nameref{ssec:hydrogen_model} in the Supplementary Information.

\subsection*{Scenarios} \label{sec:scenarios}

% Spatial scope and interconnection scenarios
Our analysis comprises 33~European countries (EU27 excl. Malta and Cyprus, the UK, Norway, Switzerland, and the Western Balkans). We analyze and compare the impact of \ac{VRE} droughts on least-cost long-duration storage capacities for four different interconnection scenarios: (1) all countries operate as energy islands without exchange of electricity or hydrogen between countries; (2) cross-border exchange of electricity in line with the assumptions on the European transmission grid in 2050 projected by the \ac{TYNDP} 2022 \cite{entso-e_tyndp2022_2022} (scenario Distributed Energy), while hydrogen exchange remains disabled; (3) cross-border electricity and hydrogen exchange according to the \ac{TYNDP} 2022 (scenario Distributed Energy); and (4) unlimited exchange of electricity and hydrogen, i.e,~all countries are perfectly integrated as pan-European copperplate. By design, the \ac{VRE} drought analysis based on renewable availability time series does not account for limited cross-border exchange. Consequently, it is conducted only for the two counterfactual interconnection scenarios (1) and (4). In the latter case, the drought identification is based on a renewable availability time series that combines all countries into a capacity-weighted single pan-European composite time series (compare Section~\nameref{sec:vre_analysis_method}).

%\begin{itemize}
%    \item electricity network draws on ENTSO-e TYNDP 2022 including West Balkans
%    \item hydrogen network draws on ENTSO-g TYNDP 2022, partially excluding West Balkans
%    \item hydrogen imports from world market, Ukraine, North Africa based on TYNDP ENTSO-g 2022, assumption baseload contracts for max import capacity
%\end{itemize}

% Motivation of interconnection scenarios
These varying degrees of interconnection allow distinguishing several effects. First, scenario (1) identifies maximal long-duration storage needs across European countries, excluding geographical balancing of \ac{VRE} droughts between countries \cite{kittel_multi-threshold_2026}. Second, comparing scenarios (2) and (3) disentangles the effects of policy-oriented electricity and hydrogen interconnection levels on long-duration storage needs. Finally, the copperplate scenario (4) identifies minimal long-duration storage needs, including unconstrained geographical balancing of \ac{VRE} droughts, which can be interpreted as unavoidable or ``no-regret'' investments in a fully renewable European energy system. We abstract from grid investment costs but impose losses on hydrogen exchange related to compression for long-distance transport via pipeline.

% Inter-annual variability
%Our objective is to assess the role of system flexibility and the impact of inter-annual variability in the context of \ac{VRE} droughts. Besides varying interconnection levels, we therefore 
We investigate a large number of weather years that differ in renewable availabilities and demand patterns. These weather years range from 1982 to 2016. Extreme \ac{VRE} droughts often occur in European winter, may last up to several weeks, and span across the turn of years \cite{kittel_multi-threshold_2026}. Hence, we use a summer-to-summer planning horizon (in line with \cite{ruhnau_storage_2022,grochowicz_using_2024}, but unlike, e.g.,~\cite{gotske_designing_2024}), comprising 8760 consecutive hours. Note that we independently optimize the capacity mix for every year.

In two additional scenarios, we exogenously add different levels of nuclear power to the generation mix, aligned with the ``Distributed Energy'' and ``Global Ambition'' scenarios from the \ac{TYNDP}~2022, respectively \cite{entso-e_tyndp2022_2022}. The first of these scenarios assumes low levels of nuclear capacity in Finland, France, Romania, Slovakia, and the UK, totaling 24~GW. The second scenario scenario assumes high levels of nuclear power, assuming additional deployment in the formerly mentioned countries as well as in Bulgaria, the Czech Republic, Hungary, Poland, Sweden, and Slovenia, totaling 102~GW.

\subsection*{Input data, technology portfolio, and capacity bounds}\label{ssec:methods_tech_cap}

% Fully renewable scenarios
To quantify the maximum impact of \ac{VRE} droughts on long-duration storage, we model fully renewable supply scenarios in our default setting, excluding carbon removal and fossil fuel-based dispatchable generation technologies. This is not enforced by binding renewable or carbon emission targets \cite{kittel_renewable_2022}, but rather by limiting the available generation technology portfolio to zero-emission options. These include solar \ac{PV}, on- and offshore wind power, bioenergy, and different types of hydroelectric power (run-of-river, reservoir, pumped-hydro). %In a complementary scenario, we add a nuclear power as firm zero-emission generation technology to the available technology portfolio.

% Capacity bounds
For policy relevance, we allow generation capacity expansion within lower and upper potentials of the \ac{TYNDP} 2024 \cite{entso-e_tyndp_2024}. Additionally, we assume an annual generation limit for bioenergy in line with the \ac{TYNDP} 2022 \cite{entso-e_tyndp_2022}. Installed power and energy capacities of the available hydro technologies are fixed to values provided in \ac{ERAA} 2021 \cite{entso-e_european_2021}.

% Storage bounds
We include underground hydrogen storage as a long-duration storage option. Underground cavern and porous storage energy potentials vary substantially across countries, including newly built and retrofitted storage facilities from natural gas infrastructure \cite{caglayan_technical_2020,talukdar_techno-economic_2024}. We constrain the expansion of long-duration storage energy capacities accordingly. For simplicity, we abstract from any differentiation between underground storage types and aggregate their potentials for each modeled country.

% Sector coupling: BEV, heat, HPs
In our analysis, we attempt to illustrate the impact of persistent \ac{VRE} droughts lasting longer than a few days on the power sector, notably regarding long-duration electricity storage. We thus abstract from an explicit representation of sector coupling options %that are used primarily for short-term flexibility 
as well as seasonal heat storage. To this end, we use near-term demand profiles, retrieved from \ac{ERAA} 2021 \cite{entso-e_european_2021} (representative for the year 2025). The profiles are scaled to the annual demand levels of the \ac{TYNDP} (scenario Distributed Energy for 2050) and are reduced by the electrical energy amount required for the generation of the exogenous hydrogen demand. 

% RES availability data
We use the Pan-European Climate Database for renewable availability factors of on- and offshore wind and solar \ac{PV} \cite{de_felice_entso-e_2022}, as well as the hydro inflow, retrieved from the \ac{ERAA} 2021 \cite{entso-e_european_2021}.
%https://iopscience.iop.org/article/10.1088/1748-9326/aca1d3/pdf

\section*{Data availability}

The input data analyzed by the \ac{VRE} drought analysis tool are a refined version of the Pan-European Climate Database (PECD 2021.3) and publicly available at Zenodo \cite{de_felice_entso-e_2022}. The input data of the capacity expansion model of the European power sector are also publicly available on GitLab at \url{https://gitlab.com/diw-evu/projects/power-sector-droughts/}. The power sector outcomes generated in this study have been deposited in a publicly available Zenodo repository \cite{kittel_minimum_2026}.

\section*{Code availability}

For transparency and reproducibility, we provide the code of the \ac{VRE} drought analysis tool in public repositories under permissive licenses available on GitLab at \url{https://gitlab.com/diw-evu/variable_renewable_energy_droughts_analyzer}. The code and manual of the capacity expansion model of the European power sector are also publicly available on GitLab at \url{https://gitlab.com/diw-evu/projects/power-sector-droughts/}.

%\newpage

%\printbibliography

%\bibliographystyle{unsrt} % unsorted bibliography
%\bibliography{references.bib}

%\putbib
%\end{bibunit}

% compiled bibliography 1 copied here:

\section*{Acknowledgments}

We thank the entire research group ``Transformation of the Energy Economy'' at the German Institute for Economic Research (DIW Berlin) for valuable inputs and discussions, as well as conference participants of the International Conference Energy \& Meteorology 2023, the International Association for Energy Economics Conference 2023, the ENERDAY Conference 2024, and the Annual Conference of the European Association of Environmental and Resource Economists 2024 for valuable comments on earlier drafts.

\section*{Funding sources}

The authors acknowledge support for the research of this work from the Einstein Foundation Berlin (grant no.~A-2020-612) and by the German Federal Ministry of Education and Research via the ``Ariadne'' projects (Fkz 03SFK5NO \& 03SFK5NO-2).

\section*{Author contributions}\label{sec:author contributions}

% https://www.elsevier.com/authors/journal-authors/policies-and-ethics/credit-author-statement

M.K. (lead), A.R. (support), W.S. (support) developed the concept of the study. M.K. developed the methodology and the time series analysis tool. M.K. (lead), A.R. (support), W.S. (support) developed the energy system model. M.K. (lead) and A.R. (support) were responsible for software development and usage. M.K. (lead), A.R. (support), W.S. (support) carried out the investigation. M.K. and A.R. curated data requirements. M.K. compiled visualizations. M.K. (lead), A.R. (lead), W.S. (support) wrote the original draft. M.K. (lead), A.R. (support), W.S. (support) reviewed and edited the manuscript. W.S. was responsible for project acquisition and administration.

%\textbf{Martin Kittel}: Conceptualization (lead), methodology, model development (lead), software (lead), investigation (lead), data curation (equal), visualization, writing - original draft (equal), review and editing (lead). \textbf{Alexander Roth}: Conceptualization (support), software (support), investigation (support), data curation (equal), writing - original draft (equal), review and editing (support). \textbf{Wolf-Peter Schill}: Conceptualization (support), model development (support), investigation (support), writing - original draft (support), writing - review and editing (support), project acquisition and administration.

\section*{Competing interests}
The authors declare no competing interests.

%%%%%%%%%%%%%%%%%%%%%%%%%%%%%%%%%%%%%%%%%%%%%%%%%%%%%%%%%%%%%%%%%%%%%%%%%%%
    
\newpage

%%% figures %%%%

%%%% alternative with more panels
\begin{comment}
\begin{figure}[h]
\centering
    \includegraphics[width=\textwidth]{figures/Figure_1.png}
    \caption{\textbf{Simulated drought events, electricity demand, and least-cost state-of-charge of long-duration storage in winter 1996/97.} Panels \textbf{a}, \textbf{c}, \textbf{e}, \textbf{g} show the identified drought patterns lasting longer than 12 hours across all color-coded thresholds, with the most extreme drought events occurring in winter (teal boxes) or throughout the year (gray boxes). Panels \textbf{b}, \textbf{d}, \textbf{f}, \textbf{h} show the associated exogenous smoothed demand profiles used in the optimization and the resulting least-cost storage state-of-charge levels. Panels \textbf{a}, \textbf{b} correspond to the pan-European copperplate scenario with unconstrained energy exchange across Europe; \textbf{c}, \textbf{d} to Germany as an island system; \textbf{e} and \textbf{f} to France as an island system; and \textbf{g}, \textbf{h} to the United Kingdom as an island system.}
    \label{fig:figure_1}
\end{figure}
\end{comment}

\begin{figure}[h]
\centering
    \includegraphics[width=\textwidth]{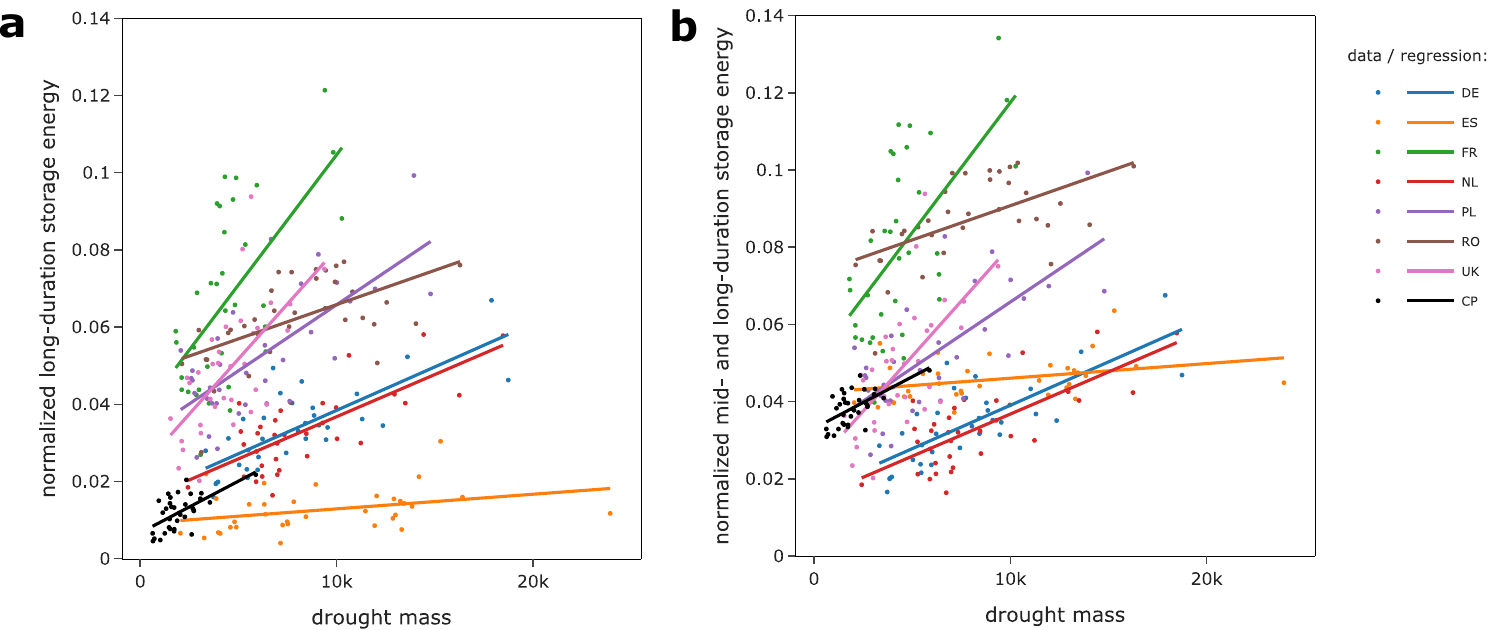}
    \caption{\textbf{Correlation of the drought mass of most extreme winter drought events and normalized storage energy capacity.} For comparison, we normalize the least-cost storage energy with the annual demand for electricity (including electrified heating) and hydrogen. For illustration, we exclude countries with least-cost storage energy below 5~TWh and countries with binding storage expansion potential constraints. Supplementary Fig.~\ref{fig:figure_si3} shows the unfiltered regression results. We further include the pan-European copperplate scenario (CP). \textbf{a} Long-duration storage only. \textbf{b} Mid- and long-duration storage. }
    \label{fig:figure_2}
\end{figure}

\begin{figure}[h]
    \centering
    \includegraphics[width=\textwidth]{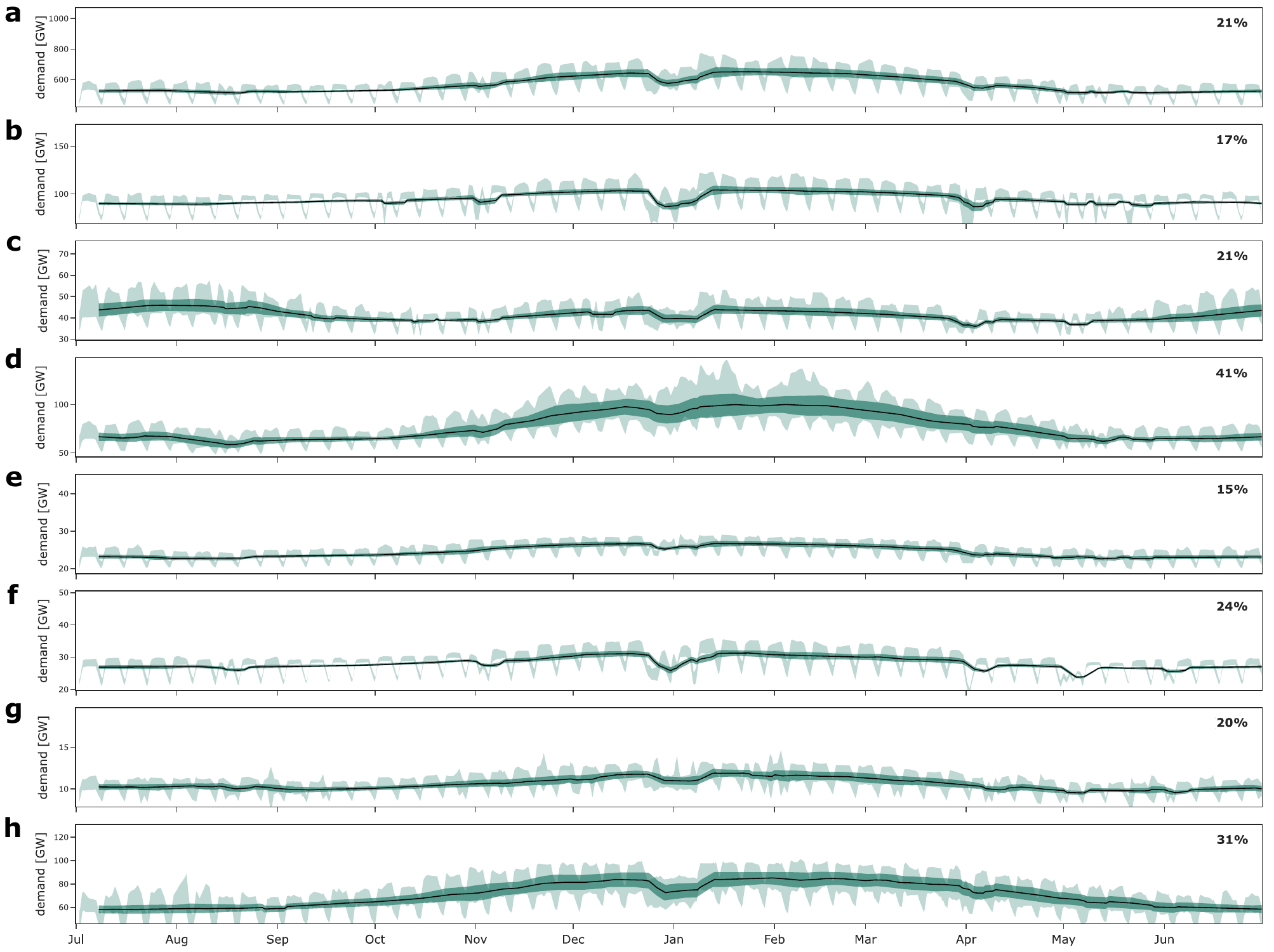}
    \caption{\textbf{Demand seasonality across regions.} The figure shows climatological mean demand as a bold line over all weather years using a moving average over a window of 168 hours (resulting in the blank first week) as a line, the standard deviation range as an area ($mean \pm std \ dev$, dark green), the difference between the climatological minimum and maximum as an area using a moving average over a window of 24 hours (light green), and regional difference between the minimum and maximum climatological mean normalized by the maximum in the upper right corner (denoted as percentage). For comparison, each vertical axis is scaled to show its range if demand seasonality in this region was as pronounced as in France. Demand seasonality is particularly pronounced in France both in terms of level but also variance during winter due to high shares of electrified heat. Panel \textbf{a} corresponds to the pan-European copperplate scenario, \textbf{b} to Germany, \textbf{c} to Spain, \textbf{d} to France, \textbf{e} to the Netherlands, \textbf{f} to Poland, \textbf{g} to Romania, and \textbf{h} to the United Kingdom.}
    \label{fig:figure_3}
\end{figure}

\begin{figure}[h] 
    \centering\includegraphics[width=.99\textwidth, keepaspectratio]{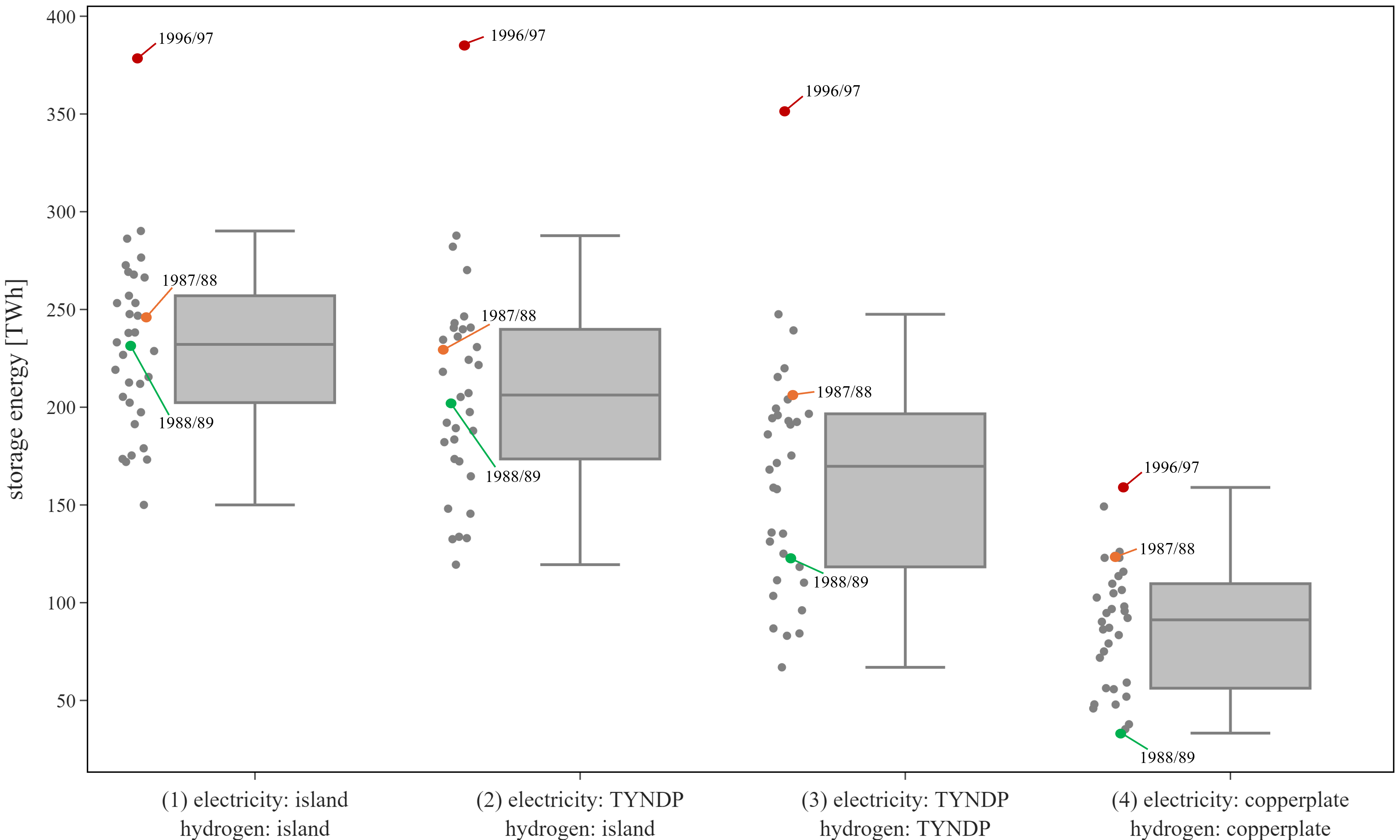}
    \caption{\textbf{Least-cost long-duration storage energy capacity aggregated across all countries for all modeled weather years and interconnection scenarios. } Each dot refers to one weather year, which is modeled independently of other weather years. The center line denotes the median, box limits indicate the interquartile range (Q1–Q3), whiskers extend to 1.5× the interquartile range, and points beyond the whiskers represent outliers. The year with the highest long-duration storage need is 1996/97 (red). The year that benefits most from rising interconnection capacity in terms of decreasing long-duration storage investments is 1988/89 (green). The year that benefits the least from increasing interconnection is 1987/88 (orange). Supplementary Fig.~\ref{fig:figure_si8} illustrates the impact of interconnection on the ranking of weather years in terms of least-cost long-duration storage energy.}
    \label{fig:figure_4}
\end{figure}

\begin{figure}[h]
\centering
    \includegraphics[width=\textwidth]{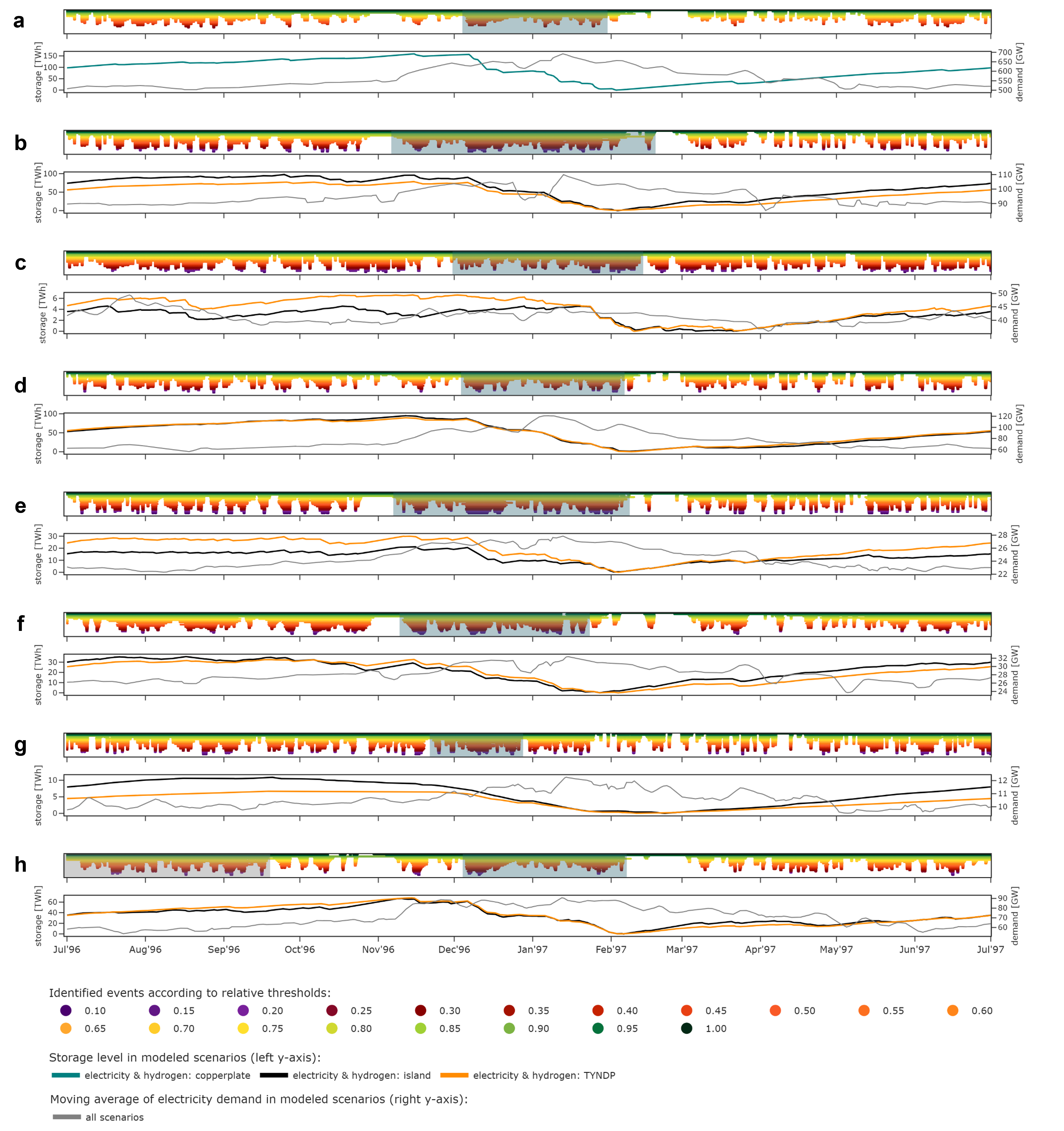}
    \caption{\textbf{Simulated drought events, electricity demand, and least-cost state-of-charge of long-duration storage in winter 1996/97 in countries with highest long-duration storage energy capacities.} The top row of each panel shows the identified drought patterns lasting longer than 12 hours across all color-coded thresholds, with the most extreme drought events occurring in winter (teal boxes) or throughout the year (gray boxes). The bottom row of each panel displays the associated exogenous smoothed demand profiles used in the optimization and the resulting least-cost storage state-of-charge levels for isolated countries modeled within the interconnection scenario (1), for policy-oriented interconnection levels in scenario (3), or the pan-European copperplate in scenario (4). Panel \textbf{a} corresponds to the pan-European copperplate scenario, \textbf{b} to Germany, \textbf{c} to Spain, \textbf{d} to France, \textbf{e} to the Netherlands, \textbf{f} to Poland, \textbf{g} to Romania, and \textbf{h} to the United Kingdom.}
    \label{fig:figure_5}
\end{figure}

%\begin{comment}

\clearpage
\newpage

%%%%%%%%%%%%%%%%%%%%%%%%%%%%%%%%%%%%%%%%%%%%%%%%%%%%%%%%%%%%%%%%%%%%%%%%%%%%%%%%
% APPENDIX
%%%%%%%%%%%%%%%%%%%%%%%%%%%%%%%%%%%%%%%%%%%%%%%%%%%%%%%%%%%%%%%%%%%%%%%%%%%%%%%%

\appendix
\nolinenumbers

\renewcommand{\theequation}{\arabic{equation}}  % SI.E.
\setcounter{secnumdepth}{0}  % No numbering
\setcounter{tocdepth}{4}     % Include paragraphs in TOC

\renewcommand{\tablename}{Supplementary Table}
\renewcommand{\figurename}{Supplementary Fig.}

%\setcounter{table}{0}

%\global\long\def\thefigure{SI.\arabic{figure}}
%\global\long\def\thetable{SI.\arabic{table}}
\global\long\def\thepage{\arabic{page}}  % SI.
\setcounter{figure}{0}
\setcounter{table}{0}
\setcounter{page}{1}
\setcounter{equation}{0}

\newcommand{\invisiblesection}[1]{%
  \phantomsection%
  \stepcounter{section}%
  \addcontentsline{toc}{section}{\protect\numberline{\thesection}#1}%
  }

%\begin{bibunit}

% PAGE CONTENT

\section{Supplementary Information}

\noindent\large{\textbf{Long-duration electricity storage needs for coping with Dunkelflaute events in Europe}}

\vspace{1cm}

\vspace{1.5cm}

\noindent\normalsize{Martin Kittel*, Alexander Roth, and Wolf-Peter Schill}

\noindent\footnotesize{*Corresponding author: mkittel@diw.de}

\newpage

% TOC
\localtableofcontents
%\locallistoffigures
%\locallistoftables

\newpage

\normalsize

%%%%%%%%%%%%%%%%%%%%%%%%%%%%%%%%%%%%%%%%%%%%%%%%%%%%%%%%%%%%%%%%%%%%%%%%%%%
%\section*{Supplementary information}\label{sec:si}
%%%%%%%%%%%%%%%%%%%%%%%%%%%%%%%%%%%%%%%%%%%%%%%%%%%%%%%%%%%%%%%%%%%%%%%%%%%

\subsection{Supplementary Methods}

\subsubsection{Formal definition of the hydrogen module}\label{ssec:hydrogen_model}

In the following, we present the model equations implementing the generation, storage, and transport of renewable hydrogen technologies as well as its re-conversion to electricity. For simplicity, we use only a single technology for each of these features.

Endogenous model variables are given in capital letters and exogenous parameters in lowercase. We denote the electricity demand for the production of green hydrogen production $G^{ely}_{n,ely,t}$ in time step $t$ and country $n$. Each electrolysis technology $ely$ has a specific efficiency $eta^{ely}_{n,ely} < 1$ and a flat availability $avail^{ely}_{n,ely}$. The electricity demand of an electrolyzer must not exceed its generation capacity $N^{ely}_{n,ely}$:

\begin{equation}\label{eq:ely_con_max}
    G^{ely}_{n,ely,t} \leq avail^{ely}_{n,ely} * N^{ely}_{n,ely}
\end{equation}

The produced hydrogen is compressed to a system-wide pressure level, constrained by the compressor capacity $N^{comp,ely}_{n,ely}$ with an efficiency $\eta^{comp,ely}_{n,ely}<1$, and fed into a hydrogen grid that links generation, storage, and re-conversion units as well as import and export pipelines: 

\begin{equation}\label{eq:ely_comp_max}
    eta^{ely}_{n,ely} * G^{ely}_{n,ely,t} \leq avail^{ely}_{n,ely} * N^{comp,ely}_{n,ely}
\end{equation}

Using a storage technology $sto$, hydrogen can be stored. Hydrogen that is added to storage $STO^{in}_{n,sto,t}$ has to be compressed. Compression losses are reflected in the charging efficiency $eta^{comp,in}_{n,sto} < 1$. A storage energy balance links the storage state-of-charge $STO^{L}_{n,sto}$ inter-temporally. While there might be self-discharge $eta^{sto,self}_{n,sto} < 1$, we assume storage discharge $STO^{out}_{n,sto,t}$ to be lossless due to the high-pressure level: 

\begin{equation}\label{eq:sto_lev}
    STO^{L}_{n,sto,t} = eta^{sto,self}_{n,sto} * STO^{L}_{n,sto,t-1} + eta^{comp,in}_{n,sto} * STO^{in}_{n,sto,t} - STO^{out}_{n,sto,t} 
\end{equation}

To avoid free lunch, we require that the storage levels are equal in the first time step $t=1$ and last time step $t=T$:

\begin{equation}\label{eq:sto_lev_start}
    STO^{L}_{n,sto,1} = eta^{sto,self}_{n,sto} * STO^{L,initial,last}_{n,sto} + eta^{comp,in}_{n,sto} * STO^{in}_{n,sto,1} - STO^{out}_{n,sto,1} 
\end{equation}

\begin{equation}\label{eq:sto_lev_end}
    STO^{L}_{n,sto,T} = STO^{L,initial,last}_{n,sto}
\end{equation}

Suppose $avail^{sto}_{n,sto}$ is the flat availability of each storage technology $sto$, the storage level cannot exceed the installed storage energy capacity $N^{sto}_{n,sto}$:

\begin{equation}\label{eq:sto_lev_max}
   STO^{L}_{n,sto,t} \leq avail^{sto}_{n,sto} * N^{sto}_{n,sto}
\end{equation}

To maintain a minimum pressure level required for the cushion gas, we impose a minimum filling state $\phi^{sto,min}_{n,sto} < 1$:

\begin{equation}\label{eq:sto_lev_min}
   STO^{L}_{n,sto,t} \geq \phi^{sto,min}_{n,sto} * avail^{sto}_{n,sto} * N^{sto}_{n,sto}
\end{equation}

Note that in this paper, we abstract from minimum filling levels to focus on required the working gas, i.e.,~$\phi^{sto,min}_{n,sto} = 0$. Hourly storage charging is constrained by the compression capacity $N^{comp,sto}_{n,sto}$:

\begin{equation}\label{eq:max_sto_in}
   STO^{in}_{n,sto,t} \leq avail^{sto}_{n,sto} * N^{comp,sto}_{n,sto}
\end{equation}

Hourly storage discharge is constrained by a maximum discharge rate $\phi^{sto,max}_{n,sto} < 1$:

\begin{equation}\label{eq:max_sto_out}
   STO^{out}_{n,sto,t} \leq \phi^{sto,max}_{n,sto} * avail^{sto}_{n,sto} * N^{sto}_{n,sto}
\end{equation}

Hydrogen can be re-converted to electricity. Suppose $avail^{recon}_{n,recon}$ is the flat availability of the re-conversion technology $recon$ and $\eta^{recon}_{n,recon} < 1$ the conversion efficiency, the electricity output $G^{recon}_{n,recon,t}$ is then limited by capacity of the re-conversion unit $N^{recon}_{n,recon}$:

\begin{equation}\label{eq:recon_max}
   G^{recon}_{n,recon,t} \leq avail^{recon}_{n,recon} * N^{recon}_{n,recon}
\end{equation}

Hydrogen can be generated domestically, imported from other world regions, or exchanged hydrogen across modeled countries based on a simple cross-border transport model. Hydrogen flows are denoted $F_{p,t}$. The set $p$ consists of all pipelines within the modeled area and those import pathways $path$ from outside the modeled area. The latter include pipeline-based imports from North Africa or Ukraine and ship-based imports from the world market. A hydrogen flow may not exceed its pipeline or shipping capacity $N^{trans}_{p}$:

\begin{equation}\label{eq:flow_max}
   F_{p,t} \leq N^{trans}_{p}
\end{equation}

We assume constant hydrogen imports $IM^{const}_{path}$ from outside the modeled area:

\begin{equation}\label{eq:const_import}
   F_{path,t} = IM^{const}_{path}
\end{equation}

% IF USED ADD IMPORT FLEXIBILIZATION
%\begin{equation}\label{eq:min_import}
%   F_{l,t} \geq (1 - import^{flex}_{path}) * IM^{const}_p
%\end{equation}
%
%\begin{equation}\label{eq:max_import}
%   F_{l,t} \geq (1 + import^{flex}_{path}) * IM^{const}_p
%\end{equation}

The sum of all imports accumulates to $IM_{path}$:

\begin{equation}\label{eq:sum_import}
   \sum_t F_{path,t} = IM_{path}
\end{equation}

To avoid infeasibility, we allow for slack hydrogen generation $INFES_{n,t}$ which needs to equal $INFES^{const}_{n}$ in all time steps to emulate constant imports in settings with disabled hydrogen flows:

\begin{equation}\label{eq:infeas}
   INFES_{n,t} = INFES^{const}_{n}
\end{equation}

Lines and countries are linked via the grid and import pathways, which are represented by the directed incidence matrices $inc^{import}_{p,n}$ and $inc^{export}_{p,n}$. Exports incur losses, represented by the transport efficiency $\eta^{trans}_p < 1$. The model considers a hydrogen demand for applications in industry, transport, or heat, which leans on \ac{TYNDP} 2022 (scenario Distributed Energy). We assume this exogenous demand $d_t$ to be constant in each time step. The hydrogen balance of each country $n$ ensures that the demand is met in each time step:

\begin{equation}\label{eq:h2_balance}
\begin{split}
    d^{h2}_t 
    &+ \sum_{sto} STO^{in}_{n,sto,t} + \sum_{recon} \frac{G^{recon}_{n,recon,t}}{\eta^{recon}_{n,recon}} + \sum_{p} inc^{export}_{p,n} * \frac{F_{p,t}}{\eta^{trans}_p} \\
    &= \sum_{ely} \eta^{ely}_{n,ely} * \eta^{comp,ely}_{n,ely} * G^{ely}_{n,ely,t} + \sum_{sto} STO^{out}_{n,sto,t} + \sum_{p} inc^{import}_{p,n} * F_{p,t} + INFES_{n,t}.
\end{split}
\end{equation}

The hydrogen grid is linked to the electricity grid through the electricity demand for electrolysis $G^{ely}_{n,ely,t}$, hydrogen compression after electrolysis $d^{comp,ely}_{n,ely}$ for injecting into the hydrogen grid, hydrogen compression for injection into storage $d^{comp,sto}_{n,sto}$, and hydrogen booster compression for exporting pipelines $d^{comp,trans}_{n,p}$. For this, we add Supplementary Equation (\ref{eq:el_balance_demand}) to the demand and Supplementary Equation (\ref{eq:el_balance_supply}) to the supply side of the electricity balance of each country $n$, respectively. The latter is documented in Zerrahn and Schill \cite{zerrahn_long-run_2017_si}:

\begin{equation}\label{eq:el_balance_demand}
    ... + \sum_{ely} (1 + \eta^{ely}_{n,ely} * d^{comp,ely}_{n,ely}) * G^{ely}_{n,ely,t} + \sum_{sto} d^{comp,sto}_{n,sto} * STO^{in}_{n,sto,t} + \sum_p d^{comp,trans}_{n,p} * inc^{export}_{p,n} * \frac{F_{p,t}}{\eta^{trans}_p}
\end{equation}

\begin{equation}\label{eq:el_balance_supply}
    ... + \sum_{recon} G^{recon}_{n,recon,t}
\end{equation}

The model endogenously determines electrolysis, storage, compression, and re-conversion capacity and operation. The operational decisions for hydrogen transport are also endogenous, while we assume exogenous transport capacities. Suppose $c^{var,sto}_{n,sto}$ are the operational costs for storage charging, $c^{var,recon}_{n,recon}$ the operational costs for re-conversion, $c^{import}_{path}$ the costs for importing hydrogen from outside the modeled area. To avoid infeasibilities, we allow unspecified imports of hydrogen incurring the costs $c^{infes}$ (here at prohibitively high 500 EUR per MWh\textsubscript{ch}). Suppose further $c^{oc}_{n}$ and $c^{fix}_{n}$ are the overnight and fixed investment costs, we impose them on the capacity of electrolysis technology $ely$, compression capacity after electrolysis $comp,ely$, compression capacity for storage injection $comp,sto$, energy capacity of storage technology $sto$, and capacity of re-conversion technology $recon$. We add these operational (Supplementary Equation (\ref{eq:opex})) and investment costs (Supplementary Equation (\ref{eq:capex})) to the model's optimization function, which minimizes total system costs. The latter is documented in Zerrahn and Schill \cite{zerrahn_long-run_2017_si}:

\begin{equation}\label{eq:opex}
    ... + \sum_{n,sto,t} c^{var,sto}_{n,sto} * STO^{in}_{n,sto,t} + \sum_{n,recon,t} c^{var,recon}_{n,recon} * G^{recon}_{n,recon,t} + \sum_{path} c^{import}_{path} * IM_{path} + \sum_{n,t} c^{infes} * INFES_{n,t}
\end{equation}

\begin{equation}\label{eq:capex}
\begin{split}
    ... 
    &+ \sum_{n,ely} (c^{oc,ely}_{n,ely} + c^{fix,ely}_{n,ely}) * N^{ely}_{n,ely} + \sum_{n,ely} (c^{oc,comp,ely}_{n,ely} + c^{fix,comp,ely}_{n,ely}) * N^{comp,ely}_{n,ely} \\
    &+ \sum_{n,sto} (c^{oc,comp,sto}_{n,sto} + c^{fix,comp,sto}_{n,sto}) * N^{comp,sto}_{n,sto} + \sum_{n,sto} (c^{oc,sto}_{n,sto} + c^{fix,sto}_{n,sto}) * N^{sto}_{n,sto} \\
    &+ \sum_{n,recon} (c^{oc,recon}_{n,recon} + c^{fix,recon}_{n,recon}) * N^{recon}_{n,recon}
\end{split}
\end{equation}

\newpage
%\subsection*{Supplementary Figures}
\subsection{Supplementary Notes}

\subsubsection{Supplementary Note 1}
\paragraph{Additional renewable drought illustrations}

\begin{figure}[H]
    \centering
    \includegraphics[width=\linewidth,height=\textheight, keepaspectratio]{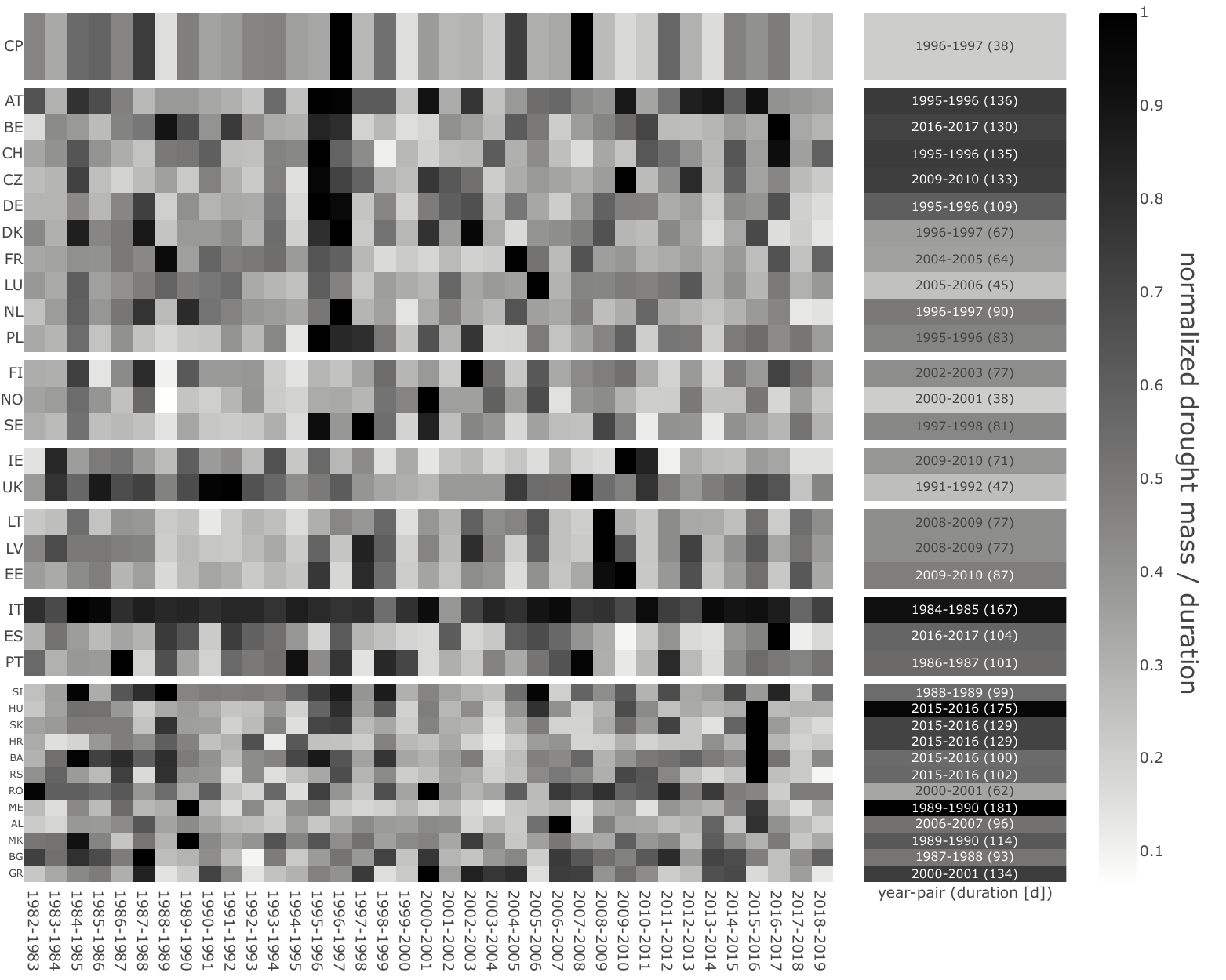}
    \caption{\textbf{Drought mass of identified most extreme simulated winter drought events.} For each country or the European copperplate (CP) in the left panel, drought mass scores are normalized using the row-specific maximum. The colors of the right panel indicate the maximum duration of the event with the highest drought mass score normalized by the column-specific maximum, i.e.,~the maximum duration across all countries. The right panel also provides the year-pair with the most severe events as identified by the drought mass score per row and its corresponding duration in days.}
    \label{fig:figure_si1}
\end{figure}

The duration and severity of most extreme winter droughts captured by the drought mass metric varies significantly across years and countries (Supplementary Fig.~\ref{fig:figure_si1}). Assuming perfect interconnection between all European countries, the most extreme event in the data occurred in the winter of 1996/97 and lasted 55~days. This European super drought was caused by pronounced and temporally overlapping events in many, yet not all, European countries (Figure~\ref{fig:figure_5}). Hence, even during this extreme event, geographical balancing remains possible to a limited extent. It is therefore substantially shorter than the most extreme droughts in nearly all countries when considered an energy island. Applying the drought mass metric to individual countries, we find the longest winter events in Eastern and Southern Europe. Further, smaller countries such as Slovenia (182~days, 2013/14) or Slovakia (182~days, 2015/16) tend to have longer extreme droughts than larger countries such as France (64~days, 2004/05), Sweden (83~days, 1997/98), Germany (109~days, 1995/96), or Spain (131~days, 1988/89). This is because smaller countries benefit less from geographical balancing within their borders.

% einschränkung abbildung: only CP considers the level of simultaneity of droughts, while country-specific droughts are computed as island. consequently, year-pair columns may appear darker because of more pronounced droughts in isolated countries, while drought in CP scenario is less pronounced because of the low temporal correlations of these drought events

\begin{figure}[H]
    \centering
    \includegraphics[width=\linewidth,height=\textheight, keepaspectratio]{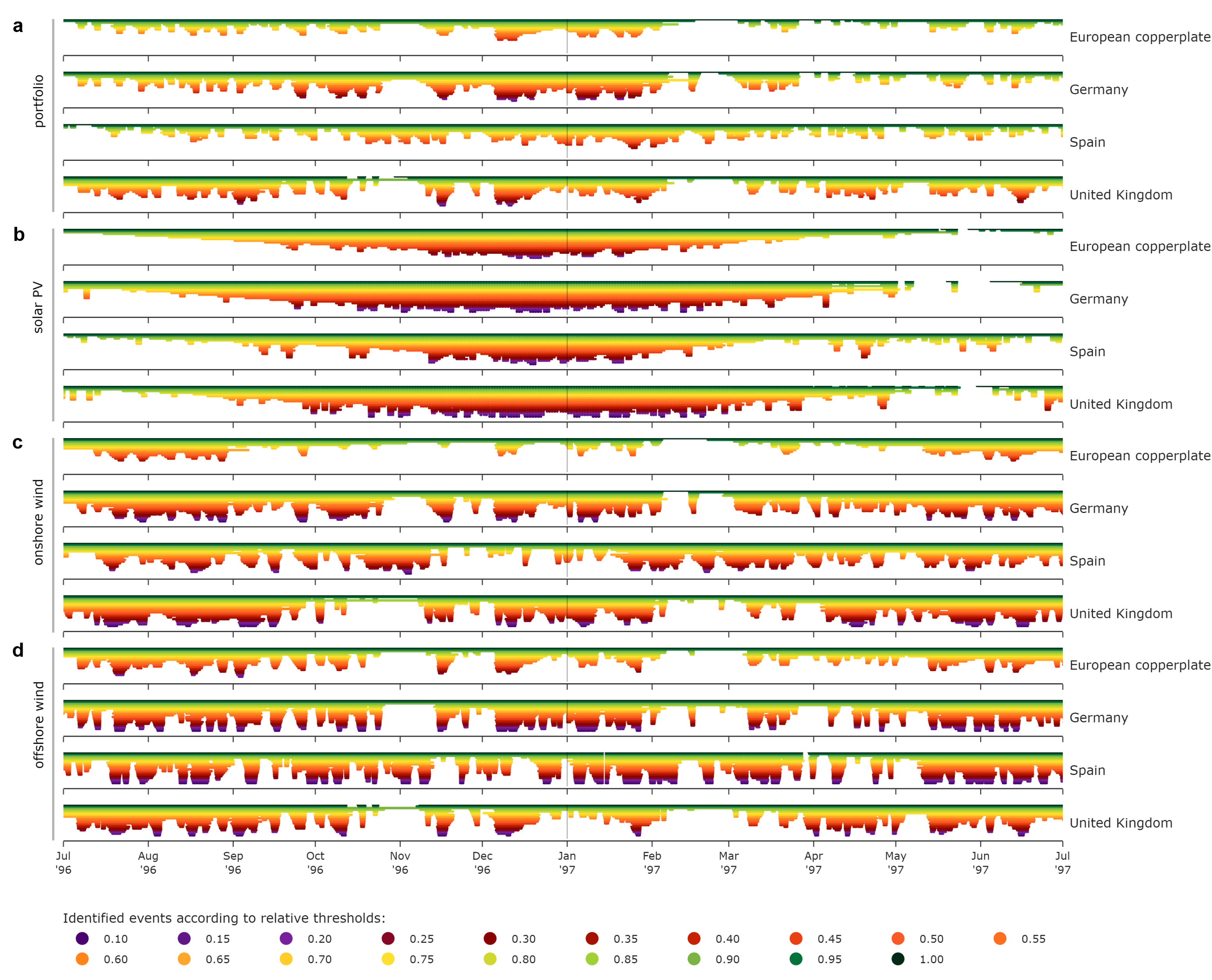}
    \caption{\textbf{Simulated drought patterns in 1996 and 1997 for all investigated relative thresholds $\tau \in [0.1, 1]$ and selected regions.} For each technology-specific panel, a horizontal band indicates drought occurrences for the color-coded threshold of one region. To illustrate persistent patterns, only droughts lasting longer than one day are displayed. Panel \textbf{a} corresponds to policy-oriented renewable technology portfolios, \textbf{b} to solar \acf{PV}, \textbf{c} to onshore wind, and \textbf{d} to offshore wind.}
    \label{fig:figure_si2}
\end{figure}

Supplementary Fig.~\ref{fig:figure_si2} illustrates drought patterns of solar \ac{PV}, onshore wind power and offshore wind power for three selected countries and the European copperplate for the year 1996/97. It can be seen that the largest droughts that emerge in renewable energy technology portfolios (panel a) are the result of coinciding drought events of individual technologies (panels b-d) in wintertime.

\subsubsection{Supplementary Note 2}
\paragraph{Additional insights on the correlation between renewable droughts and long-duration storage energy}
\label{ssec:additional_regression}

\begin{figure}[H]
    \centering
    \includegraphics[width=\linewidth,height=\textheight, keepaspectratio]{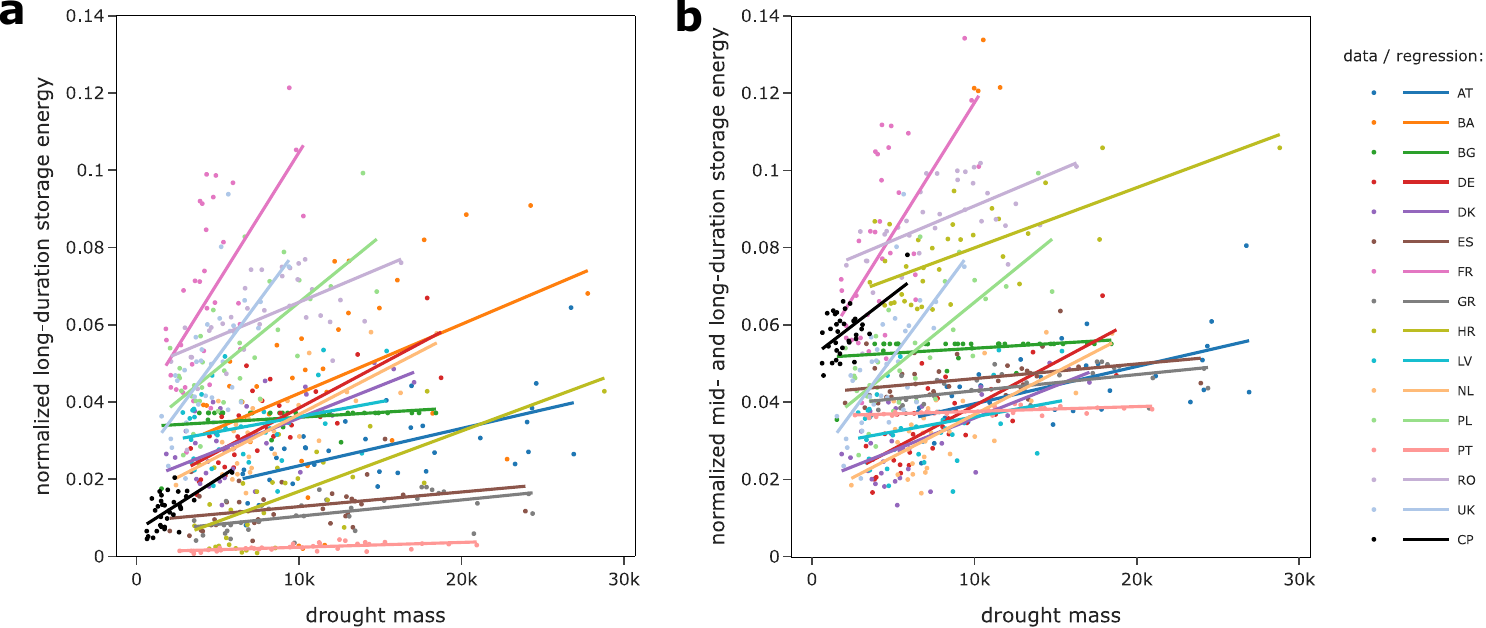}
    \caption{\textbf{Correlation of winter drought mass and different types of storage energy normalized by annual demand for electricity and hydrogen across years.} For comparison, we normalized the least-cost storage energy by annual demand for electricity (including electrified heating) and hydrogen. For illustration, we exclude countries with negligible storage energy or storage energy at potential but include the pan-European copperplate scenario (CP). \textbf{a} Long-duration storage only. \textbf{b} Mid- and long-duration storage.}
    \label{fig:figure_si3}
\end{figure}

Supplementary Fig.~\ref{fig:figure_si3} compares the correlation between winter droughts and long-duration storage needs, considering either long-duration storage only, or hydro and long-duration storage technologies combined. Including the latter in the regression shows a substantial level effect for countries with low regression coefficients, i.e.,~low sensitivity of long-duration storage needs to increasingly severe droughts, visible by the upward shift of the respective regression lines in Supplementary Fig.~\ref{fig:figure_si3}b compared to Supplementary Fig.~\ref{fig:figure_si3}b. This indicates that mid-term flexibility options can substitute long-duration storage needs for dealing with extreme droughts to a significant extent.

\begin{figure}[htbp]
    \centering
    \includegraphics[width=0.55\textwidth]{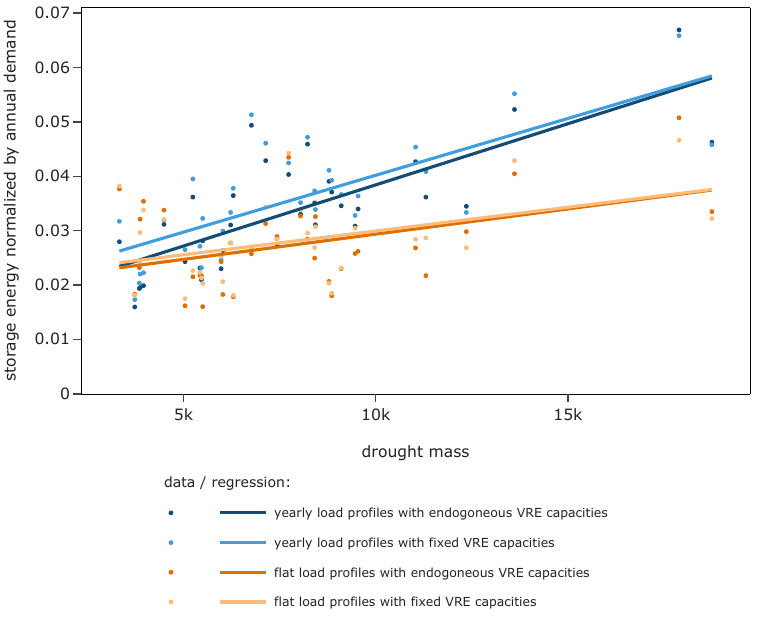}
    \caption{\textbf{Correlation of the drought mass of most extreme winter drought events and normalized storage energy capacity in Germany.} For comparison, we normalized the least-cost storage energy by annual demand for electricity (including electrified heating) and hydrogen. The renewable portfolio assumptions of the scenarios with fixed \ac{VRE} capacities align with those used for the time series-based \ac{VRE} drought analysis. In contrast, the scenarios with endogenous \ac{VRE} capacities are optimized by our power sector model.}
    \label{fig:figure_si4}
\end{figure}

Supplementary Fig.~\ref{fig:figure_si4} shows additional regression results for a complementary Germany-only setting using flat electricity demand profiles to exemplarily disentangle the storage-defining effect of renewable droughts from the storage-driving effect of demand seasonality (compare blue and orange lines in Supplementary Fig.~\ref{fig:figure_si4}). The slope of the regression line decreases with flatter electricity demand profiles, i.e.,~the storage-defining effect of droughts is less pronounced without demand seasonality.

\begin{figure}[htbp]
    \centering
    \includegraphics[width=\linewidth,height=\textheight, keepaspectratio]{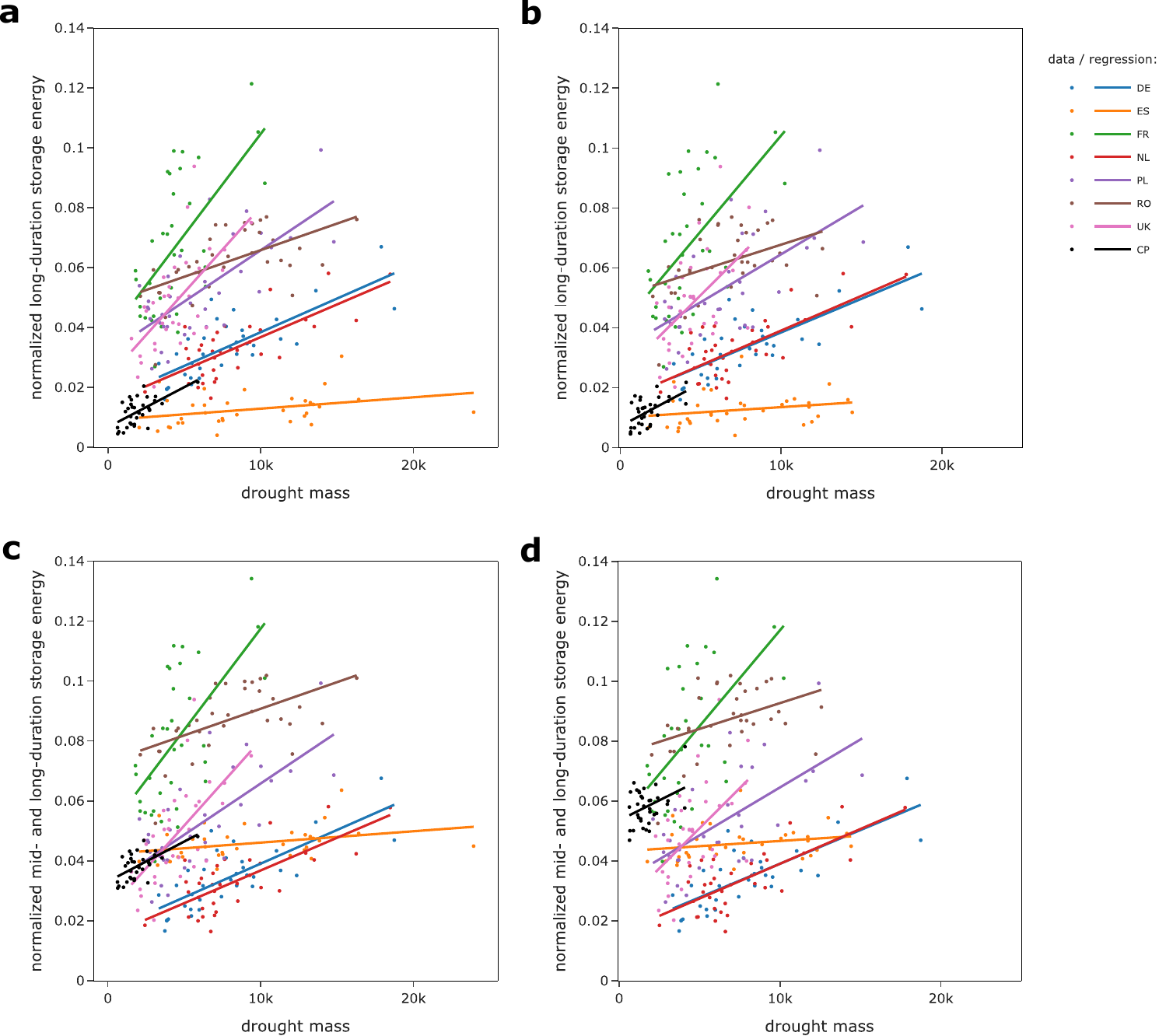}
    \caption{\textbf{Correlation of the drought mass of most extreme winter drought events and normalized storage energy capacity.} The illustrations on the left-hand side represent our default setting, which uses the wind and solar \ac{PV} capacity mix of the \ac{TYNDP} 2022 ``Distributed Energy'' scenario for drought mass computation. The regressions on the right-hand side are based on wind and solar capacity assumptions from the \ac{TYNDP} 2022 ``Global Ambition'' scenario. For comparison, we normalize the least-cost storage energy with the annual demand for electricity (including electrified heating) and hydrogen. For illustration, we exclude countries with least-cost storage energy below 5~TWh or investment at potential but include the pan-European copperplate scenario (CP). \textbf{a} Long-duration storage only (default). \textbf{b} Long-duration storage only (alternative capacity mix). \textbf{c} Mid- and long-duration storage (default). \textbf{d} Mid- and long-duration storage (alternative capacity mix).}
    \label{fig:figure_si5}
\end{figure}

In an additional sensitivity analysis, we assess how an alternative capacity mix of wind and solar \ac{PV} across Europe affects drought mass results and how this relates to least-cost long-duration storage outcomes. We compare the correlation between long-duration storage energy and drought mass values for ``Distributed Energy'' (default) and the ``Global Ambition'' scenarios of the \ac{TYNDP} 2022 (Supplementary Fig.~\ref{fig:figure_si5}). The ``Distributed Energy'' scenario features more solar \ac{PV}, while the ``Global Ambition'' scenario includes a higher share of offshore wind. In the latter, the drought mass of the most extreme events tends to decrease in some countries and the copperplate scenario. This is due to the lower solar \ac{PV} share, which alleviates the impact of solar seasonality on compound portfolio droughts in winter. However, regression slopes hardly change, indicating a limited impact of the alternative capacity mix on drought mass results. The most extreme drought for the copperplate scenario occurs in both scenarios in the winter of 1996/97.

\newpage

The scatter plots in Figure~\ref{fig:figure_2} indicate that the correlation between the most extreme winter drought event and the least-cost long-duration storage size for the same year is not perfect. Several factors explain this. First, country-specific factors such as varying portfolios of variable or firm renewable generation capacity or flexibility options, e.g.,~high shares of reservoir power plants in the Spanish capacity mix, may cause the imperfect fit. Second, demand peaks vary substantially between weather years in terms of level and temporal variation, especially in winter (compare France in Figure~\ref{fig:figure_3}). This means that similar droughts can trigger different storage needs, depending on the load situation. Third, our measurement of drought events is, by design, purely based on renewable availability time series and does not consider pre- or succeeding periods of very high availability or the seasonality of electricity demand. In contrast, the power sector optimization factors in these aspects. In addition, the drought mass metric based on the \ac{VMBT} method relies on the choice of a cut-off threshold and is solely an approximation of the cumulative energy deficit of drought events \cite{kittel_measuring_2024_si}, which is relevant for long-duration storage needs. Due to the averaging mechanism of the \ac{VMBT} method, this metric tends to underestimate solar \ac{PV} contributions within extreme droughts, particularly in countries with a less pronounced solar seasonality. These contributions generally lead to higher deployment of short-duration system flexibility and lower long-duration storage needs. Finally, \ac{VRE} portfolios differ to some extent between the drought analysis and the power sector model. While these portfolios are fixed in the former, they are endogenously optimized in the latter, yielding slightly different capacity mixes between weather years. Yet, the overall fit between the indicators appears to be reasonable, which can also be confirmed by a complementary Germany-only analysis (compare dark and bright lines in Supplementary Fig.~\ref{fig:figure_si4}).

\subsubsection{Supplementary Note 3}
\paragraph{Additional illustrations of the impact of interconnection on long-duration storage energy}

\begin{figure}[H]
\centering
    \includegraphics[width=\textwidth]{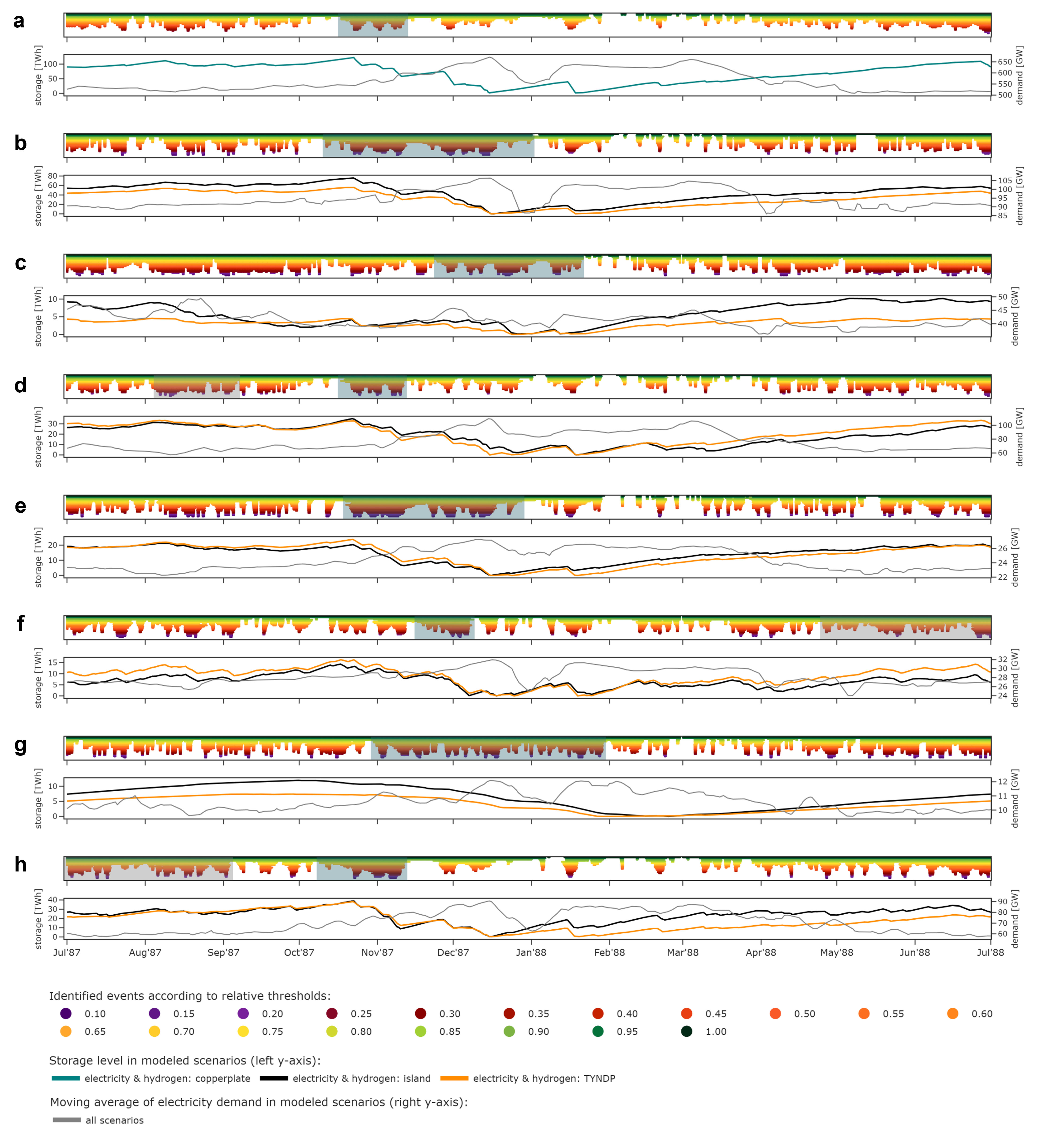}
    \caption{\textbf{Simulated drought events, electricity demand, and least-cost state-of-charge of long-duration storage in winter 1987/88 in countries with highest long-duration storage energy capacities.} The top row of each panel shows the identified drought patterns lasting longer than 12 hours across all color-coded thresholds, with the most extreme drought events occurring in winter (teal boxes) or throughout the year (gray boxes). The bottom row of each panel displays the associated exogenous smoothed demand profiles used in the optimization and the resulting least-cost storage state-of-charge levels for isolated countries modeled within the interconnection scenario (1), for policy-oriented interconnection levels in scenario (3), or the pan-European copperplate in scenario (4). Panel \textbf{a} corresponds to the pan-European copperplate scenario, \textbf{b} to Germany, \textbf{c} to Spain, \textbf{d} to France, \textbf{e} to the Netherlands, \textbf{f} to Poland, \textbf{g} to Romania, and \textbf{h} to the United Kingdom.}
    \label{fig:figure_si7}
\end{figure}

\begin{figure}[htbp]
    \centering\includegraphics[width=.99\textwidth,keepaspectratio]{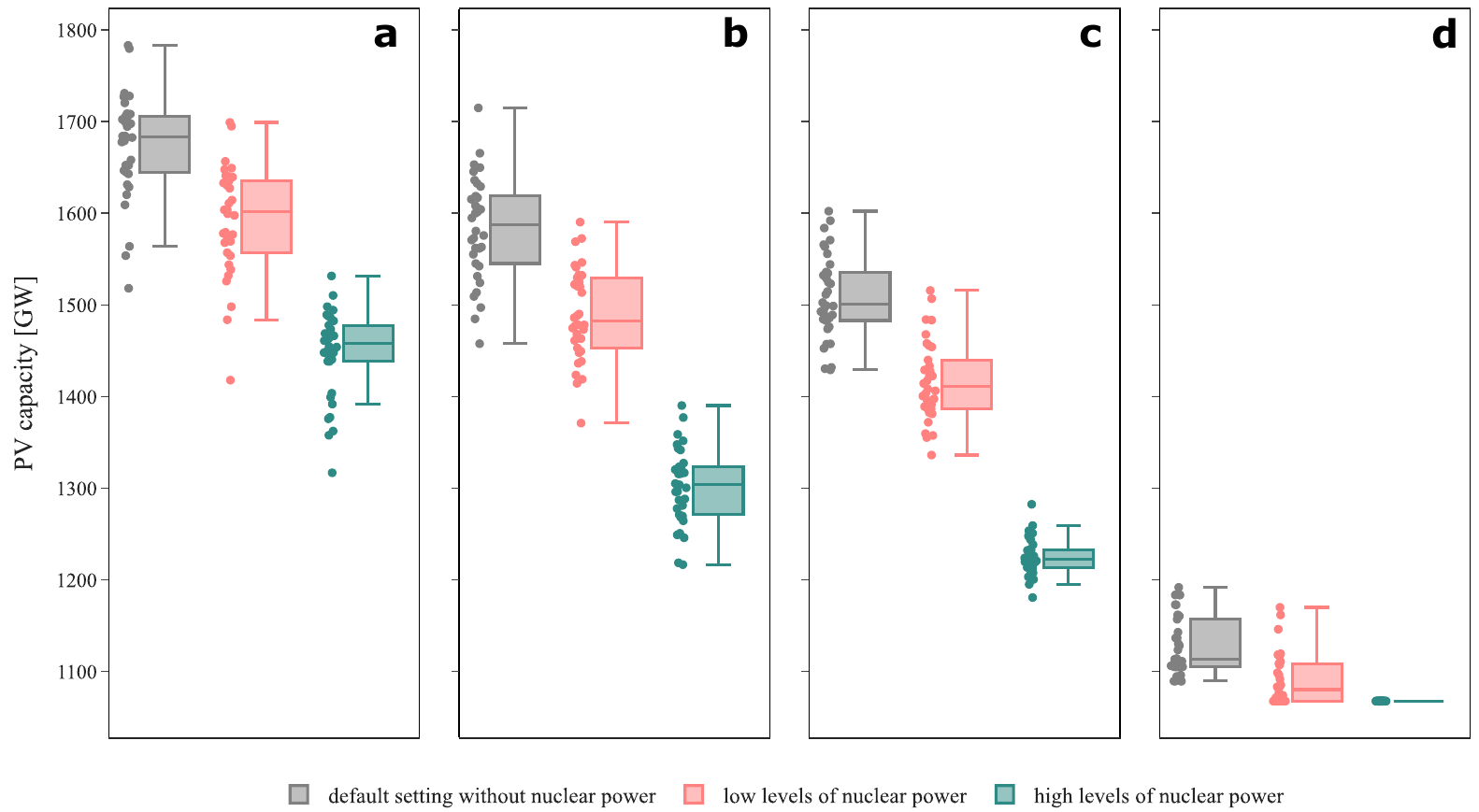}
    \caption{\textbf{Least-cost solar \ac{PV} capacity aggregated across all countries for all modeled weather years and interconnection scenarios.} Each dot refers to one independently modeled weather year. The center line denotes the median, box limits indicate the interquartile range (Q1–Q3), whiskers extend to 1.5× the interquartile range, and points beyond the whiskers represent outliers. \textbf{a} Scenario (1): no exchange of electricity nor hydrogen (island systems). \textbf{b} Scenario (2): policy-oriented exchange of electricity, no exchange of hydrogen. \textbf{c} Scenario (3): policy-oriented exchange of electricity and hydrogen. \textbf{d} Scenario (4): pan-European copperplate assuming unconstrained exchange of electricity and hydrogen.}
    \label{fig:figure_si9}
\end{figure}

\begin{figure}[htbp]
    \centering\includegraphics[width=.99\textwidth,keepaspectratio]{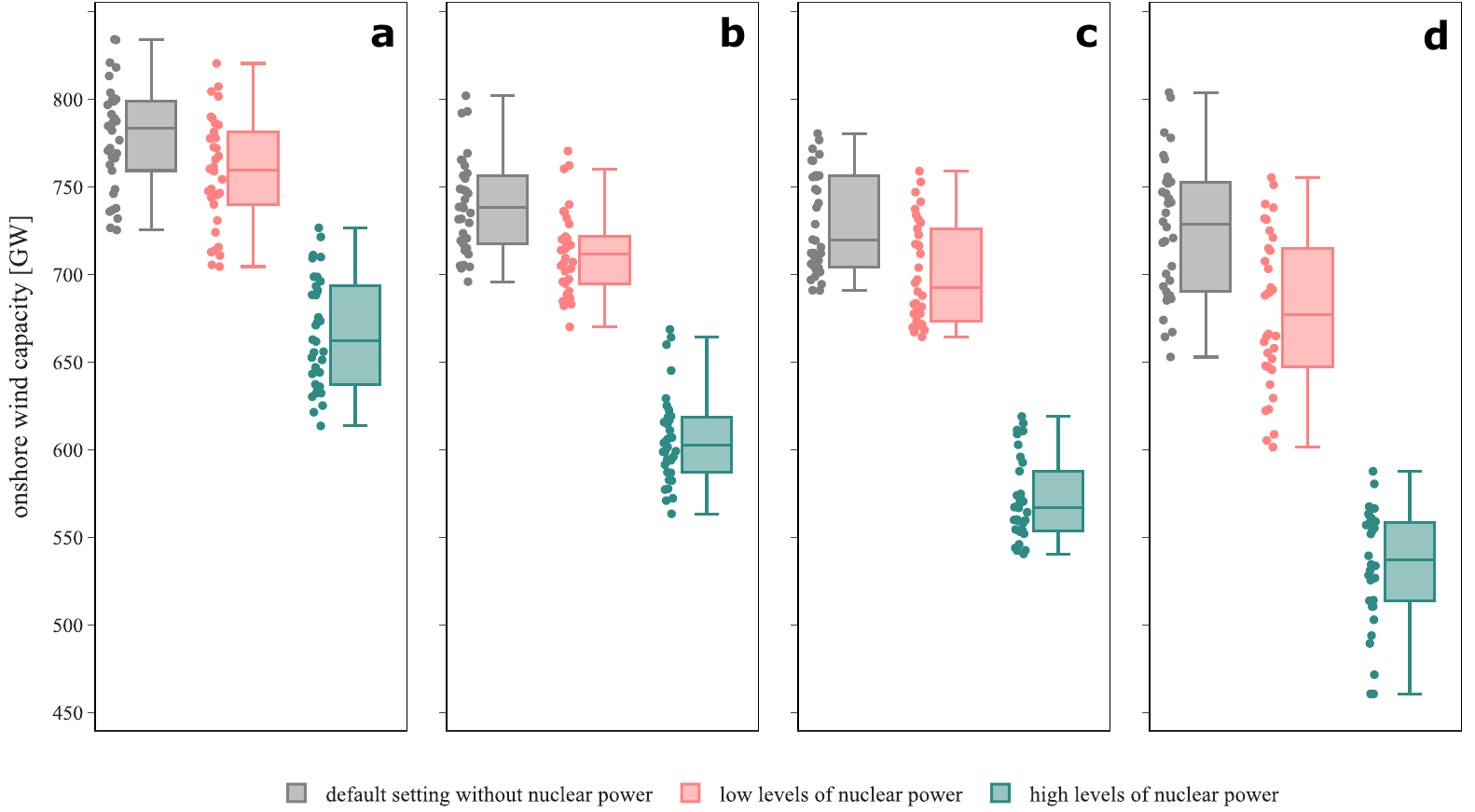}
    \caption{\textbf{Least-cost onshore wind capacity aggregated across all countries for all modeled weather years and interconnection scenarios.} Each dot refers to one independently modeled weather year. The center line denotes the median, box limits indicate the interquartile range (Q1–Q3), whiskers extend to 1.5× the interquartile range, and points beyond the whiskers represent outliers. \textbf{a} Scenario (1): no exchange of electricity nor hydrogen (island systems). \textbf{b} Scenario (2): policy-oriented exchange of electricity, no exchange of hydrogen. \textbf{c} Scenario (3): policy-oriented exchange of electricity and hydrogen. \textbf{d} Scenario (4): pan-European copperplate assuming unconstrained exchange of electricity and hydrogen.}
    \label{fig:figure_si10}
\end{figure}

\begin{figure}[htbp]
    \centering\includegraphics[width=.99\textwidth,keepaspectratio]{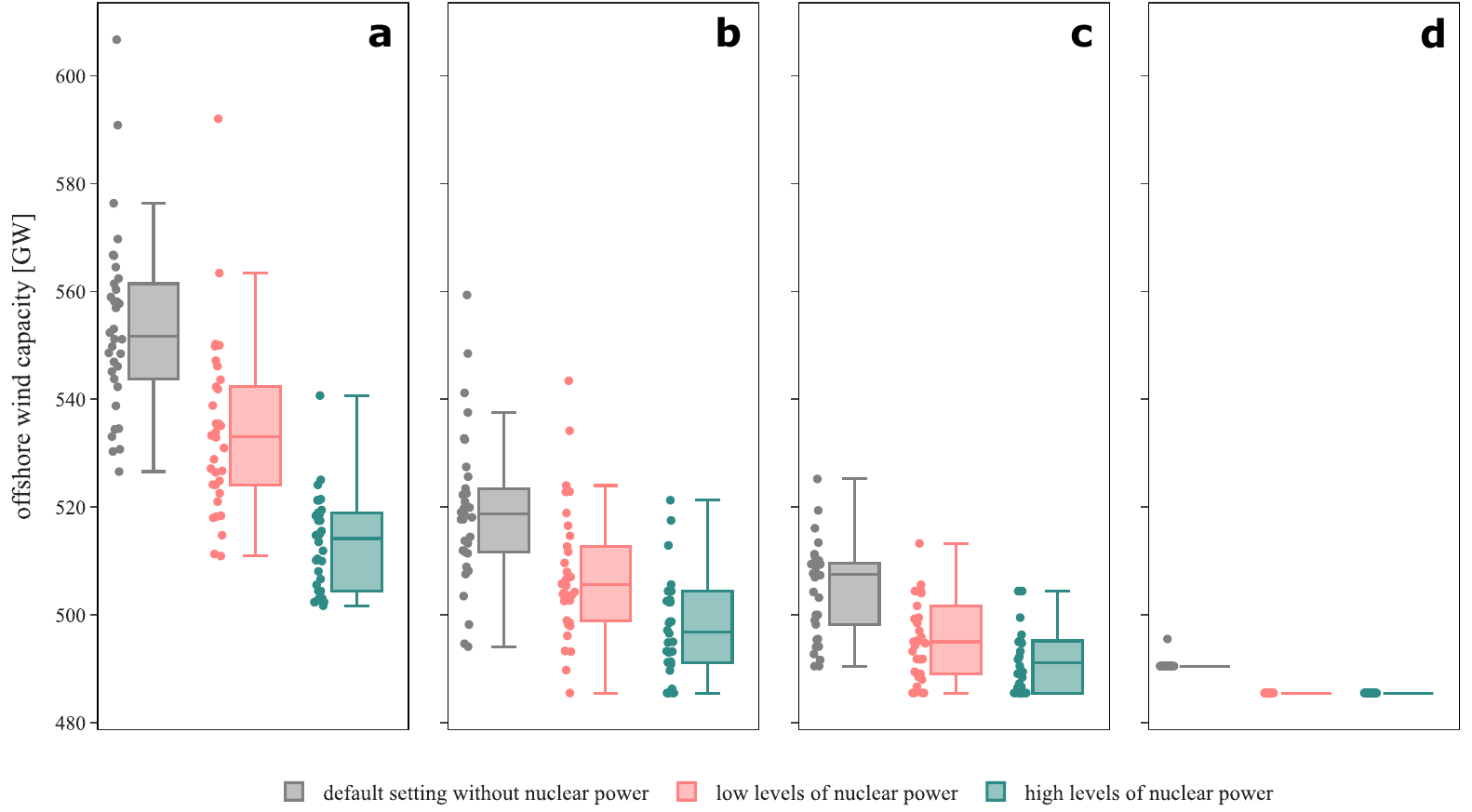}
    \caption{\textbf{Least-cost offshore capacity aggregated across all countries for all modeled weather years and interconnection scenarios.} Each dot refers to one independently modeled weather year. The center line denotes the median, box limits indicate the interquartile range (Q1–Q3), whiskers extend to 1.5× the interquartile range, and points beyond the whiskers represent outliers. \textbf{a} Scenario (1): no exchange of electricity nor hydrogen (island systems). \textbf{b} Scenario (2): policy-oriented exchange of electricity, no exchange of hydrogen. \textbf{c} Scenario (3): policy-oriented exchange of electricity and hydrogen. \textbf{d} Scenario (4): pan-European copperplate assuming unconstrained exchange of electricity and hydrogen.}
    \label{fig:figure_si11}
\end{figure}

The weather years 1987/88 and 1988/89 exhibit similar long-duration energy storage capacities for the energy island scenario (Figure~\ref{fig:figure_4}). For increasing interconnection levels, least-cost storage capacity diverges, yielding substantially lower energy storage levels for 1988/89 compared to 1987/88. This is because the most pronounced renewable droughts are temporally highly correlated with each other and, to some extent, also with high-demand periods in 1987/88 across many countries. In contrast, this is not the case in 1988/89 (compare boxes in Supplementary Fig.~\ref{fig:figure_si6} and Supplementary Fig.~\ref{fig:figure_si7}), enabling more pronounced geographical balancing of these drought events. The temporal overlap is even more pronounced in weather year 1996/97 (Figure~\ref{fig:figure_5}), which leads to the most extreme pan-European drought in the data (Supplementary Fig.~\ref{fig:figure_si1}) and, accordingly, the highest long-duration storage needs (Figure~\ref{fig:figure_4}). Supplementary Fig.~\ref{fig:figure_si9}, Supplementary Fig.~\ref{fig:figure_si10}, and Supplementary Fig.~\ref{fig:figure_si11} show the impact of additional nuclear power on the least-cost deployment of solar \ac{PV}, onshore wind, and offshore wind. Note that in the default setting with renewable generation only, which bases on the assumptions of the ``Distributed Energy'' scenario of the \ac{TYNDP}~2022~\cite{entso-e_tyndp2022_2022_si}, we adjusted the expansion bounds for variable renewable technologies accounting for the missing generation of nuclear power. This results in slightly increased lower expansion bounds compared to the complementary runs including nuclear power, which are binding for \ac{PV} and offshore wind in scenario (4).

\begin{figure}[H]
\centering
    \includegraphics[width=\textwidth]{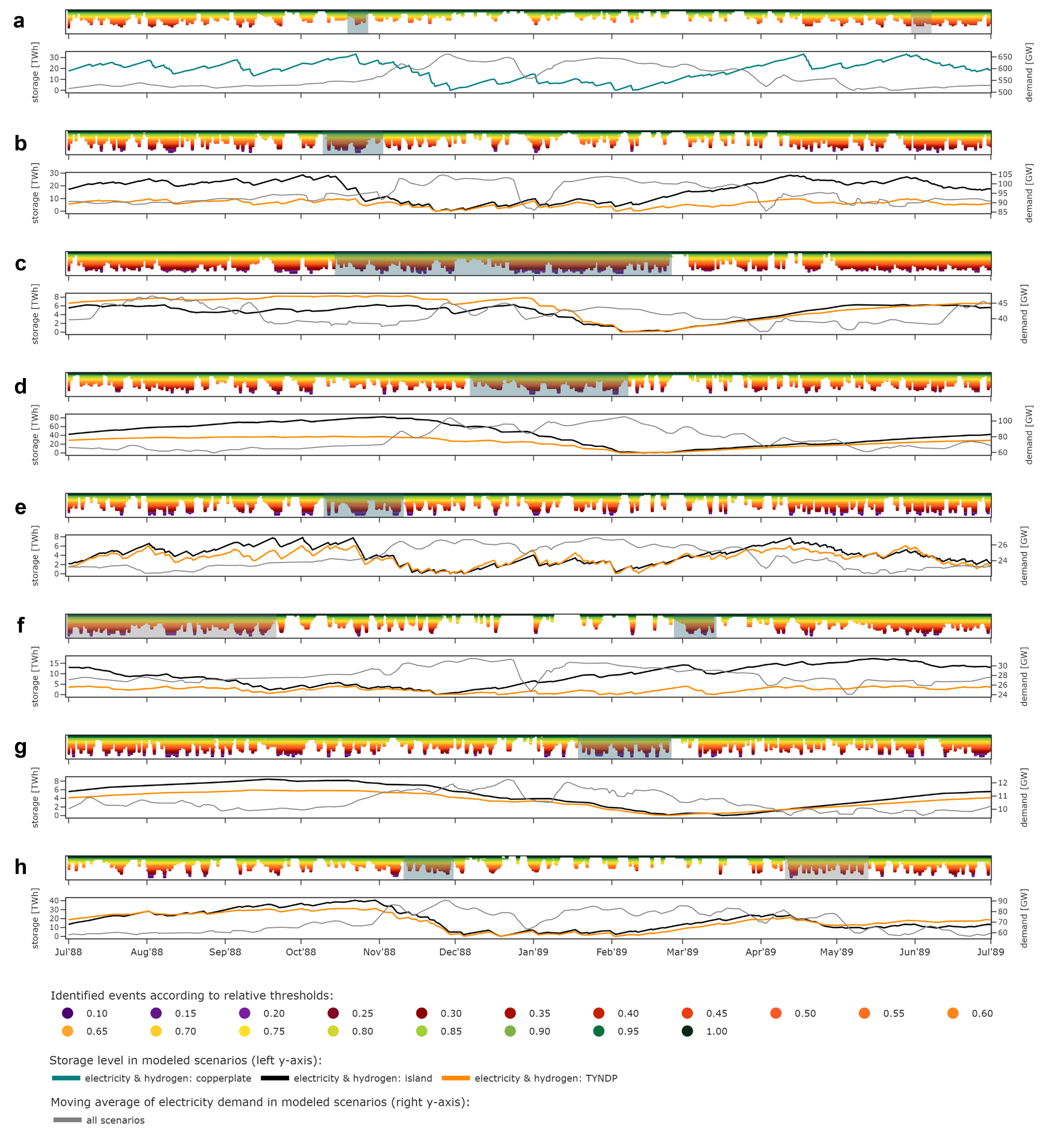}
    \caption{\textbf{Simulated drought events, electricity demand, and least-cost state-of-charge of long-duration storage in winter 1988/89 in countries with highest long-duration storage energy capacities.} The top row of each panel shows the identified drought patterns lasting longer than 12 hours across all color-coded thresholds, with the most extreme drought events occurring in winter (teal boxes) or throughout the year (gray boxes). The bottom row of each panel displays the associated exogenous smoothed demand profiles used in the optimization and the resulting least-cost storage state-of-charge levels for isolated countries modeled within the interconnection scenario (1), for policy-oriented interconnection levels in scenario (3), or the pan-European copperplate in scenario (4). Panel \textbf{a} corresponds to the pan-European copperplate scenario, \textbf{b} to Germany, \textbf{c} to Spain, \textbf{d} to France, \textbf{e} to the Netherlands, \textbf{f} to Poland, \textbf{g} to Romania, and \textbf{h} to the United Kingdom.}
    \label{fig:figure_si6}
\end{figure}

\begin{figure}[H]
    \centering
    \includegraphics[width=.99\textwidth,keepaspectratio]{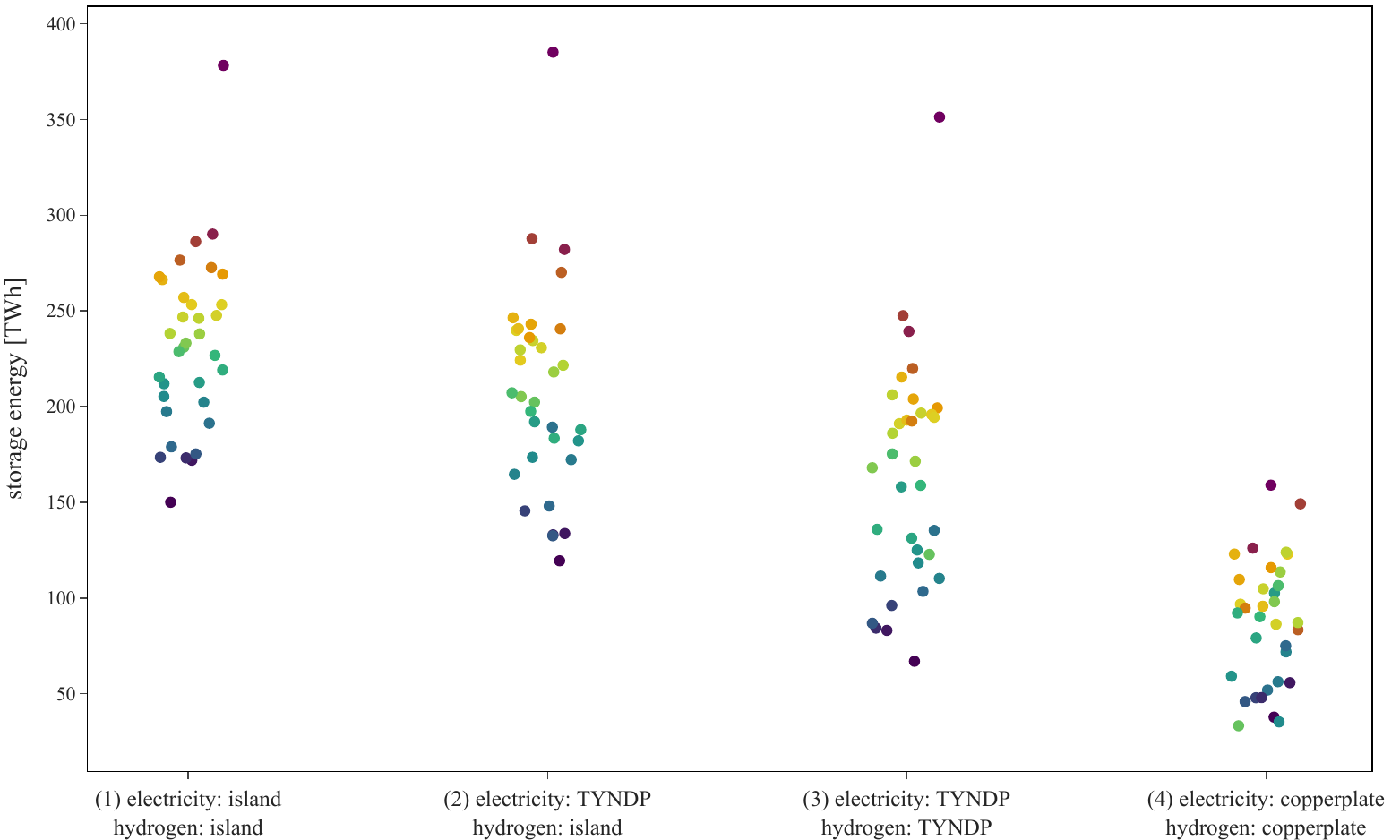}
    \caption{\textbf{Long-duration storage energy capacity aggregated across all countries for all modeled interconnection scenarios.} Every point refers to a single weather year. The coloring changes continuously in long-duration storage energy capacity according to the ranking of weather years in scenario (1). The weather year color remains consistent across the other scenarios.}
    \label{fig:figure_si8}
\end{figure}

Generally, interconnection mitigates long-duration storage energy needs (Figure~\ref{fig:figure_4}). Overall, the ranking of weather years is relatively persistent, yet the storage-mitigating effect varies between weather years (Supplementary Fig.~\ref{fig:figure_si8}), especially for higher levels of interconnection. The mechanisms behind this effect, in particular temporally correlated severe drought events, are illustrated above.

\begin{figure}[H]
    \centering
    \includegraphics[width=\linewidth]{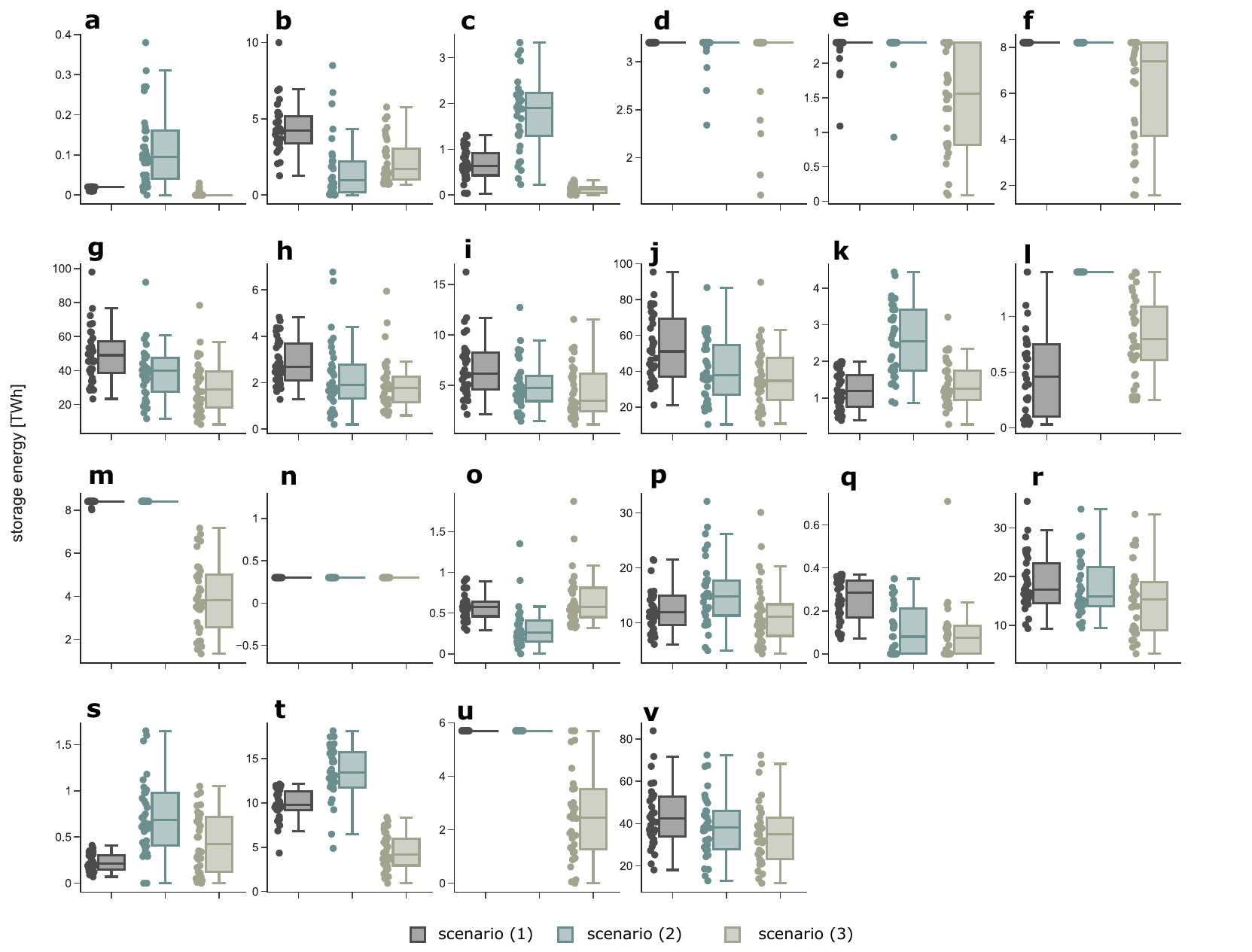}
    \caption{\textbf{Long-duration storage energy capacities across all countries with storage potential across all weather years and interconnection scenarios.} Each dot refers to one independently modeled weather year. The center line denotes the median, box limits indicate the interquartile range (Q1–Q3), whiskers extend to 1.5× the interquartile range, and points beyond the whiskers represent outliers. Note that the vertical axis are scaled for each country for illustration. Many of the effects in countries with smaller power sectors are minor relative to the aggregated effects discussed in the main body of this work. The pan-European copperplate scenario is omitted as the geographical distribution of long-duration storage capacities is arbitrary and hence results cannot be meaningful interpreted. \textbf{a}~Algeria. \textbf{b}~Austria. \textbf{c}~Bosnia-Herzegovina. \textbf{d}~Belgium. \textbf{e}~Bulgaria. \textbf{f}~Czech Republic. \textbf{g}~Germany. \textbf{h}~Denmark. \textbf{i}~Spain. \textbf{j}~France. \textbf{k}~Greece. \textbf{l}~Croatia. \textbf{m}~Hungary. \textbf{n}~Italy. \textbf{o}~Latvia. \textbf{p}~the Netherlands. \textbf{q}~Norway. \textbf{r}~Poland. \textbf{s}~Portugal. \textbf{t}~Romania. \textbf{u}~Slovakia. \textbf{v}~the United Kingdom.}
    \label{fig:figure_si12}
\end{figure}

Increasing the interconnection capacities decreases the need for long-duration storage energy capacities (Figure~\ref{fig:figure_4}) on a aggregate European level, but also individually in most countries (Supplementary Fig.~\ref{fig:figure_si12}).

\subsubsection{Supplementary Note 4}
\paragraph{Additional information on the impact of firm zero-emission generation}
\label{ssec:substitution_si} 

\begin{figure}[htbp]
    \centering
    \includegraphics[width=\linewidth,height=\textheight, keepaspectratio]{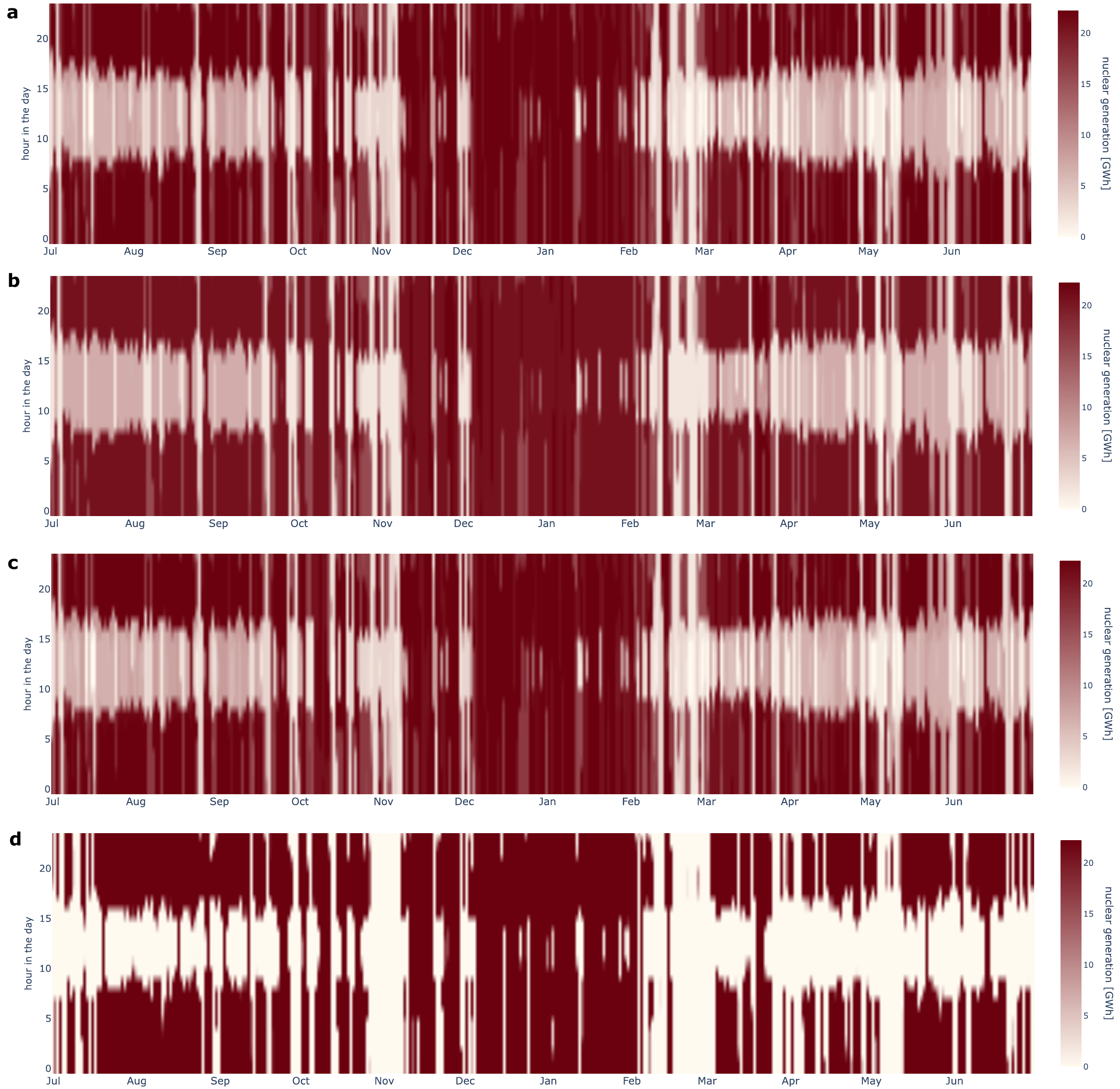}
    \caption{\textbf{Hourly and daily generation patterns of nuclear power aggregated across all countries in 1996/97 in the scenario with low levels of nuclear power.} \textbf{a} Scenario (1): no exchange of electricity or hydrogen. \textbf{b} Scenario (2): policy-oriented exchange of electricity. \textbf{c} Scenario (3): policy-oriented exchange of electricity and hydrogen. \textbf{d} Scenario (4): unconstrained exchange of electricity or hydrogen.}
    \label{fig:figure_si13}
\end{figure}

\begin{figure}[htbp]
    \centering
    \includegraphics[width=\linewidth,height=\textheight, keepaspectratio]{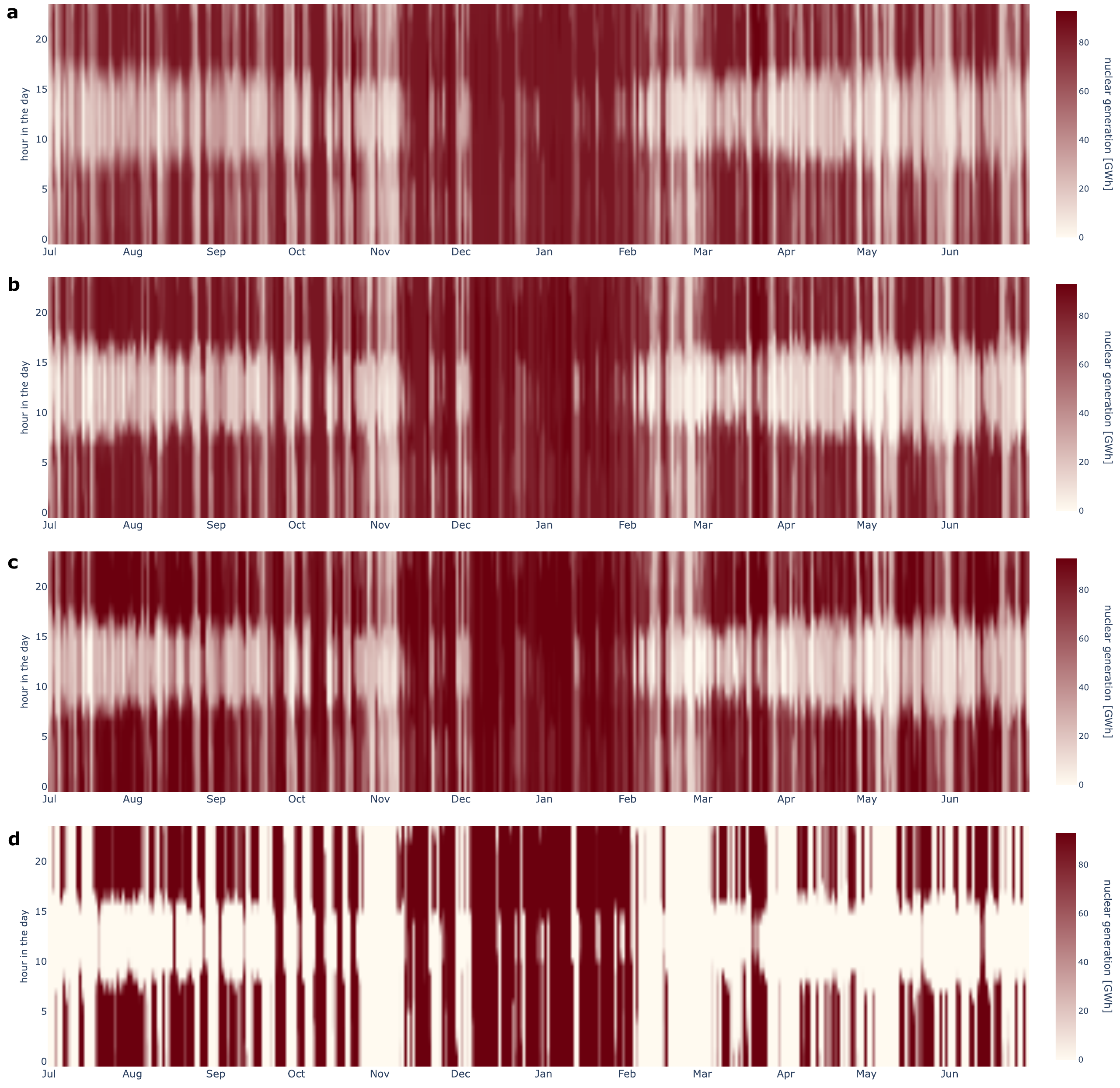}
    \caption{\textbf{Hourly and daily generation patterns of nuclear power aggregated across all countries in 1996/97 in the scenario with high levels of nuclear power.} \textbf{a} Scenario (1): no exchange of electricity or hydrogen. \textbf{b} Scenario (2): policy-oriented exchange of electricity. \textbf{c} Scenario (3): policy-oriented exchange of electricity and hydrogen. \textbf{d} Scenario (4): unconstrained exchange of electricity or hydrogen.}
    \label{fig:figure_si14}
\end{figure}

\begin{figure}[htbp]
    \centering
    \includegraphics[width=\linewidth,height=\textheight, keepaspectratio]{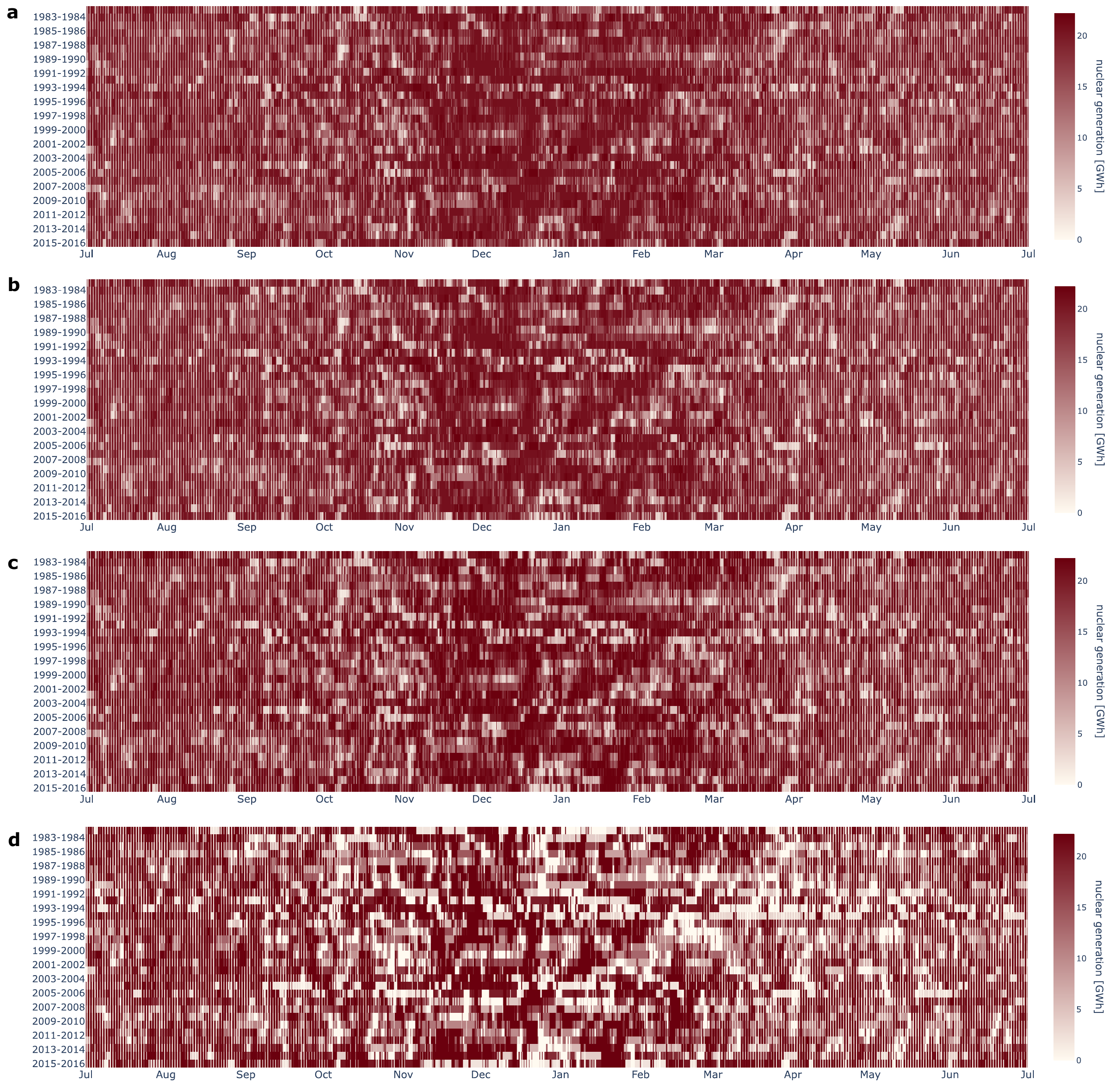}
    \caption{\textbf{Hourly and yearly generation patterns of nuclear power aggregated across all countries for all weather years in the scenarios with low levels of nuclear power.} \textbf{a} Scenario (1): no exchange of electricity or hydrogen. \textbf{b} Scenario (2): policy-oriented exchange of electricity. \textbf{c} Scenario (3): policy-oriented exchange of electricity and hydrogen. \textbf{d} Scenario (4): unconstrained exchange of electricity or hydrogen.}
    \label{fig:figure_si15}
\end{figure}

\begin{figure}[htbp]
    \centering
    \includegraphics[width=\linewidth,height=\textheight, keepaspectratio]{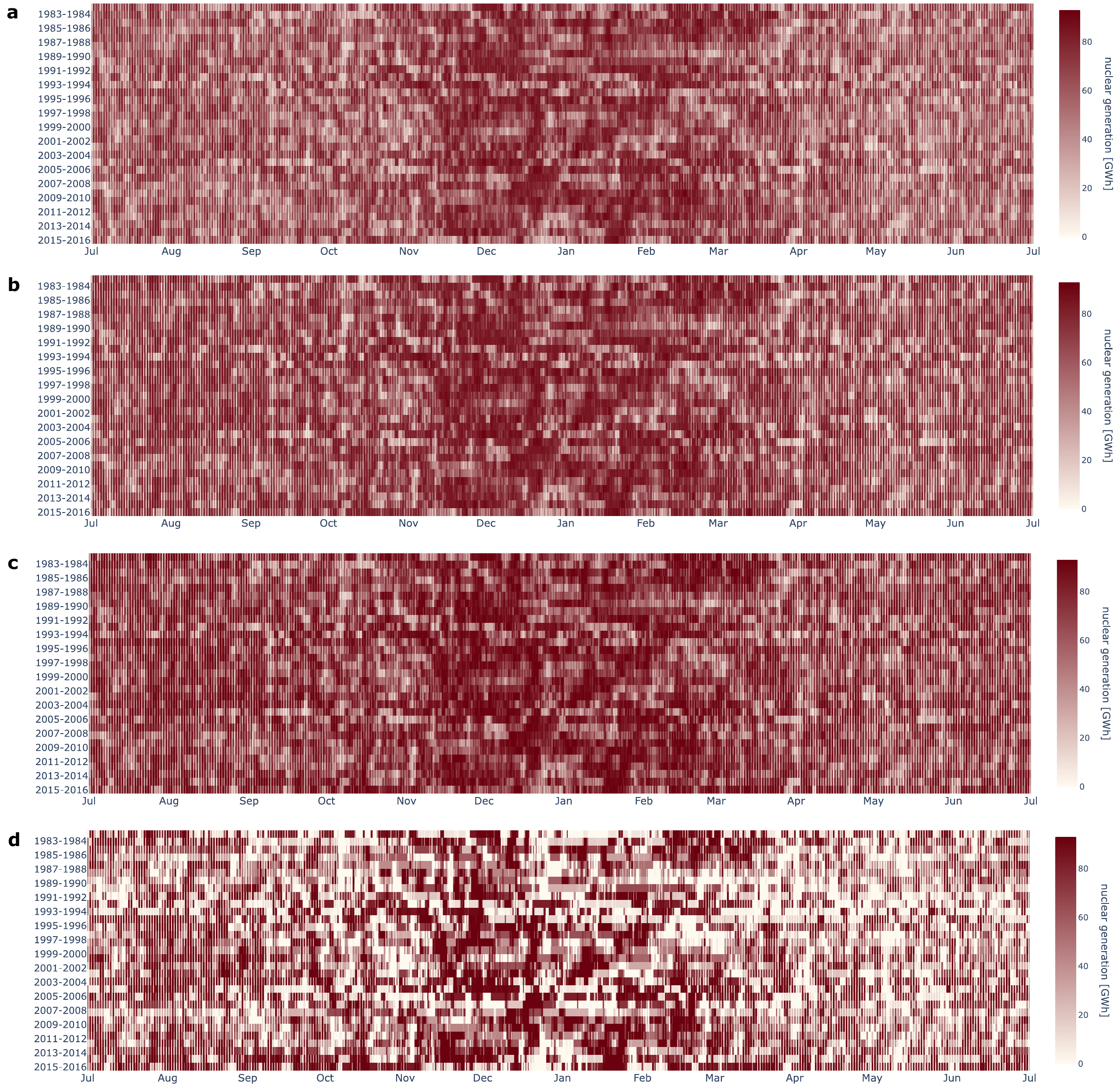}
    \caption{\textbf{Hourly and yearly generation patterns of nuclear power aggregated across all countries for all weather years in the scenarios with high levels of nuclear power.} \textbf{a} Scenario (1): no exchange of electricity or hydrogen. \textbf{b} Scenario (2): policy-oriented exchange of electricity. \textbf{c} Scenario (3): policy-oriented exchange of electricity and hydrogen. \textbf{d} Scenario (4): unconstrained exchange of electricity or hydrogen}
    \label{fig:figure_si16}
\end{figure}

% figure intro
Supplementary Fig.~\ref{fig:figure_si13} and Supplementary Fig.~\ref{fig:figure_si14} visualizes the hourly and daily nuclear generation aggregated across all countries in the winter of 1996/97 across all interconnection scenarios for low and high levels of nuclear power. Supplementary Fig.~\ref{fig:figure_si15} and Supplementary Fig.~\ref{fig:figure_si16} show the hourly operation of nuclear capacity aggregated across all countries for all interconnection scenarios and weather years for low and high levels of nuclear power.

% very high levels of zero-emission technologies in DE: intro
While limited in near-term scalability \cite{lee_ipcc_2023_si}, other firm zero-emission generation technologies may become viable in the longer run, such as advanced nuclear fission or fusion or advanced geothermal power generation. Such technologies are expected to have very high capital costs but low operational costs, which implies that they would optimally operate at very high full-load hours \cite{stoecker_2025_si}. Based on the weather year 1996/97, which includes a very pronounced renewable drought in many European countries, we analyze the impact of a generic dispatchable zero-emission technology on long-duration storage for the illustrative example of Germany, modeled as an energy island to limit the computation burden. In a series of 51~model runs, we iteratively increase the exogenous generation capacity of the zero-emission capacity by 1~GW increments as dispatch and investment decisions of all other generation and storage technologies remain endogenous. The firm zero-emission technology can continuously generate electricity not only during extreme droughts but also throughout the entire modeled weather year. This reduces the reliance on variable wind and solar power, which decreases the need for system flexibility. 

\begin{figure}[htbp]
    \centering
    \includegraphics[width=\linewidth]{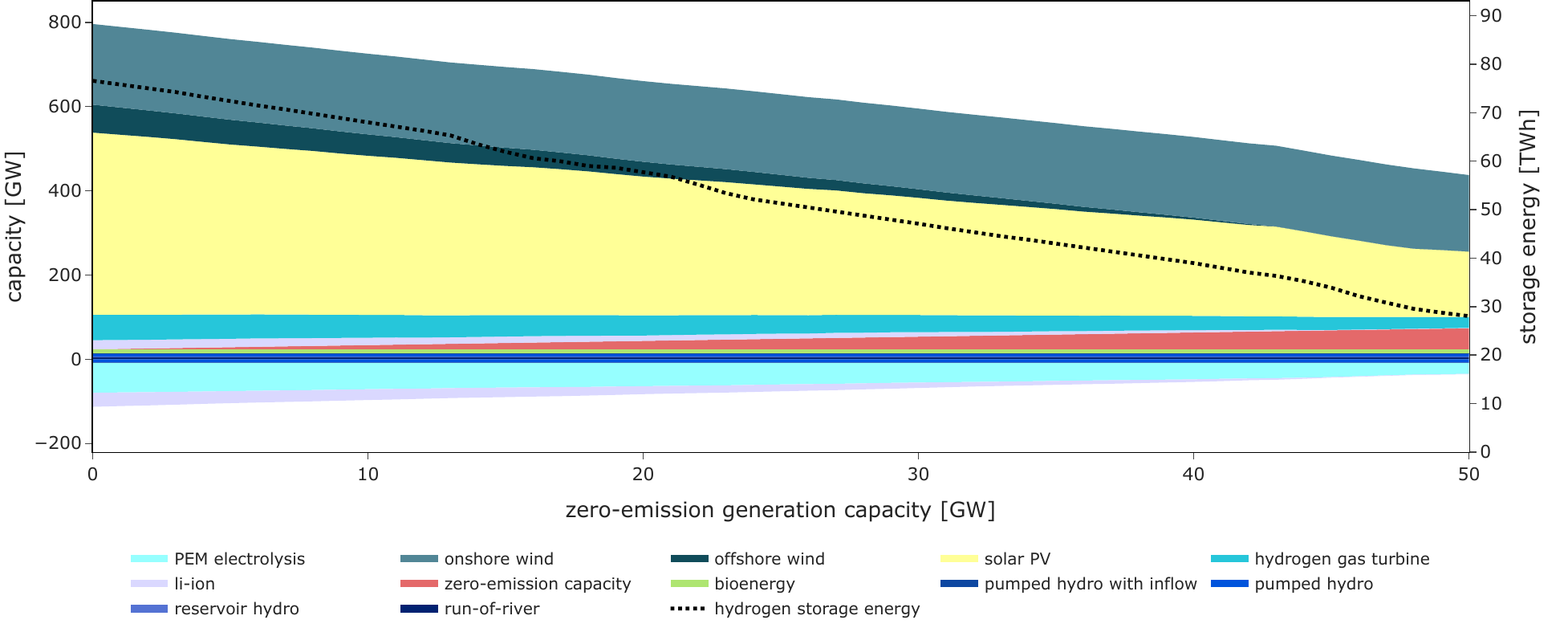}
    \caption{\textbf{Least-cost wind and solar capacity as well as short- and long-duration flexibility options for increasing zero-emission generation capacity in a Germany only setting.} The positive part of the left y-axis relates to generation and storage discharge, and its negative part to storage charge. The right y-axis refers to the long-duration storage energy. The expansion of solar \ac{PV} as well as on- and offshore wind is limited according to the upper bounds of the \ac{TYNDP} 2024, while we remove the lower expansion bounds.}
    \label{fig:figure_si17}
\end{figure}

% very high levels of zero-emission technologies in DE: results I
Supplementary Fig.~\ref{fig:figure_si17} illustrates these substitution effects. In addition to disproportionally displacing \ac{VRE} capacity due to higher full-load hours, the increasing zero-emission generation capacity also reduces the need for battery storage, hydrogen gas turbines, and long-duration storage energy capacity. Yet, long-duration storage remains required as long as variable renewables are still part of the energy mix. This holds true even for very high levels of generation capacity of the zero-emission technology, going far beyond the peak nuclear power capacity ever reached in Germany.
% "storage elasticity of firm cap" 

% very high levels of zero-emission technologies in DE: results II
The figure indicates a near-linear reduction of long-duration storage capacity for increasing levels of zero-emission generation capacity. Varying ratios of \ac{VRE} technologies in the capacity mixes explain irregularities in the negative slope. For instance, for a firm zero-emission capacity of more than 43~GW, the model abstains from deploying cost-intensive offshore wind power. Each additional gigawatt of the zero-emission technology has now a significantly higher substitution rate to onshore wind and solar \ac{PV} compared to scenarios with less capacity. This is because of the difference in full-load hours of these \ac{VRE} technologies. In Germany, offshore wind has typically around twice the full-load hours of onshore wind and four times as many as \ac{PV}. In scenarios without offshore wind, additional dispatchable capacity therefore replaces much more onshore wind and \ac{PV}, which causes a more pronounced decrease in long-duration energy storage capacity. 

\subsubsection{Supplementary Note 5}
\paragraph{Additional information on a sensitivity with varying values of lost load}
\label{ssec:voll_sensitivity_si}

\begin{figure}[t]
    \centering
    \includegraphics[width=\linewidth]{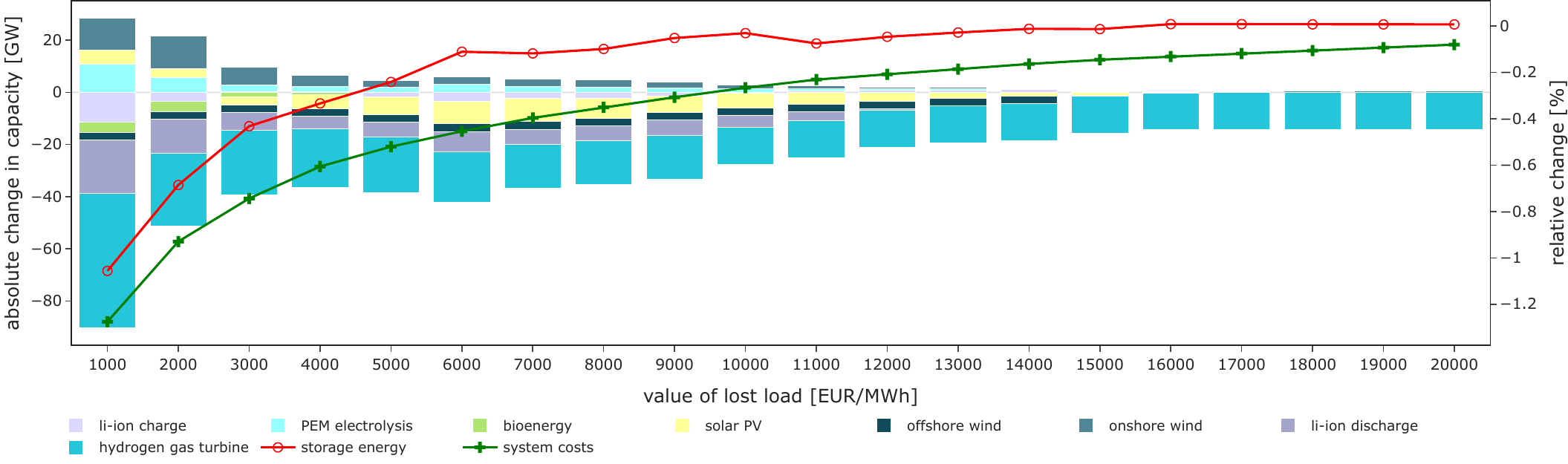}
    \caption{\textbf{Change in capacities and total system costs for varying assumptions on the value of lost load.} Absolute changes in capacities (bars, left y-axis) and relative change of long-duration energy storage capacity and total system costs (lines, right y-axis) aggregated across all countries in Europe in the interconnection scenario (3) compared to the default setting in 1996/97 for varying assumptions on the value of lost load. For readability, we show the zero line of the left y-axis in gray.}
    \label{fig:figure_si18}
\end{figure}

% impact of VOLL
Another option for dealing with extreme renewable droughts could be load shedding by system operators. This is considered as a last resort, as it can have substantial economic and societal implications. The value of lost load metric is often used to approximate the socio-economic costs of unmet electricity demand. Leaning on estimates by Kachirayil et al. \cite{kachirayil_trade-offs_2025}, Supplementary Fig.~\ref{fig:figure_si18} shows the impact load shedding at values of lost load ranging between 1,000 and 20,000 EUR per MWh for the interconnection scenario (3) using the weather year 1996/97. At lower values of lost load, load shedding mainly displaces long-duration discharging capacity and shorter-duration flexibility such as bioenergy or batteries. This effect diminishes for higher, and more plausible, values of lost load. Least-cost long-duration storage energy capacity and total systems costs prove to be very robust. They decrease between zero and 1\%, i.e., they are hardly affected by load shedding.

\newpage
\subsection{Supplementary Tables}

\subsubsection{Additional information on the cost sensitivity analysis}
\label{ssec:cost_sensitivity_si}

\begin{table}[h!]
\centering
\caption{Overnight investment costs in the default setting as well as the lower and upper end of the cost range investigated in sensitivity analyses.}
\label{tab:cost_assumptions}
\begin{adjustbox}{max width=\textwidth}
\begin{tabular}{l|cccc}
\toprule
 & default & variation range & lowest costs & highest costs \\
\midrule
Solar \ac{PV} [EUR/MW$_{el}$] & 305,600 & $\pm$ 50\% & 152,800 & 458,400 \\
Onshore wind [EUR/MW$_{el}$] & 920,000 &  $\pm$ 50\% & 460,000 & 1,380,000 \\
Offshore wind [EUR/MW$_{el}$ ] & 2,048,760 & $\pm$ 50\% & 1,024,380 & 3,073,140 \\
Long-duration storage [EUR/MWh$_{th}$]  & 1,276 & -50\%, +400\% & 638 & 5,104 \\
\bottomrule
\end{tabular}
\end{adjustbox}
\end{table}

\newpage

\renewcommand{\refname}{Supplementary References}  % For article class
%\putbib

% bibliograpy 2 directly copied into main file:

%\end{bibunit}
%\end{comment}

\end{document}